\documentclass[11pt]{article}

\makeindex

\date{}
\usepackage{fullpage}
\usepackage{amssymb}
\usepackage{amsmath}
\usepackage{amsthm}
\usepackage{multirow}
\usepackage{enumerate}
\usepackage{graphicx}
\usepackage{mathrsfs}
\usepackage[utf8]{inputenc}
\usepackage{cite}
\usepackage{hyperref}
\usepackage{cleveref}
\usepackage{mathtools}
\usepackage{amsfonts}
\usepackage[T1]{fontenc}
\usepackage{mathpazo}
\usepackage{bm}
\usepackage{isomath}
\usepackage{verbatim}
\usepackage{changepage}
\usepackage[usenames,dvipsnames,svgnames,table]{xcolor}

\definecolor{blueviolet}{RGB}{60,50,200}
\definecolor{oliveg}{RGB}{40,200,30}
\hypersetup{colorlinks=true,       
    linkcolor=blueviolet,
    citecolor=oliveg,
}

\usepackage{tikz}
\usepackage{tikz-cd}
\usetikzlibrary{calc}

\usepackage{algorithmicx,algorithm}
\usepackage{algpseudocode}

\usepackage{color}

\usepackage{todonotes}

\theoremstyle{definition}
\newtheorem{definition}{Definition}[section]

\theoremstyle{plain}
\newtheorem{lemma}{Lemma}[section]
\newtheorem{theorem}{Theorem}
\newtheorem{proposition}{Proposition}

\newtheorem{corollary}{Corollary}[section]
\newtheorem{claim}{Claim}[section]

\newcommand{\F}{\mathbb{F}}
\newcommand{\K}{\mathbb{K}}
\newcommand{\R}{\mathbb{R}}

\newcommand{\B}{\bf}

\makeatletter
\newcommand\footnoteref[1]{\protected@xdef\@thefnmark{\ref{#1}}\@footnotemark}
\makeatother

\def\DEBUG{true} 
	\ifdefined\DEBUG

	    \def\rem#1{{\marginpar{\raggedright\scriptsize #1}}}
	    \newcommand{\vin}[1]{\rem{\textcolor{Red}{$\bullet$ #1}}}
	    \newcommand{\vis}[1]{\rem{\textcolor{OliveGreen}{$\bullet$ #1}}}
	    \newcommand{\gau}[1]{\rem{\textcolor{NavyBlue}{$\bullet$ #1}}}
	\else

        \newcommand{\vin}[1]{}
	    \newcommand{\vis}[1]{}
	    \newcommand{\gau}[1]{}
	 
	\fi
\begin{document}

\title{Efficient reconstruction of depth three arithmetic circuits with top fan-in two}
 
\author{
{Gaurav Sinha\footnote{Adobe Research Bangalore, India, email: gasinha@adobe.com}}
}




\maketitle
\setcounter{page}{0}

\begin{abstract}
 In this paper we develop efficient randomized algorithms to solve the black-box reconstruction problem for polynomials over finite fields, computable by depth three arithmetic circuits with alternating addition/multiplication gates, such that output gate is an addition gate with in-degree two. Such circuits naturally compute polynomials of the form $G\times(T_1 + T_2)$, where $G,T_1,T_2$ are product of affine forms computed at the first layer in the circuit, and polynomials $T_1,T_2$ have no common factors. Rank of such a circuit is defined to be the dimension of vector space spanned by all affine factors of $T_1$ and $T_2$. For any polynomial $f$ computable by such a circuit, $rank(f)$ is defined to be the minimum rank of any such circuit computing it.
Our work develops randomized reconstruction algorithms which take as input black-box access to a polynomial $f$ (over finite field $\F$), computable by such a circuit. Here are the results. 
 
  \begin{itemize}
     \item $[$Low rank$]:$ When $5\leq rank(f) = O(\log^3 d)$, it runs in time $(nd^{\log^3d}\log |\F|)^{O(1)}$, and, with high probability, outputs a depth three circuit computing $f$, with top addition gate having in-degree $\leq d^{rank(f)}$.
     \item $[$High rank$]:$ When $rank(f) = \Omega(\log^3 d)$, it runs in time $(nd\log |\F|)^{O(1)}$, and, with high probability, outputs a depth three circuit computing $f$, with top addition gate having in-degree two.
  \end{itemize}

Prior to our work, black-box reconstruction for this circuit class was addressed in \cite{Shpilka2007, Karnin2009, Sinha2016}. Reconstruction algorithm in \cite{Shpilka2007} runs in time quasi-polynomial in $n,d,|\mathbb{F}|$ and that in \cite{Karnin2009} is quasi-polynomial in $d,|\mathbb{F}|$. Algorithm in \cite{Sinha2016} works only for polynomials over characteristic zero fields. Thus, ours is the first blackbox reconstruction algorithm for this class of circuits, that runs in time polynomial in $\log |\mathbb{F}|$. This problem has been mentioned as an open problem in \cite{Gupta2012} (STOC 2012). In the high rank case, our algorithm runs in $(nd\log|\mathbb{F}|)^{O(1)}$ time, thereby significantly improving the existing algorithms in \cite{Shpilka2007, Karnin2009}.

 \end{abstract}

\newpage

\section{Introduction}
\label{section:introduction} 
Arithmetic circuits (Definition $1.1$ in \cite{Shpilka2010}) are Directed Acyclic Graphs (DAG), describing succinct ways of computing multivariate polynomials. Analogous to the exact learning problem for boolean circuits \cite{Angluin1988}, black-box reconstruction problem (Section $5$, \cite{Shpilka2010}) has been asked for arithmetic circuits:

\emph{Given oracle (also known as black-box) access to a multivariate polynomial computable by an arithmetic circuit of size $s$, construct an explicit circuit (ideally $poly(s)$ sized) that computes the same polynomial.}

In it's most general setting, this problem is believed to be hard, as illustrated in Section $1.4$ of \cite{Garg2020} via an analogy with the boolean world. This is because the exact learning \cite{Angluin1988} of boolean circuits from membership queries is closely related to the Minimum Circuit Size Problem (MCSP), which, under certain cryptographic assumptions\footnote{assuming the existence of cryptographically secure one-way functions} was shown in \cite{Kabanets2000} to not be in {\bf P}. In fact, under the same assumptions \cite{Allender2019} showed that even approximating the minimum circuit size was not in {\bf P}. Drawing an analogy from this, approximating the minimum circuit size for general arithmetic circuits might not be in {\bf P} as well, implying the hardness of black-box reconstruction. We refer the reader to \cite{Garg2020} for more details on the analogy. As a result of this, most of the research on black-box reconstruction has focused on restricted but interesting sub-classes of arithmetic circuits. One such natural restriction is that of depth three circuits which we study in this paper. These are layered circuits with three layers of alternating plus($\Sigma$) gates and product($\Pi$) gates. Reconstruction of $\Pi\Sigma\Pi$ circuits amounts to black-box polynomial factorization into sparse factors and efficient randomized algorithms that solve this are known \cite{Kaltofen1990}. However no such algorithm is known for $\Sigma\Pi\Sigma$ circuits\footnote{from here on wards by depth three circuits we mean $\Sigma\Pi\Sigma$ circuits only} (Definition \ref{defn:sps}). First non-trivial algorithm for this class, which takes exponential time in the fan-in of the multiplication gates, was given in \cite{Klivans2003}. In fact, in a recent work, \cite{Kayal2018a} (Section $1.2$) discuss that efficient reconstruction algorithms for depth three circuits will imply super-polynomial lower bounds for them which is a long standing open problem in Arithmetic complexity \cite{Shpilka1999, Wigderson2006}. Therefore, even for the class of depth three circuits, reconstruction problem appears to be very challenging. Current state of the art reconstruction algorithms for this class either work in the average case \cite{Kayal2018a} or puts further restrictions such as restricting the circuit class to be (set)-multilinear \cite{Shpilka2007, Karnin2009, Bhargava2021}, or restricting the fan-in of the top addition gate (also called top fan-in) \cite{Shpilka2007, Karnin2009, Sinha2016}.
In this paper we are interested in the latter i.e. depth three circuits where fan-in of the top addition gate is assumed to be $k=O(1)$. When $k=1$, the polynomial computed by the circuit is a product of linear forms and black-box reconstruction can be easily performed using black-box factorization algorithm in \cite{Kaltofen1990}. However, the problem seems to become very challenging as soon as we go to circuits with $k>1$. For $k=2$, \cite{Shpilka2007} designed a randomized reconstruction algorithm which was generalized\footnote{algorithm in \cite{Karnin2009} is deterministic} in \cite{Karnin2009} to circuits with  $k=O(1)$.
An important point to note is that while the algorithm in \cite{Shpilka2007} is proper\footnote{when rank (Definition \ref{defn:rank}) of the input polynomial is $\Omega(\log^2 d)$}, i.e., output also has top fan-in $2$, the one in \cite{Karnin2009} is improper and output might have much larger top fan-in. Both these algorithms use fairly sophisticated techniques and have
time complexity quasi-polynomial in $d, |\F|$\footnote{$d$ is degree of $\Pi$ gates and $|\F|$ is size of the underlying field} (even for $k=2$ in \cite{Karnin2009}). Note that ideally we would want the time complexity to depend polynomially on $\log|\F|$, since $O(\log|\F|)$ bits can represent any scalar in the circuit. Therefore, even for $k=2$, designing algorithms which run in time polynomial in $n$\footnote{$n$ is the number of variables in the circuit}, $d$ and $\log|\F|$ are not known. This was asked as an open problem in \cite{Gupta2012} (STOC 2012). In a recent work, \cite{Sinha2016} also considered the top fan-in $2$ case, but over characteristic zero fields, and rank of input polynomial being $\Omega(1)$. Their algorithm runs in time polynomial in $n,d$, but their techniques do not work over finite fields.
Based on the above, the following questions seem very natural to ask.
\par
{\bf (Q1)} Does there exist a reconstruction algorithm for depth $3$ circuits with top fan-in $2$ (over a finite field $\F$), whose run-time is polynomial in $\log |\F|$? \emph{This was asked as an  open problem in \cite{Gupta2012} (STOC 2012)}. {\bf (Q2)} Can such an algorithm be fully polynomial time (at-least in high rank case) i.e. runs in time $(nd \log |\F|)^{O(1)}$? \emph{This will substantially improve results in \cite{Shpilka2007, Karnin2009} for $k=2$}. In this paper we resolve both of these questions.

\subsection{Our Results}
\paragraph{Notation and Preliminaries:} Let $n,d$ denote positive integers and $\F$ be a finite field. We denote the sets $\{1,\ldots,n\}$ and $\{m,m+1,\ldots,n\}$ by $[n]$ and $[m,n]$ respectively. $\B x$ denotes the tuple (or set) of variables $(x_1,\ldots,x_n)$ and $\F[\B x]$ denotes the ring of multivariate polynomials. For a set of linear forms $\ell_1,\ldots, \ell_k \in \F[\B x]$, we use $ \mathbb{V}(\ell_1,\ldots, \ell_k)$ to denote the subspace $\{a\in \F^n : \ell_1(a)=\ldots = \ell_k(a)=0\}$. For a subset of variables $x_{i_1}, \ldots, x_{i_k}$, by $f_{|_{x_{i_1}=\alpha_{i_1}, \ldots , x_{i_k}=\alpha_{i_k}}}$ we denote the polynomial obtained on setting $x_{i_1}=\alpha_{i_1}, \ldots , x_{i_k}=\alpha_{i_k}$ in $f$. As given in Lemma $3.5$ of \cite{Dvir2005}, every depth three circuit $C$ of rank $r$, computing an $n-$variate, degree $d$ polynomial $f$ can be converted into a homogeneous depth three circuit $C_{hom}$ over $\leq n+1$ variables and rank $\leq r+1$, such that it's multiplication gates have in-degree $d$. Section $1.5$ of \cite{Sinha2016b} implies that black-box access to $C_{hom}$ can be simulated efficiently using black-box access to $f$ and integers $n,d$. Also there is a simple and efficient algorithm to obtain $C$ from $C_{hom}$. Hence, from now onwards we only consider homogeneous depth three circuits ($\Sigma\Pi\Sigma(k,n,d,\F)$, Definition \ref{defn:homogeneous-sps}). Also, for any polynomial $f$, $rank(f)$ (Definition \ref{defn:rank-of-polynomial})
will be the minimum rank of any $\Sigma\Pi\Sigma(2,n,d,\F)$ circuit computing it.  Here are our results.

\begin{theorem}[Low rank reconstruction]
\label{theorem:low-rank-reconstruction}
There exists a randomized algorithm which takes as input integers $n,d$ and black-box access to a polynomial $f$ computable by a $\Sigma\Pi\Sigma(2,n,d,\F)$ circuit $(5\leq rank(f) = O(\log^3 d))$, runs in time $(nd^{\log^3d}\log |\F|)^{O(1)}$ and, with probability $1-o(1)$, outputs a $\Sigma\Pi\Sigma(k,n,d,\F)$ $(k\leq d^{rank(f)})$ circuit computing $f$. 
\end{theorem}

\begin{theorem}[High rank reconstruction]
\label{theorem:high-rank-reconstruction}
There exists a randomized algorithm which takes as input integers $n,d$ and black-box access to a polynomial $f$ computable by a $\Sigma\Pi\Sigma(2,n,d,\F)$ circuit $(rank(f) = \Omega(\log^3{d}))$, runs in time $(nd\log |\F|)^{O(1)}$ and, with probability $1-o(1)$, outputs a $\Sigma\Pi\Sigma(2,n,d,\F)$ circuit computing $f$.
\end{theorem}
We allow algorithms to query input polynomial at points in a $(nd)^{O(1)}$ sized extension $\K$ of $\F$. Here are some remarks on the above results.
\begin{itemize}
    \item Theorems \ref{theorem:low-rank-reconstruction} and \ref{theorem:high-rank-reconstruction} completely resolve {\bf (Q1)}. Therefore we solve an open problem from \cite{Gupta2012}. Theorem \ref{theorem:high-rank-reconstruction} resolves {\bf (Q2)} in the high rank case ($\Omega(\log^3 d)$) and thus both theorems substantially improve the overall reconstruction time complexity for this circuit class (as compared to \cite{Shpilka2007} and \cite{Karnin2009}).
    
    \item A crucial component of our proofs is a new structural result, which might be of independent interest. We show that for $f$ computable by a $\Sigma\Pi\Sigma(2,n,d,\F)$ circuit $(rank(f) \geq 5)$, the set of co-dimension $2$ subspaces of $\F^n$ on which the ``non-linear'' part (Definition \ref{defn:lin-nonlin-part}) of $f$ vanishes, has size $d^{O(1)}$, and can be computed efficiently. We give a formal statement in Proposition \ref{propn:annihilating-planes}.
    
    \item In order to prove Theorem \ref{theorem:high-rank-reconstruction}, we develop an interesting result related to Sylvester Gallai (SG) type configurations (Definition \ref{defn:sg}) and present it in Proposition \ref{propn:ordinary-line-dim-main}. We believe it might be of independent interest. Similar results called Quantitative SG theorems are known (Theorem $5.1.2$ and Section $5.3$ in \cite{Dvi12}). These quantitative versions prove bounds on number of ordinary lines through a point, whereas our theorem considers dimension of the space spanned by the union of ordinary lines through a point.
    
    \item When $rank(f) = 1$, $f$ factors into a product of linear forms and can be reconstructed efficiently using Lemma \ref{lemma:sps(k)-pit}. So only $rank(f)=2,3,4$ are not covered by the algorithms above.
    
    \item We note that when char($\F$) $>d$ or $0$, Lemma \ref{lemma:variable-reduction} (\cite{Carlini2006, Kayal2006}) gives an algorithm for Theorem \ref{theorem:low-rank-reconstruction} i.e. low rank reconstruction. But, this only works for fields with large characteristic, whereas Algorithm \ref{algorithm:low-rank-reconstruction} in our paper is independent of the characteristic of the field.
    
    \item We would like to highlight that derandomizing our algorithms seems rather difficult. Theorem $5$ in  \cite{Volkovich16} implies that any proper and efficient reconstruction algorithms for (set)-multilinear $\Sigma\Pi\Sigma(2, n, d, \F)$ circuits with running time polynomial in $\log|\F|$ can be deterministically converted\footnote{in time polynomial in $\log|\F|$.} into a square root oracle over $\F$. This is a well studied problem \cite{Lenstra1982, Goldwasser1984, Shoup1991, Gao2004, Rabin2005, Kayal2006, Gathen2013} and till date no deterministic algorithm with running time having polynomial dependence in $\log|\F|$ is known.
    
    \item  We conjecture that, for $k=O(1)$, our algorithms can be generalized to proper\footnote{at least in the high rank case.} and efficient\footnote{with $\log |\F|$ dependence on field size.} reconstruction algorithms for $\Sigma\Pi\Sigma(k,n,d,\F)$ circuits. Some crucial parts that would need generalization/refinement are $(a)$ Proposition \ref{propn:annihilating-planes} to higher co-dimension sub-spaces, and $(b)$ The ``gluing'' algorithm (Algorithm $5$ in \cite{Shpilka2007}) used in Algorithm \ref{algorithm:general-case}, which merges factors of restrictions of the input polynomial and reconstructs one of the product gates. Recall that the known algorithms for this class are either exponential time in in-degree of product gates \cite{Klivans2003} or are improper and run in quasi-polynomial time in $d,|\F|$ \cite{Karnin2009}.
    
    \item Note that as proved in Corollary $7$ of \cite{Shpilka2007}, $\Sigma\Pi\Sigma(2,n,d,\F)$ circuit for a polynomial $f$ is unique when $rank(f) = \Omega(\log^2 d)$. In fact, for smaller ranks, it is easy to construct example polynomials computable by multiple $\Sigma\Pi\Sigma(2,n,d,\F)$ circuits. Therefore, for low rank polynomials, in the absence of uniqueness, proper reconstruction might be far fetched. Moreover, many of our techniques such as construction of a candidate set of linear forms (Algorithm \ref{algorithm:computing-candidate-set-of-linear-forms}) that help in proper reconstruction only work in the high rank case. Due to these reasons we needed to break our results into the low rank and high rank cases.
    
    \item A $(nd)^{O(1)}$ time algorithm for reconstructing $\Sigma\Pi\Sigma(2,n,d,\R)$ was designed in \cite{Sinha2016, Sinha2016b} when $rank(f) = \Omega(1)$. They construct a set of linear forms modulo which the polynomial factorizes completely into linear forms. This is done using  Brill's equations \cite{Gelfand1995} which construct a system of polynomial equations whose solutions characterize polynomials that decompose into product of linear forms. As derived in Appendix $B$ of \cite{Sinha2016b}, computation of Brill's equations involve division by multiples of $d$ and therefore they are not likely to work over finite fields\footnote{of general characteristic}
    To the best of our knowledge, analogous equations for polynomials over finite fields are not well studied. On the other hand, we construct a set of candidate linear forms in a much simpler way by looking at co-dimension $2$ subspaces where $f$ vanishes. Another difference between the two techniques is during the ``gluing'' process of Algorithm \ref{algorithm:general-case}. In \cite{Sinha2016, Sinha2016b} the gluing is done using $\delta-SG_k$ theorems \cite{Barak2011} which prove existence of many ``ordinary'' $k$-flats. On the other hand we construct a large independent set of linear forms dividing one of the product gates and use it along with the ``gluing'' technique from \cite{Shpilka2007} which depends on lower bounds for locally decodable codes.

\end{itemize}
Next, we state our proposition regarding the number of co-dimension $2$ spaces on which the non-linear part of $f$ vanishes. In order to do so we refer the reader to $(a)$ definition of non-linear part ($NonLin(f)$) of a polynomial $f$ (Definition \ref{defn:lin-nonlin-part}), $(b)$ definition of vanishing of a polynomial on a co-dimension $2$ subspace and the set $\mathcal{S}(f)$ of all such co-dimension $2$ spaces (Definition \ref{defn:polynomial-vanishing-codim}).

\begin{proposition}
\label{propn:annihilating-planes}
Let $f \in \F[\B x]$ be a polynomial computable by a $\Sigma\Pi\Sigma(2,n,d,\F)$ circuit with $rank(f) \geq 5$. The following are true.
\begin{enumerate}
    \item \label{item:annihilating-subspaces-size} $|\mathcal{S}(NonLin(f))| \leq 3d^7$.
    \item \label{item:compute-annihilating-subspaces} 
There exists a randomized algorithm that takes as input black-box access to $f$ along with integers $n,d$, runs in time $(nd\log |\F|)^{O(1)}$ and, outputs a set $\mathcal{S}$ (of size $\leq 3d^7$) containing tuples of independent linear forms in $\F[\B x]$ such that with probability $1-o(1)$,
\[
\{\mathbb{V}(\ell_1, \ell_2) : (\ell_1,\ell_2)\in \mathcal{S}\} = \mathcal{S}(NonLin(f)).
\]
\end{enumerate}
\end{proposition}
Next we state Proposition \ref{propn:ordinary-line-dim-main}  about ordinary lines and the space spanned by them, that was mentioned in remarks following the theorems. This requires definitions of proper sets (Definition \ref{defn:proper-set}), ordinary lines and the set $\mathcal{O}(t, \mathcal{S})$ of all ordinary lines from a point $t$ to set $\mathcal{S}$ (Definition \ref{defn:ordinary-line}).

\begin{proposition}
\label{propn:ordinary-line-dim-main}
Let $\mathcal{S} \subset \F^n$ be a proper set (Definition \ref{defn:proper-set}) and $\mathcal{T}\subset \F^n$ be any linearly independent set of size $\geq \log|\mathcal{S}|+2$. Then there exists $t\in \mathcal{T}$, such that union of all elements of $\mathcal{O}(t, \mathcal{S})$ spans a high dimensional space. More precisely,
\[
dim(\sum_{W \in \mathcal{O}(t, \mathcal{S})} W) \geq \frac{dim(sp(\mathcal{S}))}{\log |\mathcal{S}| + 2}.
\]
\end{proposition}

\subsection{Ideas and analysis of main algorithms}
\label{subsection:overview-of-our-techniques}
The algorithms mentioned in Theorems \ref{theorem:low-rank-reconstruction}, \ref{theorem:high-rank-reconstruction} and Proposition \ref{propn:annihilating-planes} are given in Algorithms \ref{algorithm:low-rank-reconstruction}, \ref{algorithm:high-rank-reconstruction} and \ref{algorithm:compute-annihilating-subspaces} respectively. Proofs of Propositions \ref{propn:annihilating-planes} and \ref{propn:ordinary-line-dim-main} are provided in Sections \ref{section:codim-two-subspaces-f-vanishes} and \ref{section:ordinary-lines} respectively.  In this section we discuss key technical ideas required for proving these results. Missing details are supplied in the subsequent sections. As described in Definition \ref{defn:rank}, we write $f = G\times(T_1+T_2)$ where $G, T_1, T_2$ are product of linear forms and $gcd(T_1,T_2)=1$. We know that
\[
Lin(f) \times NonLin(f) = f = G\times (T_1+T_2).
\]

\subsubsection{Theorem \ref{theorem:low-rank-reconstruction}: Key ideas for Algorithm \ref{algorithm:low-rank-reconstruction}}
\label{subsubsection:algorithm-idea-low-rank-reconstruction}

The algorithm mentioned in Theorem \ref{theorem:low-rank-reconstruction} is presented in Algorithm \ref{algorithm:low-rank-reconstruction} and it's correctness/complexity is discussed in Section \ref{section:low-rank-reconstruction}. We describe the main ideas now.
Since $NonLin(f)$ has no linear factors and $Lin(f), G$ are product of linear forms, $NonLin(f)$ divides $T_1+T_2$ implying that $NonLin(f) = h(y_1,\ldots,y_r)$, for some homogeneous polynomial $h$ over $\F$ and independent linear forms $y_1, \ldots,y_r$ spanning the set of linear factors of $T_1\times T_2$ (here $r=rank(f)$). Clearly
$NonLin(f)$ is non-constant, otherwise rank of $f$ would not be $>=5$. Using Algorithm \ref{algorithm:sps(1)-reconstruction}, with high probability, we get black-box access to $NonLin(f)$ and it's degree $t$. If we also had access to $(a)$ the integer $r=rank(f)$, and $(b)$ a $d^{O(1)}$ sized set $\mathcal{L}$ of linear forms containing required $y_1,\ldots,y_r$, then we could just iterate over all $r$ sized subsets $\{y_1,\ldots,y_r\}$ of $\mathcal{L}$ and using deterministic multivariate black-box interpolation (Lemma \ref{lemma:interp}) compute polynomial $h(y_1,\ldots,y_r)$ as a sum of degree $t$ monomials in $y_1,\ldots,y_r$ which is trivially computed by a  $\Sigma\Pi\Sigma(t^r, n, t, \F)$ circuit. We can then multiply all linear factors of $Lin(f)$, obtained using Algorithm \ref{algorithm:sps(1)-reconstruction}, to all multiplication gates of this circuit resulting in a $\Sigma\Pi\Sigma(t^r, n, d, \F)$ circuit for $f$. So we only need to argue about the required access described above. We do not know $rank(f)$ but we know that $rank(f) = O(\log^3 d)$. Therefore, we try all values of $r$ in $[O(\log^3 d)]$.  To get access to the set $\mathcal{L}$, we use results in Proposition \ref{propn:annihilating-planes}. It guarantees that the set of co-dimension $2$ subspaces on which $NonLin(f)$ vanishes, has size $d^{O(1)}$ and also efficiently constructs a set $\mathcal{S}$ that comprises of tuples of linear forms representing such co-dimension $2$ spaces. Using $\mathcal{S}$, we define,
\[
\mathcal{L} = \{\ell_1 : \exists \ell_2 \text{ such that }(\ell_1,\ell_2)\in \mathcal{S} \text{ or }(\ell_2,\ell_1)\in \mathcal{S}\}
\]
$\mathcal{L}$ is easily constructed from $\mathcal{S}$. Also $|\mathcal{S}| = d^{O(1)}$ implies $|\mathcal{L}| = d^{O(1)}$. In Lemma \ref{claim:L-is-good}, we show that $\mathcal{L}$ contains an independent set $\{y_1,\ldots,y_r\}$ of linear forms that spans the set of linear factors of $T_1\times T_2$. Basically, for any linear form $\ell_1$ dividing $T_1$, we show there is a linear form $\ell_2$ dividing $ T_2$ (and vice versa) such that $NonLin(f)$ vanishes on $\mathbb{V}(\ell_1, \ell_2)$. This gives rise to a tuple $(\ell_1^\prime, \ell_2^\prime)\in \mathcal{S}$ (i.e. $\ell_1^\prime, \ell_2^\prime \in \mathcal{L}$) such that $sp\{\ell_1, \ell_2\} = sp\{\ell_1^\prime, \ell_2^\prime\}$. Let $\mathcal{L}^\prime$ be the collection of all such $\ell_1^\prime, \ell_2^\prime$. By construction $\mathcal{L}^\prime \subset \mathcal{L}$ and $sp\{\mathcal{L}^\prime\} = sp\{\text{linear form }\ell : \ell \mid T_1\times T_2\}$. Now we can take 
$y_1, \ldots,y_r$ to be any basis of $\mathcal{L}^\prime$. At the end we perform a randomized polynomial identity test to check whether the reconstructed circuit computes the input polynomial or not. This guarantees that with probability $1-o(1)$, no incorrect reconstruction is returned. At the same time, for correct $r$ and $\mathcal{L}$, by the above technique, with probability $1-o(1)$, we recover the correct circuit which will pass the test.  
Our algorithm takes $(nd^{\log^3 d}\log|\F|)^{O(1)}$ time. Full details can be found in Section \ref{section:low-rank-reconstruction}.

\paragraph{Comparison with algorithm in \cite{Shpilka2007}}
The broad idea for low rank\footnote{their low rank case assumes $rank(f) = O(\log^2d)$. we assume $rank(f) = O(\log^3d)$.} reconstruction given in Algorithm $3$ of \cite{Shpilka2007} is similar to ours. However, their algorithm runs in time quasi-polynomial in $n,d$ and $|\F|$. The main reason is that they search for the required basis $\{y_1,\ldots, y_r\}$ of linear forms (Step $2$ of Algorithm $3$ in \cite{Shpilka2007}) by iterating over the entire set of linear forms in $O(\log^2 d)$ many variables. This makes their algorithm quasi-polynomial time with respect to $|\F|$, since this set has size $|\F|^{O(\log^2 d)}$. As described above, our algorithm performs a more efficient search by searching within the $d^{O(1)}$ sized set $\mathcal{L}$, that is efficiently constructed. This leads to a polynomial time dependence on $\log|\F|$ which is ideal as $O(\log|\F|)$ bits can represent each scalar in the circuit.

\subsubsection{Theorem \ref{theorem:high-rank-reconstruction}: Key ideas for Algorithm \ref{algorithm:high-rank-reconstruction}}
\label{analysis-high-rank-reconstruction}

The algorithm mentioned in Theorem \ref{theorem:high-rank-reconstruction} is presented in Algorithm \ref{algorithm:high-rank-reconstruction}. It's correctness and time complexity are discussed in Section \ref{section:high-rank-reconstruction}. Our algorithms crucially utilize the set of ``candidate linear forms'' which are defined in Definition \ref{definition:candidate-linear-form}. This definition further requires us to define what it means for a polynomial to factorize into non-zero linear forms on a co-dimension $1$ subspace which is defined in Definition \ref{defn:factorize-codim-1-subspace}. Next, we present a reconstruction algorithm solving a corner case, where one of $T_1, T_2$ is power of a linear form (up to scalar multiplication). Then we discuss the general case algorithm which is run if the corner case fails to reconstruct. In this case, linear factors of both $T_1, T_2$ span at least a two dimensional space.

 \paragraph{Corner case - One of $T_1,T_2$ is power of a linear form: } 
 Formal statement is provided in Lemma \ref{lemma:corner-case-reconstruction-lemma} and corner case reconstruction algorithm is given in Algorithm \ref{algorithm:corner-case}. We sketch the idea here. If one of $T_1,T_2$ is power of a linear form, then we prove in Claim \ref{claim:corner-case-factors} that $Lin(f)=G$ and $NonLin(f) = T_1+T_2$. The basic idea is that if $T_1 + T_2$ has a non trivial linear factor $\ell$, then span of any any linear factor of $T_1$ and $\ell$ will contain some linear factor of $T_2$. This can be used to show that dimension of $sp\{\text{linear form }\ell : \ell\mid T_1\}$ and $sp\{\text{linear form }\ell : \ell\mid T_2\}$ can differ by at most $1$. Since $rank(f) =\Omega(\log^3 d)$, we arrive at a contradiction to our assumption in this case. Therefore, using Algorithm \ref{algorithm:sps(1)-reconstruction} we get black-box access to $T_1+T_2$, it's degree $t$, and the entire list of linear factors (with multiplicity) of $G$. Let's assume that for some $i\in [2]$, $T_i$ is power of some linear form. If we also had access to
 $(a)$ a linear factor $\ell_1$ of $T_i$, and $(b)$ a $d^{O(1)}$ sized set $\mathcal{X}$ of scalars such that $T_i = \delta \ell_1^t$ for some $\delta\in \mathcal{X}$, then we could just go over all scalars $\delta\in \mathcal{X}$ and try to factorize black-box of  $T_1+T_2-\delta \ell_1^t$, using Algorithm \ref{algorithm:sps(1)-reconstruction}. If factorization gives all linear factors, we would have obtained a $\Sigma\Pi\Sigma(2,n,t,\F)$ circuit for $T_1+T_2$. Combining this with linear factors of $G$ gives a $\Sigma\Pi\Sigma(2,n,d,\F)$ circuit for $f$. So we only need to argue about the required access. In Claim \ref{claim:corner-case-candidate}, we show that a linear factor $\ell_1$ of $T_i$ belongs to $\mathcal{L}(NonLin(f))$ that we defined in Definition \ref{definition:candidate-linear-form}. To see this, notice that since $NonLin(f) = T_i+T_{3-i}$, it will factorize into a non-zero product of linear forms on $\mathbb{V}(\ell_1)$ for any linear factor $\ell_1$ of $T_i$. Since rank of $T_i+T_{3-i}$ is $\Omega(\log^3 d)$, we easily obtain linear factors $\ell_2, \ell_2$ of $T_{3-i}$ such that $\ell_1, \ell_2, \ell_3$ satisfy conditions required by Definition \ref{definition:candidate-linear-form} $\Rightarrow \ell_1 \in \mathcal{L}(NonLin(f))$. Definition \ref{definition:candidate-linear-form} and Proposition \ref{propn:annihilating-planes} imply that  $\mathcal{L}(NonLin(f))$ has size $d^{O(1)}$ and Algorithm \ref{algorithm:computing-candidate-set-of-linear-forms} efficiently constructs it. So we search for $\ell_1$ in this set. To construct set $\mathcal{X}$ containing $\delta$ where $T_i = \delta \ell_1^t$, we restrict $T_i+T_{3-i}$ to $\mathbb{V}(\ell_1)$, and obtain two linearly independent factors $\ell_2, \ell_3$ of the restriction of ${T_{3-i}}$ using Algorithm \ref{algorithm:sps(1)-reconstruction}. These factors will exists since $rank(f) = \Omega(\log^3 d)$. For simplicity map  $\ell_1\mapsto x_1, \ell_2\mapsto x_2, \ell_3\mapsto x_3$. Our polynomial has the following form,
 \[
 NonLin(f) = \delta x_1^t + (x_2-\beta x_1)(x_3-\gamma x_1)T_{3-i}^\prime,
 \]
 for some scalars $\beta, \gamma$ and product of linear forms $T_{3-i}^\prime$. To find $\delta$, we observe that this polynomial depends on $x_3$ but becomes independent on plugging $x_2=\beta x_1$. We first set $x_4,\ldots, x_n$ to random values in $\F$ and use multivariate interpolation from Lemma \ref{lemma:interp}, to represent this new polynomial as a degree $t$ polynomial in $\F[x_1,x_2,x_3]$. Then we solve for a fresh variable $\beta$ such that setting $x_2=\beta x_1$, makes this polynomial independent of $x_3$. This is done by collecting all coefficients ($\in \F[\beta]$) of monomials containing $x_3$ and solving the system of equations they define. This system has $d^{O(1)}$ many solutions, since all polynomials are univariate with degree $d^{O(1)}$. All solutions to this system are computed using algorithm given in Lemma \ref{lemma:polynomial-equations}. We plug these $\beta$s back into coefficient of $x_1^t$ and obtain a $d^{O(1)}$ sized set $\mathcal{X}$ containing $\delta$. At the end, using polynomial identity testing algorithm in Lemma \ref{lemma:sps(k)-pit}, we deterministically check whether the reconstruction is correct or not. Thus, for choices of $\ell_1, \mathcal{X}$, where the circuit was not correct, we don't output anything and for the right values of $\ell_1, \mathcal{X}$, by the algorithm described above, we correctly reconstruct the circuit. Our algorithm takes $(nd\log|\F|)^{O(1)}$ time.

 \paragraph{General case - Both $T_1,T_2$ have at least $2$ independent linear factors:}
  This is the more general case of our algorithm and is tried after the above mentioned corner case fails to provide a reconstruction. Our algorithm tries to find an $\Omega(\log d)$ sized set of linear forms such that all linear forms in this set divide the same $T_i$. Once such a set is found we use it to reconstruct all linear forms dividing $G\times T_{3-i}$ and using this the entire circuit. We break down our key ideas below.
  \begin{itemize}
  \item We first explain, how one can complete the reconstruction given access to such a set. Formal statement is given in Lemma \ref{lemma:large-li-set-known} and algorithm is provided in Algorithm \ref{algorithm:general-case}. The basic idea is as follows. Without loss of generality, we assume the independent set of linear forms is the set of variables $x_1, \ldots, x_t$ where $t=\Omega(\log d)$ and that all of these divide $T_1$. Therefore,
  \[
  Lin(f)\times NonLin(f) = f = G\times (x_1\ldots x_t T_1 ^\prime + T_2)
  \]
  where $T_1^\prime$ is a product of linear forms and $gcd(T_1^\prime, T_2)=1$. Without loss of generality we also assume that no $x_i$ divides $f$ since we can divide $f$ by largest power of all the $x_i$\footnote{we add them back after reconstruction of this new polynomial is complete}. The idea is to construct all linear factors of $G\times T_2$ by first computing all linear factors of $(G\times T_2)_{|_{x_i=0}}$ for $i\in [t]$ and then gluing these factorizations together. Linear factors of $(G\times T_2)_{|_{x_i=0}}$ can be easily computed by applying Algorithm \ref{algorithm:sps(1)-reconstruction} to the black-box computing $f_{|_{x_i=0}}$. Clearly for each $i$ the multi-sets of linear factors will have the same (i.e. $deg(f)$) number of elements. These multi-sets are glued using Algorithm $5$ from \cite{Shpilka2007}. The idea behind this algorithm is to find a linear form $\ell_1$ dividing $(G\times T_2)_{|_{x_1=0}}$ (with multiplicity say $k$), and an integer $2\leq i\leq t$ such that there are exactly $k$ linear factors $\ell_i^1, \ldots, \ell_i^k$ (could be multiples of each other) of $(G\times T_2)_{|_{x_i=0}}$ such that ${\ell_1}_{|_{x_i=0}}$ and $ {\ell_i}_{|_{x_1=0}}$ are scalar multiples. Once such $\ell_1, i$ and $\ell_i^j$, $j\in [k]$ are found, $\ell_1$ is glued with each $\ell_i^j$ by comparing coefficients and $k$ glued linear forms dividing $G\times T_2$ are obtained. Then $\ell_1$ (with all it's multiplicity) and all $\ell_i^j$, $j\in [k]$ are removed from their respective multi-sets. This process is repeated until the multi-sets are empty. When the multi-sets are non-empty, such $\ell_1$ and $i$ always exist. If not, then in Theorem $33$ of \cite{Shpilka2007}, they show that a lower bound on length of linear $2$-query locally decodable codes gets violated. Details are provided inside proof of Theorem $29$ in \cite{Shpilka2007} and for cleaner presentation we do not repeat it here. At the end, all linear factors (with multiplicity) of $G\times T_2$ are known. To know $G\times T_2$ completely, we still need to know the appropriate constant to multiply to the product of these linear factors. For this, we restrict all linear forms in our computed multi-set to $x_1=0$ and compare with the multi-set of linear factors of $(G\times T_2)_{|_{x_1=0}}$ which we had already computed earlier. Now we can factorize the black-box for $f-G\times T_2$ and recover all linear factors of $G\times T_1$ and construct a $\Sigma\Pi\Sigma(2,n,d,\F)$ circuit for $f$. Finally using polynomial identity test in Lemma \ref{lemma:sps(k)-pit}, we can check whether this circuit correctly computes $f$ or not, and only output a correct circuit.
  
  \item Now, we come back to our process of finding the linearly independent set utilized above.  We use the set of candidate linear forms $\mathcal{L}(NonLin(f))$ efficiently constructed using Algorithm \ref{algorithm:computing-candidate-set-of-linear-forms}. In Parts $1$ and $2$ of Lemma \ref{lemma:general-case-search} (which uses Proposition \ref{propn:ordinary-line-dim-main}), we show existence of a linear form $\ell\in \mathcal{L}(NonLin(f))$ and a linearly independent set $\mathcal{B}\subset \mathcal{L}(NonLin(f))$ of size $\Omega(\log d)$
  such that $\ell$ and all linear forms in $\mathcal{B}$ divide $T_1\times T_2$.
  Moreover, for all $\ell^\prime\in \mathcal{B}$, $sp\{\ell, \ell^\prime\}$ does not contain any other\footnote{apart from $\ell, \ell^\prime$} linear factor of $T_1\times T_2$, and any linear factor of $T_1+T_2$. Using this, in Part $3$ of Lemma \ref{lemma:general-case-search}, we show that for $\ell^\prime\in \mathcal{B}$, $NonLin(f)$ vanishes on $\mathbb{V}(\ell,\ell^\prime)$ if and only if $\ell_1, \ell_2$ divide different $T_i$. This is used to split $\mathcal{B}$ into two parts, with linear forms in each part dividing the same $T_i$. One of these would be $\Omega(\log d)$ in size giving us the required linearly independent set. Full details can be found in Part $3$ of Lemma \ref{lemma:general-case-search}. We use the existence of $\ell, \mathcal{B}$ in Algorithm \ref{algorithm:general-case} in the following way. For every $\ell$ in $\mathcal{L}(NonLin(f))$ using the construction of $\mathcal{B}$ in parts $1,2$ of Lemma \ref{lemma:general-case-search}, we construct a $O(rank(f))$ sized collection of sets containing $\mathcal{B}$. For each $\mathcal{B}$ in this collection, we apply the test (from part $3$ of Lemma \ref{lemma:general-case-search}) mentioned above to divide it into two parts $\mathcal{U}, \mathcal{V}$. The larger set is provided to the previous algorithm (details in Algorithm \ref{algorithm:general-case}) to reconstruct the circuit. In the end, we use deterministic polynomial identity test to reject incorrect constructions. The existence of $\ell, \mathcal{B}$ and the test above, make sure that for the correct choices, we will output the correct circuit.

  \end{itemize}
  
  \paragraph{Comparison with algorithm in \cite{Shpilka2007}:} As described above, if we have access to an $\Omega(\log d)$ sized set of linear forms such that all of them divide the same $T_i$, our algorithm exactly matches the one given in Algorithm $5$ of \cite{Shpilka2007}. The main difference\footnote{also we assume rank to be $\Omega(\log^3 d)$ whereas \cite{Shpilka2007} assumes it to be $\Omega(\log^2 d)$ for high rank reconstruction} is in the way such a set is created. In Steps $1,2$ of Algorithm $4$ in \cite{Shpilka2007}, they iterate over all possible $\Omega(\log d)$ sized sets of linear forms inside an $\Omega(\log^2 d + \log^2 n)$ sized random subspace of $\F^n$. Such a brute force search considers $|\F|^{\Omega(\log d(\log^ d + \log^2 n))}$ many sets leading to a quasi polynomial time complexity in $n,d$ and $|\F|$. Using $\mathcal{L}(NonLin(f))$, in Lemma \ref{lemma:general-case-search} we are able to create a small collection of sets of independent linear forms, such that at least one set in this collection has size $\Omega(\log d)$ and comprises of linear forms all of which divide the same $T_i$. Construction of this collection has required new structural techniques from Proposition \ref{propn:annihilating-planes} and Proposition \ref{propn:ordinary-line-dim-main}. Searching through this small collection and rejection of incorrect reconstructions by a deterministic polynomial identity test lead to an overall running time of $(nd\log|\F|)^{O(1)}$ which is a huge improvement compared to \cite{Shpilka2007}. 
  
  \subsubsection{Proposition \ref{propn:annihilating-planes}: Key ideas}
  Part $1$ of Proposition \ref{propn:annihilating-planes} is proved in Section \ref{section:codim-two-subspaces-f-vanishes} and algorithm for Part $2$ is provided in Algorithm \ref{algorithm:compute-annihilating-subspaces}. We describe the main ideas involved in both these parts now. Let $W$ be a co-dimension $2$ subspace of $\F^n$ on which $NonLin(f)$ vanishes. This implies that $T_1 + T_2$ also vanishes on $W$. Now there are two cases, either both $T_1, T_2$ vanish on $W$ or both do not. When both vanish, we get independent linear forms $\ell_1$ dividing $T_1$ and $\ell_2$ dividing $T_2$ such that $\ell_1, \ell_2$ vanish on $W$ i.e. $ W = \mathbb{V}(\ell_1, \ell_2)$. There are $\leq d^2$ such pairs $(\ell_1, \ell_2)$ implying there are $\leq d^2$ such $W$s. When both do not vanish, in Lemma \ref{lemma:one-dimension-found}, we create a $d^4+d^6$ sized set $\mathcal{A}$ of co-dimension $1$ spaces, such that $W$ is contained in some $V = \mathbb{V}(\ell^{\star})\in \mathcal{A}$. Since $NonLin(f)$ vanishes on $W$, writing $W = \mathbb{V}(\ell^{\star}, \ell^{\star\star})$, would imply that $\ell^{\star\star}$ is a linear factor of $NonLin(f)$ restricted to $\mathbb{V}(\ell^{\star})$. Thus, there are $\leq d$ possible $\ell^{\star\star}$ (up to scalar multiplication) for every possible $V$, implying that there are $\leq d^5 + d^7$ such $W$s. Putting all bounds together we get an overall size bound of $3d^7$ completing the proof of Part $1$ of Proposition \ref{propn:annihilating-planes}. We now explain the creation of $\mathcal{A}$ (Lemma \ref{lemma:one-dimension-found}). Let $W = \mathbb{V}(\ell, \ell^\prime)$ and $U = sp\{\ell, \ell^{\prime}\}$. Since $T_1, T_2$ don't vanish on $W$, using unique factorization (after restricting to $W$), we get that for every $\ell_1$ dividing $T_1$, there is an $\ell_2$ dividing $T_2$ such that $\ell_1, \ell_2, \ell, \ell^{\prime}$ are linearly dependent $\Rightarrow sp\{\ell_1, \ell_2\}$ intersects $U$ non-trivially. Let $T_1 = \ell_{1,1}\ldots \ell_{1,t}$ and $T_2 = \ell_{2,1}\ldots \ell_{2,t}$ with $\ell_{i,j}$ being linear forms. Without loss of generality, we assume that $U_j = sp\{\ell_{1,j}, \ell_{2,j}\}$ intersects $U$ non-trivially. Since $T_1, T_2$ don't vanish on $W$, none of the $U_j$s equals $U$. Now we again have two possibilities. $(a)$ Either $U\cap U_j = U\cap U_k$ for some $j\neq k \in [t]$, or $(b)$ For all $j,k\in [t]$, $U\cap U_j \neq U\cap U_k$. To explain, the construction of $\mathcal{A}$, we now give a geometric intuition. Think of $U, U_j, U_k$ as lines $\vec{L}, \vec{L}_j, \vec{L}_k$ (respectively) in the projective space. Then $(a)$ implies that for some $j,k$, line $\vec{L}$ intersects both the lines $\vec{L}_j, \vec{L}_k$ at the same point, which can be identified as intersection of lines $\vec{L}_j$ and $\vec{L}_k$ only. This intersection point on $\vec{L}$ corresponds to a linear form $\ell^{\star}$ implying that $\ell^{\star}\in U \Rightarrow W\subset V = \mathbb{V}(\ell^{\star})$. We add this $V$ to our set $\mathcal{A}$.  There are $\leq d^2$ possible lines and therefore $\leq d^4$ possible pairs $\vec{L}_j, \vec{L}_k$ implying that at most $d^4$ elements have been added to $\mathcal{A}$. In case of $(b)$, we know that for all distinct $\vec{L}_i, \vec{L}_j, \vec{L}_k$, line $\vec{L}$ intersects them at different points. Also, since $rank(f)\geq 5$, there exist distinct $\vec{L}_i, \vec{L}_j, \vec{L}_k$ that span at least a $4$ dimensional projective space. Since $\vec{L}$ intersects $\vec{L}_i, \vec{L}_j$ at different points, it completely lies in the plane spanned by $\vec{L}_i, \vec{L}_j$. The line $\vec{L}_k$ is outside the plane and therefore intersects it at a point on $\vec{L}$. This point gets identified using $\vec{L}_i,\vec{L}_j,\vec{L}_k$ and gives us a linear form $\ell^{\star}$ on $\vec{L}$ for which we add $V = \mathbb{V}(\ell^{\star})$ to $\mathcal{A}$.
  There are $\leq d^2$ lines and therefore $\leq d^6$ such triplets $(\vec{L}_i,\vec{L}_j,\vec{L}_k)$. Thus size of $\mathcal{A}$ is now upper bounded by $d^4 + d^6$ and we have covered all cases.
  
  Now we sketch the key ideas involved in Algorithm \ref{algorithm:compute-annihilating-subspaces} which constructs this $d^{O(1)}$ sized set $\mathcal{S}(NonLin(f))$. We first apply a random invertible linear transformation to the variables in black-box computing $f$ giving polynomial $g$ having non-degeneracies (with high probability) required in the steps that follow. In Part $2$ of Lemma \ref{lemma:annihilating-spaces-details-1}, we show our problem is equivalent to finding co-dimension $2$ subspaces on which $NonLin(g)$ vanishes. Using Algorithm \ref{algorithm:sps(1)-reconstruction}, we get black-box access to $NonLin(g)$ and it's degree $t$. In Part $5$ of Lemma \ref{lemma:annihilating-spaces-details-1}, we show that with high probability, any co-dimension $2$ subspace on which $NonLin(g)$ vanishes can be written as $\mathbb{V}(x_1-\ell_1, x_2-\ell_2)$ with $\ell_1, \ell_2 \in \F[x_3,\ldots,x_n]$, so we focus on constructing these kind of subspaces. Next, for each $i\in [5,n]$ we restrict $g$ to $5$ variables $x_1,x_2,x_3,x_4,x_i$ and obtain $5$-variate polynomials $g_i$. In Part $3$ of Lemma \ref{lemma:annihilating-spaces-details-1}, we show that with high probability these $g_i$ exhibit $\Sigma\Pi\Sigma(2,5,d,\F)$ circuits of rank $5$ and therefore $|\mathcal{S}(NonLin(g_i))| \leq 3d^7$ (by Part $1$ of Proposition \ref{propn:annihilating-planes}). In Part $6$ of Lemma \ref{lemma:annihilating-spaces-details-1}, we show that $NonLin(g_i)$ vanishes on subspace $\mathbb{V}(x_1-\ell_1^i, x_2-\ell_2^i)$ whenever $NonLin(g)$ vanishes on subspace $\mathbb{V}(x_1-\ell_1, x_2-\ell_2)$\footnote{$\ell_1, \ell_2 \in \F[x_3,\ldots,x_n]$.}, where $\ell_j^i$ is obtained by setting all variables except $x_1, x_2,x_3,x_4,x_i$ to $0$ in $\ell_j$, $j\in [2]$. So we now compute the set $\mathcal{S}_i$\footnote{containing tuples of independent linear forms representing the co-dimension $2$ spaces.} of co-dimension $2$ subspaces of the form $\mathbb{V}(x_1-\ell_1^i, x_2-\ell_2^i)$ on which these $NonLin(g_i)$s vanish and glue them together. We compute these sets by interpolating $NonLin(g_i)$ as degree $t$ polynomials in the monomial basis of $\F[x_1,x_2,x_3,x_4,x_i]$. Then we substitute $x_1 = y_3x_3+y_4x_4+y_ix_i, x_2=z_3x_3+z_4x_4+z_ix_i$  in $NonLin(g_i)$ for fresh variables $y_3,y_4,y_i, z_3,z_4,z_i$, and solve for the system of equations defined by all coefficient polynomials ($\in \F[y_3,y_4,y_i, z_3,z_4,z_i]$), using Lemma \ref{lemma:polynomial-equations}. Since the number of solutions are bounded ($rank(g_i)\geq 5\Rightarrow |\mathcal{S}(NonLin(g_i))|\leq 3d^7$ by Part $1$ of Proposition \ref{propn:ordinary-line-dim-main}) and the number of variables are constant (i.e. $5$) we are able to solve these systems in $(nd\log|\F|)^{O(1)}$ time. Next, we glue tuples across the $\mathcal{S}_i$, $i\in [5,n]$. As mentioned earlier for every co-dimension $2$ space $\mathbb{V}(x_1-\ell_1, x_2-\ell_2)$ on which $NonLin(g)$ vanishes, we would have recovered the tuples $(x_1-\ell_1^i, x_2-\ell_2^i)$ inside the set $\mathcal{S}_i$. For every tuple, $(x_1-\ell_1^5, x_2-\ell_2^5) \in \mathcal{S}_5$, we go over all $i\in [6,n]$ and try to find the tuple $(x_1-\ell_1^i, x_2-\ell_2^i)$. This is done by comparing coefficients of $x_3, x_4$. Using all such tuples, we can combine the linear forms and arrive at tuple $(x_1-\ell_1, x_2-\ell_2)$ on which $NonLin(g)$ vanishes. However, for some $i$, we might have two tuples that can be glued to $(x_1-\ell_1^5, x_2-\ell_2^5)$. To avoid this situation, we apply another random invertible transformation to linear forms in each tuple in every $\mathcal{S}_i$, $i\in [5,n]$. This transformation maps $x_j\mapsto x_j$, $j\in [1,4]$ and $x_j\mapsto x_j + \alpha_{j,3} x_3 + \alpha_{j,4}x_4$, where $\alpha_{j,k}$ ($j\in [5,n], k\in [2]$) are independently chosen from the uniform distribution on $\F$. In Lemma \ref{lemma:annihilating-spaces-details-2}, we show that after this transformation, with high probability, linear forms in no two distinct tuples in $\mathcal{S}_i$ have identical coefficients of $x_3,x_4$ i.e. the gluing process will find only one candidate in every $\mathcal{S}_i$. Since all $\mathcal{S}_i$ are of size $\leq 3d^7$, this gluing process gives a set (of tuples of linear forms) of size $\leq 3d^7$. Then, using randomized polynomial identity test from Lemma \ref{lemma:randomized-polynomial-identity-test}, we check whether $NonLin(g)$ actually vanishes on the glued co-dimension $2$ subspaces or not and only keep the ones where it does. Our algorithm runs in $(nd\log|\F|)^{O(1)}$ time. Complete details can be found in Algorithm \ref{algorithm:compute-annihilating-subspaces} in Section \ref{section:codim-two-subspaces-f-vanishes}.
   \subsubsection{Proposition \ref{propn:ordinary-line-dim-main}: Key ideas}
    Proposition \ref{propn:ordinary-line-dim-main} is proved in Section \ref{section:ordinary-lines}. We describe the main ideas involved now. Proposition \ref{propn:ordinary-line-dim-main} basically implies that if $\mathcal{S}, \mathcal{T}$ are proper sets with $\mathcal{T}$ independent and large, then there exists $t\in \mathcal{T}$ such that span of the union of ordinary lines from $t$ into $\mathcal{S}$ i.e. $\sum_{W\in \mathcal{O}(t,\mathcal{S})}W$, is high dimensional. For the rest of the argument, we define the vector space $V = \sum_{t\in \mathcal{T}}\sum_{W\in \mathcal{O}(t,\mathcal{S})}W$.
    \begin{itemize}
        \item In Lemma \ref{lemma:ordinary-line-dim}, we show that $V$ contains $sp(\mathcal{S})$ inside it i.e. span of the union ordinary lines from $\mathcal{T}$ into $\mathcal{S}$ covers the set $\mathcal{S}$ completely. Assuming this, a simple union bound gives the required result. In order to show $sp(S) = V$, we show that the complement set $\mathcal{S}^\prime = \mathcal{S}\setminus V$ is empty. The basic idea is to observe that $\mathcal{S}^\prime$ does not have ordinary lines coming from any $t\in \mathcal{T}$. If $\mathcal{S}^\prime$ has an ordinary line from some $t\in \mathcal{T}$, we show that the line is contained in $V$ which is not possible as $\mathcal{S}^\prime \cap V = \phi$.
        \item Once we have shown this, Lemma \ref{lemma:two-sets-ordinary-lines} comes to our rescue. This lemma says that in order to have no ordinary lines from $\mathcal{T}$ into $\mathcal{S}^\prime$, the size of $\mathcal{T}$ should be small i.e. $|\mathcal{T}|\leq \log |\mathcal{S}^\prime| + 1$, which is contradictory to what is assumed in the proposition and therefore we have a contradiction $\Rightarrow \mathcal{S} \subset V$.
        \item Therefore, the only thing left to explain is Lemma \ref{lemma:two-sets-ordinary-lines}. This lemma essentially says that if there are no ordinary lines from $\mathcal{T}$ into $\mathcal{S}$, then $\mathcal{T}$ spans a low dimensional space. We prove\footnote{we are thankful to Neeraj Kayal for sharing this simple proof.} this using Claim \ref{claim:one-one-map} which uses the non existence of ordinary lines to construct a one to one function from subsets of $[|\mathcal{T}|-1]$ into $\mathcal{S}$ implying that $|\mathcal{S}| \geq 2^{|\mathcal{T}|-1} \Rightarrow |\mathcal{T}| \leq \log |\mathcal{S}| +1$. Complete details are provided in Section \ref{section:ordinary-lines}.

    \end{itemize}

\section{Preliminaries}
\label{section:definitions-and-notation}

\subsection{Notations and definitions}
Throughout the paper $[n]$ will denote the set $\{1,\ldots,n\}$, $[m,n]$ will denote the set $\{m, m+1, \ldots,n-1, n\}$ and $\F$ will denote a finite field. We use calligraphic letters like $\mathcal{B,P,Q,R,S,T,X}$ to denote sets. Bold small letters ${\bf x, y, u}$ are used to represent column vectors or tuples of variables. Unless otherwise specified, ${\bf x}$ will denote the tuple $(x_1,\ldots,x_n)$. Bold capital letters ${\bf A,B}$ are used to represent matrices. $\F[{\bf x}]$ denotes the ring of polynomials in variables ${\bf x} = (x_1,\ldots,x_n)$ with coefficients in field $\F$. Capital letters like $G, H, T_1,T_2, S_1, S_2, U, U_i$ are either used to denote polynomials that are a product of linear forms. Small letters $f,g,h, u, \ell$ are also used to denote polynomials and linear forms. Let $g,f$ be any two polynomials, then, $g$ divides $f$ is denoted by $g\mid f$ and $g$ does not divide $f$ is denoted by $g\nmid f$.

\begin{definition}[Depth $3$ circuit, $\Sigma\Pi\Sigma$]
\label{defn:sps}
A depth $3$ circuit is a layered arithmetic circuit with three layers of nodes labelled by arithmetic operations, defined on a finite number of variables. First and third ($\Sigma$)  layers have addition nodes and second ($\Pi$) layer has multiplication nodes. Top layer has a single addition node. 
\end{definition}

\begin{definition}[Homogeneous Depth $3$ circuit, $\Sigma\Pi\Sigma(k,n,d,\F)$]
\label{defn:homogeneous-sps}
A $\Sigma\Pi\Sigma(k,n,d,\F)$ circuit is a depth three circuit such that the first ($\Sigma$) layer computes linear forms on $n$ variables, there are $k$ multiplication nodes at the second $(\Pi)$ layer all having in-degree $d$, and the addition node at third($\Sigma$) layer can only have incoming edges from the $k$ multiplication nodes at second layer.
Any circuit belonging to this class naturally computes an $n-$variate polynomial $f = M_1 + \ldots + M_k$,
where $M_i, i\in [k]$ are product of linear forms computed at the multiplication gates and $deg(M_1) = \ldots = deg(M_k) = d$.
\end{definition}

\begin{definition}[Simple $\Sigma\Pi\Sigma(k, n ,d, \F)$ circuit]
\label{defn:simple}
Let $C$ be a $\Sigma\Pi\Sigma(k,n,d,\F)$ circuit computing polynomial $f = M_1 + \ldots + M_k$ as described in Definition \ref{defn:homogeneous-sps}. We say that $C$ is simple if $gcd(M_1, \ldots, M_k)=1$.
\end{definition}

\begin{definition}[Minimal $\Sigma\Pi\Sigma(k, n ,d, \F)$ circuit]
\label{defn:minimal}
Let $C$ be a $\Sigma\Pi\Sigma(k,n,d,\F)$ circuit computing the polynomial $f = M_1 + \ldots + M_k$ as described in Definition \ref{defn:homogeneous-sps}. We say that $C$ is minimal if no proper sub collection of polynomials $M_1, \ldots,M_k$ sums to zero.
\end{definition}

\begin{definition}[Rank of $\Sigma\Pi\Sigma(2,n,d,\F)$ circuit, Section $1.3$ in \cite{Shpilka2007}]
\label{defn:rank}
Let $C$ be a $\Sigma\Pi\Sigma(2,n,d,\F)$ circuit computing the polynomial $f = M_1 + M_2$ as described in Definition \ref{defn:homogeneous-sps}. If $G = gcd(M_1, M_2)$, then $f$ can be written as $f = G\times T_1 + G\times T_2$
where $G,T_1,T_2$ are product of linear forms with $gcd(T_1,T_2)=1$. \emph{Rank} of $C$ is then defined as 
\[
rank(C) = dim(sp\{\text{linear form }\ell\in \F[{\bf x}] : \ell \mid T_1\times T_2\})
\]
\end{definition}

\begin{definition}[Rank of polynomial]
\label{defn:rank-of-polynomial}
For any polynomial $f\in \F[{\bf x}]$ computable by a $\Sigma\Pi\Sigma(2,n,d,\F)$ circuit, it's rank, called $rank(f)$ is defined as the minimum of $rank(C)$ over all $\Sigma\Pi\Sigma(2,n,d,\F)$ circuits computing $f$.
\end{definition}

\begin{definition}[Proper set, Section $5.3$, \cite{Dvi12}]
\label{defn:proper-set}
We call a set of points $v_1,\ldots,v_m \in \F^n$ \emph{proper} if no two points are a constant multiple of each other and the zero point is not in the set (i.e. it is a subset of the projective space).
\end{definition}

\begin{definition}[Ordinary line, Section $5.1$, \cite{Dvi12}]
\label{defn:ordinary-line}
Let $\mathcal{S}\subset \F^n$ be a proper set. For any $t\in \F^n$ and $s\in \mathcal{S}$, such that $t\notin sp\{s\}$, the vector space $sp\{t,s\}$ is called an ordinary line from $t$ into $\mathcal{S}$, if and only if $sp\{t,s\}\cap \mathcal{S} \subseteq \{t,s\}$. Define $\mathcal{O}(t,\mathcal{S})$ to be the set of ordinary lines from $t$ into $\mathcal{S}$.
\end{definition}

\begin{definition}[Linear and Non-linear parts]
\label{defn:lin-nonlin-part}
Let $f\in \F[\B x]$. We define $Lin(f)$, called the linear part of $f$ to be the product (with multiplicity) of all linear polynomials dividing $f$ and $NonLin(f)$ called the non-linear part of $f$ as $NonLin(f) = \frac{f}{Lin(f)}$\footnote{$Lin(f), NonLin(f)$ are unique up to scalar factors which are constrained such that $f = Lin(f)\times NonLin(f)$.}.
\end{definition}

\begin{definition}
\label{defn:polynomial-vanishing-codim}
Let $f\in \F[\B x]$. For any co-dimension $2$ space $W =\mathbb{V}(\ell_1, \ell_2)\subset \F^n$, we say that $f$ vanishes on $W$, if, for isomorphism $\Phi : \F[\B x]\rightarrow\F[\B x]$ mapping $\ell_1\mapsto x_1, \ell_2\mapsto x_2$, the polynomial $\Phi(f)_{|_{x_1=0, x_2=0}}$ is identically zero. This is well defined, i.e. if we take some other linear forms $\ell_1^\prime, \ell_2^\prime$ such that $W = \mathbb{V}(\ell_1^\prime, \ell_2^\prime)$ and some other isomorphism $\Phi^\prime$ mapping $\ell_1^\prime\mapsto x_1, \ell_2^\prime \mapsto x_2$, then $
\Phi(f)_{|_{x_1=0, x_2=0}} = 0 \Leftrightarrow \Phi^\prime(f)_{|_{x_1=0, x_2=0}} = 0$.
For any polynomial $f$, we define $\mathcal{S}(f)$ to be the set of all co-dimension $2$ sub-spaces $W\subset \F^n$ such that $f$ vanishes on $W$.
\end{definition}

\begin{definition}
\label{defn:factorize-codim-1-subspace}
Let $f\in \F[\B x]$. For any co-dimension $1$ space $W\subset \F^n$, we say that $f$ factorizes into non-zero linear forms on $W$, if, for linear form $\ell_1$ such that $W = \mathbb{V}(\ell_1)$, and isomorphism $\Phi : \F[\B x]\rightarrow\F[\B x]$ mapping $\ell_1\mapsto x_1$, the polynomial $\Phi(f)_{|_{x_1=0}}$ is a non-zero product of linear forms in $\F[x_2,\ldots,x_n]$. It's easy to see that this is well defined, i.e. if we take some other linear form $\ell_1^\prime$ such that $W = \mathbb{V}(\ell_1^\prime)$ and some other isomorphism $\Phi^\prime$ mapping $\ell_1^\prime\mapsto x_1$ then $\Phi(f)_{|_{x_1=0}}$ is a non-zero product of linear forms $\Leftrightarrow \Phi^\prime(f)_{|_{x_1=0}}$ is a non-zero product of linear forms.
\end{definition}

\begin{definition}[Candidate linear forms]
\label{definition:candidate-linear-form}
Let $f\in \F[\B x]$. Let $\ell$ be a linear form and $W = \mathbb{V}(\ell)$. Suppose $f$ factorizes into non-zero linear forms on $W$, and there exist linear forms $\ell_1,\ell_2$ with $\ell,\ell_1, \ell_2$ being linearly independent, such that $f$ vanishes on co-dimension $2$ subspaces $\mathbb{V}(\ell, \ell_1), \mathbb{V}(\ell, \ell_2)$.
Then, $\ell$, considered as a point in the projective space, is called a candidate linear form. The set of candidate linear forms is denoted by $\mathcal{L}(f)$. It's easy to see that $|\mathcal{L}(f)|\leq |\mathcal{S}(f)|^2$.
\end{definition}

\begin{definition}[Sylvester Gallai (SG) configuration, Definition $5.3.1$, \cite{Dvi12}]
\label{defn:sg}
A proper set $\mathcal{S} = \{s_1,\ldots,s_m\}\subset \F^n$ is called an SG configuration if for every $i\neq j \in [n]$, $\exists k\in [n]\setminus \{i,j\}$ with $s_i, s_j, s_k$ linearly dependent.
\end{definition}

\begin{definition}[Number of essential variables, restated from \cite{Bhargava2021}]
Let $f(\B x) \in \F[\B x]$. We say that $f(\B x)$ has $m$ ($\leq n$) essential variables if there exist an invertible matrix $A \in \F^{(n\times n)}$ such that $f(A\cdot \B x)$ depends only on $m$ variables.
\end{definition}

\subsection{Known results}
In this subsection, we list a few known results that are used in the paper.

\begin{lemma}[\cite{Carlini2006, Kayal2011}]
\label{lemma:variable-reduction}
Let $n,d$ be positive integers and $\F$ be a field with char($\F$) $>d$ or $0$. There is a randomized algorithm that takes as input black-box access to an $n$-variate degree $d$ polynomial $f(\B x) \in \F[\B x]$ having $m$ essential variables and computable by a circuit of size $s$, that runs in time $(nds)^{O(1)}$ and outputs an invertible matrix $A \in \F^{(n×n)}$
such that $f(A\cdot \B x)$ depends only on the first $m$-variables.
\end{lemma}

\begin{lemma}[Solving polynomial equations, Implied from \cite{Ierardi1989, Lazard2001}]
\label{lemma:polynomial-equations}
There is a randomized algorithm that takes as input $n$ variate polynomials $f_1,\ldots,f_m$ each of degree $\leq d$. If the system of equations defined by setting all these polynomials simultaneously to zero, has finitely many solutions in $\bar{\F}$ and all solutions are in $\F^n$, then the algorithm computes all solutions with probability $1-exp(-mnd\log|\F|)$. Running time of the algorithm is $(md^n\log|\F|)^{O(1)}$.
\end{lemma}

\begin{lemma}[Schwartz Zippel Lemma, \cite{Schwartz1980, Zippel1979}]
\label{lemma:szlemma}
Let $p(x_1,\ldots,x_n)$ be a polynomial of total degree $d$ such that it is not identically zero. Let $S\subset\F$ be any finite set. For  $s_1,\ldots,s_n$ picked independently and uniformly from $S$,
\[
Pr[p(s_1,\ldots,s_n)=0]\leq \frac{d}{|S|}.
\]
\end{lemma}
This immediately gives the following randomized polynomial identity test.
\begin{lemma}[Randomized polynomial identity test, Section $1$, Lemma $1.2$ in \cite{Sax09}]
\label{lemma:randomized-polynomial-identity-test}
There exists a randomized algorithm that takes as input integer $n$ and black-box access to a degree $d$, $n-$variate polynomial $f$ with coefficients in $\F_q$, runs in time $(nd\log q)^{O(1)}$ and outputs either $`yes'$ or $`no'$ such that,
\[
	\begin{array}{ll}
		\text{output is }`yes'  & \mbox{if } f \equiv 0 \\
		Pr[\text{output is }`no']\geq 1-o(1) & \mbox{if } f \not \equiv 0
	\end{array}
\]
\end{lemma}

\begin{lemma}[$\Sigma\Pi\Sigma(k,n,d, \F)$ deterministic polynomial identity test, Theorem $1$ in \cite{SaxSes11}]
\label{lemma:sps(k)-pit}
There exists a deterministic algorithm that takes as input black-box access to a degree $d$, $n-$variate polynomial $f$ computable by a $\Sigma\Pi\Sigma(k,n,d, \F)$ circuit, runs in time $(nd^k\log |\F|)^{O(1)}$ and, outputs $`yes'$ if $f\equiv 0$ and $`no'$ if $f\not \equiv 0$.

\end{lemma}

\begin{lemma}[$\Sigma\Pi\Sigma(k,n,d, \F)$ Rank bound, Theorem $1.7$ in \cite{Saxena2013}]
\label{lemma:sps(k)-rank-bound}
Let $C$ be a $\Sigma\Pi\Sigma(k,n,d, \F)$ circuit, over an arbitrary field $\F$, that is simple, minimal and
zero. Then, $rank(C) < 3k^2 + \frac{k^2}{4}\log d$.
\end{lemma}

\begin{lemma}[Black-box multivariate polynomial interpolation, Theorem $11$ in \cite{Klivans2001}]
\label{lemma:interp}
Let $n,m,d$ be parameters and $\F$ be a finite field. There exists a deterministic algorithm that runs in time $(nmd\log|\F|)^{O(1)}$, and outputs a set $S$ of points in $\F^n$, such that given black-box access to any polynomial $f\in \F[x_1,\ldots,x_n]$ with at most $m$ monomials, the coefficients of all monomials can be recovered in $(nmd\log|\F|)^{O(1)}$ time using evaluations from the set $\{f(s) : s\in S\}$.
\end{lemma}

\begin{lemma}[Effective Hilbert irreducibility / Quantitative Bertini theorem, Corollary $2$ \cite{Kal91}, Remarks $11.5.33$, $11.5.66$ \cite{MulPa13}, Theorem $1.1$ \cite{KSS14}]\label{lemma:effectivehilbert}
 Let $\F$ be a perfect field and $g({\bf x})\in \F[{\bf x}]$ be a degree $d$ irreducible polynomial. Pick tuples, ${\bf a} = (a_2,\ldots, a_n)$,  ${\bf b} = (b_1,\ldots,b_n)$, ${\bf c} = (c_1,\ldots,c_n)$
 such that every $a_i,b_j,c_k$ is chosen uniformly randomly and independently from a set $S\subset \F$.
 Consider the bi-variate restriction
\[
\hat{g}(X,Y) = g(X + b_1Y + c_1, a_2X + b_2Y + c_2, \ldots a_nX + b_n Y + c_n)                                                                                              
\]
Then, 
\[
 P[ ({\bf a, \bf  b,\bf  c}) \in S^{n-1}\times S^n \times S^n  :  \hat{g}(X,Y) \text{ is not irreducible }] \leq \frac{2d^4}{|S|}
\]
\end{lemma}

\begin{lemma}[Black-box multivariate polynomial factorization, \cite{Kaltofen1990}]
\label{lemma:black-box-multivariate-polynomial-factorization}
There exists a randomized algorithm that takes as input black-box access to a degree $d$, $n-$variate polynomial $f$ with coefficients in $\F$, runs in time $(nd\log |\F|)^{O(1)}$ and outputs black-box access to polynomials $f_1, \ldots, f_m$ ($m\leq d$) along with integers $e_1,\ldots,e_m$ such that,
\[
Pr[f\equiv f_1^{e_1}\ldots f_m^{e_m} \bigwedge f_1,\ldots,f_m\text{ are irreducible}] \geq 1-o(1).
\]
\end{lemma}

\begin{corollary}[Decomposition into linear and non-linear factors]
\label{corrolary:decomposition-into-linear-and-non-linear-factors}
There exists a randomized algorithm that takes as input black-box access to a degree $d$, $n-$variate polynomial $f$ with coefficients in $\F$, runs in time $(nd\log |\F|)^{O(1)}$ and outputs a list $\{\ell_1,\ldots,\ell_s\}$ ($s\leq d$) of affine forms along with black-box access to a polynomial $NonLin(f)$ such that,
\[
Pr[f\equiv l_1\ldots l_s NonLin(f) \bigwedge NonLin(f)\text{ has no linear factors}] \geq 1-o(1).
\]
\end{corollary}
\begin{proof}
We give the algorithm in Algorithm \ref{algorithm:sps(1)-reconstruction}. Correctness and time complexity proofs are pretty straight-forward using Lemma \ref{lemma:black-box-multivariate-polynomial-factorization} and Lemma \ref{lemma:randomized-polynomial-identity-test}.
\begin{algorithm}
    \caption{Decomposition into linear and non-linear factors}
    \label{algorithm:sps(1)-reconstruction}
    \hspace*{\algorithmicindent} \textbf{Input - } Black-box access to polynomial $f$, integers $n, d$. \\
 \hspace*{\algorithmicindent} \textbf{Output - } List of affine forms $L$ and black-box access to polynomial $NonLin(f)$.
    
    \begin{enumerate}
        \item Using algorithm in Lemma \ref{lemma:black-box-multivariate-polynomial-factorization} on black-box computing $f$, obtain black-box access to polynomials $f_1,\ldots,f_m$ along with integers $e_1,\ldots,e_m$. Initialize lists $L, B\gets \phi$.
        \item For every $i\in [m]$, construct linear form $\ell_i = \sum\limits_{j=1}^n (f_i({\bf e_j}) - f_i({\bf 0}))x_j + f_i({\bf 0})$, where ${\bf e_j} \in \F^n$ is the vector with $1$ in $j^{th}$ co-ordinate and $0$ elsewhere and ${\bf 0} = (0,\ldots,0)\in \F^n$. Using randomized polynomial identity test in Lemma \ref{lemma:randomized-polynomial-identity-test}, check if $f_i-\ell_i\equiv 0$. If yes, add $e_i$ copies of $\ell_i$ to $L$. Otherwise add $e_i$ copies of black-box computing $f_i$ to $B$.
        \item Simulate black-box $\mathcal{B}$ computing polynomial $NonLin(f) = \prod\limits_{h\in B}h$. {\bf Return $L, \mathcal{B}$}.
    \end{enumerate}
\end{algorithm}
\end{proof}

\section{Low Rank Reconstruction: Proof of Theorem \ref{theorem:low-rank-reconstruction}}
\label{section:low-rank-reconstruction}

\begin{algorithm}
    \caption{Low rank reconstruction}
    \label{algorithm:low-rank-reconstruction}
    \hspace*{\algorithmicindent} \textbf{Input - } Black-box access to $f$, integers $n, d$. \\
 \hspace*{\algorithmicindent} \textbf{Output - } $\Sigma\Pi\Sigma$ circuit $C$ or $\#$.
\begin{enumerate}
\item Using Algorithm \ref{algorithm:sps(1)-reconstruction} with inputs as black-box access to $f$ and integers $n,d$, compute list of linear factors $\ell_1,\ldots, \ell_s$ and black-box access to $NonLin(f)$. Compute degree of $NonLin(f)$ as $t=d-s$. Using this black-box and integers $n,t$ as input to Algorithm \ref{algorithm:compute-annihilating-subspaces}, obtain set $\mathcal{S}(NonLin(f))$ containing tuples of linear forms representing co-dimension $2$ subspaces of $\F^n$ on which $NonLin(f)$ vanishes.

\item Construct set $\mathcal{L}$ of linear forms $\ell$, such that for some $\ell^\prime$ either $(\ell, \ell^\prime)$ or $(\ell^\prime, \ell)$ is in $\mathcal{S}(NonLin(f))$.  For each $r\in [O(\log^3 d)]$, iterate over all $r$ sized linearly independent subsets $\{y_1,\ldots,y_r\}\subset \mathcal{L}$. Construct isomorphism $\Gamma$ mapping $y_i\mapsto x_i, i\in [r]$. Simulate black-box for $\Gamma(NonLin(f))$ and using Lemma \ref{lemma:interp} interpolate it as a linear combination of degree $t$ monomials in $\F[x_1,\ldots,x_r]$, obtaining a polynomial $h(x_1,\ldots,x_r)$.

\item By creating appropriate multiplication/addition gates, construct a $\Sigma\Pi\Sigma(t^r, n, d, \F)$ circuit $C$ that computes polynomial
\[
f^\prime = \ell_1\times\ldots\times\ell_s\times h(y_1,\ldots,y_r)
\]
Using randomized polynomial identity test from Lemma \ref{lemma:randomized-polynomial-identity-test}, check if $f -f^\prime = 0$. If yes, {\bf Return $C$}. If no, try the next $r$ sized subset in Step $2$. If all $r$ sized subsets have been tried, $r=r+1$. 
\end{enumerate}        		
\end{algorithm}

We present the low rank reconstruction algorithm required by Theorem \ref{theorem:low-rank-reconstruction} in Algorithm \ref{algorithm:low-rank-reconstruction}. We analyze it's correctness and running-time here.  Using correctness of Algorithm \ref{algorithm:sps(1)-reconstruction} and Algorithm \ref{algorithm:compute-annihilating-subspaces}, at the end of step $1$, with probability $1-o(1)$, we have obtained a black-box computing $NonLin(f)$, degree $t$ of $NonLin(f)$, and all linear factors $\ell_1,\ldots,\ell_s$ (with multiplicity) of $f$. Next, we show that, for some $r\leq rank(f)$ and linear forms $y_1,\ldots,y_r$, Step $2$ computes a polynomial $h(x_1,\ldots,x_r)$ such that $NonLin(f) = h(y_1,\ldots,y_r)$. In order to do so we prove the following lemma.
\begin{lemma}
Let $r = rank(f)$. There exists linearly independent subset $\{y_1,\ldots,y_r\} \subset \mathcal{L}$ such that it spans the set of linear factors of $T_1\times T_2$, implying the existence of the polynomial $h$. 
\end{lemma}
\label{claim:L-is-good}
\begin{proof}
Since $rank(f) \geq 5$, we know that $NonLin(f)$ is a non-constant polynomial. Consider any linear form $\ell\mid T_i$ for some $i\in [2]$. We will show that there is some $\ell^\prime\mid T_{3-i}$ such that $NonLin(f)$ vanishes on the co-dimension $2$ subspace $\mathbb{V}(\ell, \ell^\prime)$. Assuming this is true, we know there is a tuple $(\ell_1, \ell_2) \in \mathcal{S}(NonLin(f))$ such that $\mathbb{V}(\ell, \ell^\prime) = \mathbb{V}(\ell_1, \ell_2) \Rightarrow sp(\{\ell, \ell^\prime\}) = sp(\{\ell_1, \ell_2\})$. By construction of set $\mathcal{L}$, $\ell_1, \ell_2 \in \mathcal{L}$. By going over different $\ell$ dividing $T_1\times T_2$ this process would give a list of $2m$ ($m = deg(T_1)=deg(T_2)$) linear forms $\{\ell_1,\ldots,\ell_{2m}\}\subset \mathcal{L}$ such that
\[
sp(\{\text{linear form }\ell : \ell \mid T_1\times T_2\}) \subset sp(\{\ell_1,\ldots,\ell_{2m}\})\subset  sp(\{\text{linear form }\ell : \ell \mid T_1\times T_2\})
\]
Since $sp(\{\text{linear form }\ell : \ell \mid T_1\times T_2\})$ is $rank(f)$ dimensional we get that there are $r$ linearly independent linear forms $y_1,\ldots,y_r\in \{\ell_1,\ldots,\ell_{2m}\}\subset \mathcal{L}$ and the proof would be complete. So we only need to show that there exists $\ell^\prime\mid T_{3-i}$ such that $NonLin(f)$ vanishes on the co-dimension $2$ subspace $\mathbb{V}(\ell, \ell^\prime)$. To see this, let $L$ be the product of all linear factors of $T_1+T_2$. Let $\Phi$ be an isomorphism mapping $\ell\mapsto x_1$. On setting $x_1=0$, we get,
\[
\Phi(L)_{|_{x_1=0}}\times \Phi(NonLin(f))_{|_{x_1=0}} = \Phi(T_{3-i})_{|_{x_1=0}} \neq 0.
\]
The non zeroness comes from the fact that $gcd(T_1,T_2)=1$. The above equation implies\footnote{by using unique factorization in the ring $\F[x_2,\ldots,x_n]$} that there is some linear form $\ell^\prime \mid T_{3-i}$ such that $\Phi(\ell^\prime)_{|_{x_1=0}}$ divides $\Phi(NonLin(f))_{|_{x_1=0}}$. Now, define the isomorphism $\Delta$ mapping $x_1\mapsto x_1, \Phi(\ell^\prime) \mapsto x_2$\footnote{$x_1$ and $\Phi(\ell^\prime)$ are linearly independent, otherwise $\ell$ divides $\ell^\prime$ violating $gcd(T_1,T_2)=1$. }. Applying $\Delta$ to the fact that $\Phi(\ell^\prime)_{|_{x_1=0}}$ divides $\Phi(NonLin(f))_{|_{x_1=0}}$, we get that $\Delta(\Phi(\ell^\prime)_{|_{x_1=0}})\mid \Delta(\Phi(NonLin(f))_{|_{x_1=0}})$. Since $\Delta$ fixes $x_1$, we get $ \Delta(\Phi(\ell^\prime))_{|_{x_1=0}}\mid \Delta(\Phi(NonLin(f)))_{|_{x_1=0}}$. So there is polynomial $g$ such that 
\[
\Delta(\Phi(NonLin(f)))_{|_{x_1=0}} = \Delta(\Phi(\ell^\prime))_{|_{x_1=0}} \times g.
\]
Now setting $x_2=0$ on both sides will send the right hand side to $0$ since $\Delta\circ \Phi$ maps $\ell\mapsto x_1, \ell^\prime\mapsto x_2$. Therefore $\Delta(\Phi(NonLin(f)))_{|_{x_1=0, x_2=0}}=0$, and so using Definition \ref{defn:polynomial-vanishing-codim} one can see that $NonLin(f)$ vanishes on the co-dimension $2$ subspace $\mathbb{V}(\ell, \ell^\prime)$.
\end{proof}

$h(y_1, \ldots,y_r)$ naturally exhibits a $\Sigma\Pi\Sigma(t^r, n,t,\F)$ circuit. This can be seen as follows. Addition gates at the bottom layer will compute linear forms $y_1,\ldots,y_r$. For each monomial, there will be one multiplication gate. If $x_j^k$ is the largest power of $x_j$ dividing some monomial, then there will be $k$ connections from $y_j$ to the multiplication gate corresponding to this monomial. Finally, the top layer is connected to all the multiplication gates and weight on such an edge is equal to the coefficient of the monomial the multiplication gate corresponded to. 
Step $3$ just multiplies this circuit with all the linear factors and therefore computes a candidate $\Sigma\Pi\Sigma(t^r, n, d, \F)$ circuit for $f$. Randomized polynomial identity test in Step $3$ ensures that with high probability we output a correct $\Sigma\Pi\Sigma(d^r, n, t, \F)$ circuit for $f$. If for some $r$ and linear forms $y_1,\ldots,y_r$, an incorrect circuit gets constructed, probability that it will be outputted is $o(1)$. There are at most $(d^{\log^3 d} \log^3 d )^{O(1)}$ many such bad settings of $r$ and $y_1,\ldots,y_r$. Using boosting with independent runs of randomized polynomial identity test, we can make error exponentially small in $nd$ so that overall the probability of error still remains $o(1)$ by union bound $\Rightarrow$ with probability $1-o(1)$ all these bad settings will be rejected. For $r=rank(f)$ and the correct linearly independent set $\{y_1,\ldots,y_r\}$ (i.e. one spanning all linear factors of $T_1\times T_2$), we have seen that with probability $1-o(1)$, a correct circuit will be constructed which will always pass the randomized polynomial identity test and will be returned. So overall with probability $1-o(1)$, a correct $\Sigma\Pi\Sigma(t^r, n, d, \F)$ circuit for $f$ will be returned. Next we discuss the time complexity of the above algorithm.

\begin{lemma}
Algorithm \ref{algorithm:low-rank-reconstruction} takes $(nd^{\log^3 d}\log|\F|)$ time.
\end{lemma}
\begin{proof}
Time complexity of Algorithm \ref{algorithm:sps(1)-reconstruction} and Algorithm \ref{algorithm:compute-annihilating-subspaces} imply that Step $1$ takes $(nd\log|\F|)^{O(1)}$ time. $\mathcal{L}$ can be constructed in $(nd\log|\F|)^{O(1)}$ time since it involves iterating over the $d^{O(1)}$ sized set $\mathcal{S}(NonLin(f))$. Our search for the correct $r=rank(f)$ and linear forms $y_1,\ldots,y_r$ takes $(nd^{\log^3 d}\log|\F|)^{O(1)}$ time in the worst case and multivariate interpolation (Lemma \ref{lemma:interp}) also takes the same amount of time in the worst case. Step $3$ multiplies linear factors to all the gates in the circuit for $NonLin(f)$ and therefore takes  $(nd^{\log^3 d}\log|\F|)^{O(1)}$ time and therefore overall time complexity is $(nd^{\log^3 d}\log|\F|)^{O(1)}$.
\end{proof}

\section{High Rank Reconstruction: Proof of Theorem \ref{theorem:high-rank-reconstruction}}
\label{section:high-rank-reconstruction}
The algorithm in Theorem \ref{theorem:high-rank-reconstruction} is presented below in Algorithm \ref{algorithm:high-rank-reconstruction}. This algorithm further calls Algorithms \ref{algorithm:computing-candidate-set-of-linear-forms}, \ref{algorithm:corner-case} and \ref{algorithm:general-case}. We present and analyze them in Sections \ref{subsection:candidate-linear-forms}, \ref{subsection:corner-case} and \ref{subsection:reconstruction-li-set-given} respectively. Correctness of our algorithm heavily relies on Lemma \ref{lemma:general-case-search}, which we prove in Section \ref{subsection:identify-li-set}. We first give the full algorithm in Algorithm \ref{algorithm:high-rank-reconstruction} and then discuss it's correctness and time complexity.

\begin{algorithm}
    \caption{High rank reconstruction}
    \label{algorithm:high-rank-reconstruction}
    \hspace*{\algorithmicindent} \textbf{Input - } Black-box access to $f$, integers $n, d$. \\
 \hspace*{\algorithmicindent} \textbf{Output - } $\Sigma\Pi\Sigma(2, n, d, \F)$ circuit $C$ or $\#$.
\begin{enumerate}

\item Run Algorithm \ref{algorithm:corner-case} with inputs as black-box access to $f$ along with integers $n,d$. If output is a circuit $C$, {\bf Return $C$}. If output was $\#$, go to the next step. 

\item  Using Algorithm \ref{algorithm:sps(1)-reconstruction} with input as black-box access to $f$ and integers $n,d$, compute list of linear factors $\ell_1,\ldots, \ell_s$ and black-box access to $NonLin(f)$. Compute the degree of $NonLin(f)$ as $t=d-s$. 

\item Using Algorithm \ref{algorithm:computing-candidate-set-of-linear-forms} with inputs as black-box access to $f$ and integers $n,d$, construct the set $\mathcal{L}(NonLin(f))$. For each $\ell\in \mathcal{L}(NonLin(f))$ consider all linear forms $\ell^\prime \in \mathcal{L}(NonLin(f))\setminus\{\ell\}$ such that $sp\{\ell, \ell^\prime\}$ does not intersect $\mathcal{L}(NonLin(f))$ at any point other than $\ell, \ell^\prime$. Find a maximal independent set $\mathcal{X}$ of such $\ell^\prime$s and continue if $|\mathcal{X}| = \Omega(\log^2d)$. If no such $\ell$ exists, {\bf Return $\#$}. Otherwise, partition $\mathcal{X}$ into equal parts of size $\Omega(\log d)$ each and iterate over all parts $\mathcal{B}$.
\begin{enumerate}
    \item Initialize sets $\mathcal{U}, \mathcal{V} \gets \phi$. Iterate over all linear forms $\ell^\prime \in \mathcal{B}$. Define an isomorphism $\Phi$ mapping $\ell\mapsto x_1, \ell^\prime \mapsto x_2$ and using Lemma \ref{lemma:randomized-polynomial-identity-test}, check if $\Phi(NonLin(f))_{|_{x_1=0,x_2=0}}\equiv 0$. If yes, add $\ell^\prime$ to $\mathcal{U}$ else add it to $\mathcal{V}$. Select $r = 60\log d+61$ linear forms $y_1,\ldots,y_r$ from the larger of $\mathcal{U}, \mathcal{V}$. 
    
    \item Run Algorithm \ref{algorithm:general-case} with inputs as black-box access to $f$, integers $n, d$ and linear forms $y_1,\ldots,y_r$. If it returns a $\Sigma\Pi\Sigma(2,n,d,\F)$ circuit $C$, {\bf Return $C$}. Else, go to the next partition $\mathcal{B}$ and then to the next linear form $\ell$ in the search.
\end{enumerate}
\item {\bf Return $\#$}
\end{enumerate}     
\end{algorithm}
We first prove the correctness of the above algorithm. Step $1$ first tries to solve the corner case where one of $T_1,T_2$ is power of a linear form. By correctness of Algorithm $6$, we know that, if this corner case is satisfied, then with probability $1-o(1)$, the correct $\Sigma\Pi\Sigma(2,n,d,\F)$ circuit is returned. Also Algorithm $6$ never returns an incorrect circuit. Therefore with high probability Step $1$ will complete the reconstruction if the corner case condition holds. If it does not hold this algorithm will always proceed to Step $2$. Also, if it does not return a circuit we can assume that with high probability the corner case does not hold and therefore linear factors of each $T_i$ span at least a two dimensional space. By correctness of Algorithm \ref{algorithm:sps(1)-reconstruction}, we know that with probability $1-o(1)$, Step $2$ correctly obtains a black-box computing $NonLin(f)$, it's degree $t$ and correctly identifies all linear factors of $f$ with multiplicity. Correctness of the next step is proved in the following lemma.

\begin{lemma}
If outputs of Steps $1$ and $2$ are correct, then with probability $1-o(1)$, Step $3$ computes a $\Sigma\Pi\Sigma(2,n,d,\F)$ circuit computing $f$.
\end{lemma}
\begin{proof}
By correctness of Algorithm \ref{algorithm:computing-candidate-set-of-linear-forms}, we know that the set $\mathcal{L}(NonLin(f))$ is correctly computed. Our algorithm goes through all linear forms $\ell \in \mathcal{L}(f)$ and for each such linear form goes through $\Omega(\log d)$ sized sets which are parts of a partition of the set $\mathcal{X}$ defined using $\ell$. In Step $3(b)$, correctness of Algorithm \ref{algorithm:general-case} ensures that if a circuit is returned for any choice of $\ell, \mathcal{B}$, it is always correct. So all we need to show is that for some choice of $\ell, \mathcal{B}$, Algorithm \ref{algorithm:general-case} will return the correct circuit with high probability. We know from correctness of Algorithm \ref{algorithm:general-case} that if the linear forms $y_1, \ldots,y_r$ (that are given as input to it), all divide the same $T_i$ and are independent, then with high probability a correct circuit will be returned. Therefore, now all we need to show is that there is some choice of $\ell, \mathcal{B}$, for which the constructed $y_1,\ldots,y_r$ are independent linear forms dividing the same $T_i$. Since we have assumed that output of Step $1$ is correct, $f$ does not satisfy the corner case implying that linear factors of each $T_i$ span at least a two dimensional space and therefore Lemma \ref{lemma:general-case-search} can be applied. Parts $1,2,3$ of Lemma \ref{lemma:general-case-search} prove that such $\ell, \mathcal{B}$ exist for which the test in Step $3(a)$ creates a partition $\mathcal{U}\cup \mathcal{V} = \mathcal{B}$ such that linear forms in $\mathcal{U}$ divide $T_j$ and linear forms in $\mathcal{V}$ divide $T_{3-j}$ for some $j\in [2]$. Since $|\mathcal{B}| = \Omega(\log d)$, one of $\mathcal{U}, \mathcal{V}$ has size $\Omega(\log d)$. By construction $\mathcal{B}$ is linearly independent and thus both $\mathcal{U}, \mathcal{V}$ are linearly independent. Therefore $y_1,\ldots,y_r$ with $r=\Omega(\log d)$ are independent linear forms dividing the same $T_i$. This completes the proof.
\end{proof}
Now we discuss the time complexity of the above algorithm.
\begin{lemma}
Algorithm \ref{algorithm:high-rank-reconstruction} takes $(nd\log|\F|)^{O(1)}$ time.
\end{lemma}
\begin{proof}
Time complexity of Algorithm \ref{algorithm:corner-case} and Algorithm \ref{algorithm:sps(1)-reconstruction} imply that Steps $1$ and $2$ take $O(nd\log|\F|)^{O(1)}$ time. By Algorithm \ref{algorithm:computing-candidate-set-of-linear-forms} we know that the set $\mathcal{L}(f)$ has $d^{O(1)}$ size. We iterate over all $\ell\in \mathcal{L}(NonLin(f))$ and for each $\ell^\prime$ check if $sp\{\ell, \ell^\prime\}$ intersects $\mathcal{L}(NonLin(f))$ at any other point. This can be done in $(nd\log|\F|)^{O(1)}$ time. From these $\ell^\prime$, we can simply check linear independence of linear forms and create a maximal set in $\mathcal{X}$ in $(nd\log|\F|)^{O(1)}$ time. Creating a partition of $\mathcal{X}$, iterating over all parts $\mathcal{B}$, and isomorphism can be created in $(nd\log|\F|)^{O(1)}$ time. Isomorphism can be efficiently applied to the black-box computing $NonLin(f)$ by taking every input through $\Phi$ before applying the black-box. By time complexity of algorithm in Lemma \ref{lemma:randomized-polynomial-identity-test}, the check in Step $3(a)$ takes $(nd\log|\F|)^{O(1)}$ time. Time complexity of Step Algorithm \ref{algorithm:general-case} implies that Step $3(b)$ takes $(nd\log|\F|)^{O(1)}$ time. Therefore overall Algorithm \ref{algorithm:high-rank-reconstruction} takes $(nd\log|\F|)^{O(1)}$ time.
\end{proof}
In the next subsection, we explain construction of the candidate linear forms (Definition \ref{definition:candidate-linear-form}).

\subsection{Computing Candidate Linear forms}
\label{subsection:candidate-linear-forms}

Here is a lemma summarizing the construction of set $\mathcal{L}(NonLin(f))$ of candidate linear forms (Definition \ref{definition:candidate-linear-form}).
\begin{lemma}
\label{lemma:candidate-linear-forms}
There exists a randomized algorithm that takes as input integers $n,d$ and black-box access to $f$, runs in time $(nd\log|\F|)^{O(1)}$, and outputs a set $\mathcal{L}$ of linear forms such that,
\[
Pr[\mathcal{L} =\footnote{up to scalar multiplication of linear forms in the sets} \mathcal{L}(NonLin(f))] = 1-o(1).
\]
\end{lemma}
\begin{algorithm}
    \caption{Candidate linear forms}
    \label{algorithm:computing-candidate-set-of-linear-forms}
    \hspace*{\algorithmicindent} \textbf{Input - } Black-box access to polynomial $f$, integers $n, d$. \\
    \hspace*{\algorithmicindent} \textbf{Output - } A set of linear forms $\mathcal{L}$.
        \begin{enumerate}
            \item Using Algorithm \ref{algorithm:sps(1)-reconstruction} with inputs as black-box access to $f$ and integers $n,d$, obtain list of linear factors $\ell_1,\ldots,\ell_s$ and access to black-box computing $NonLin(f)$. Compute degree of $NonLin(f)$ as $t=d-s$. Using Algorithm \ref{algorithm:compute-annihilating-subspaces}, compute the set $\mathcal{S}$ of tuples of linear forms representing co-dimension $2$ subspaces on which $NonLin(f)$ vanishes.
            
            \item  Initialize $\mathcal{L}\gets \phi$. For all pairs of tuples $(p_1,q_1), (p_2, q_2) \in \mathcal{S}$, check if $sp\{p_1,q_1\}\cap sp\{p_2,q_2\} $is one dimensional. For this we construct the $n\times 4$ matrix $M$ with it's columns containing coefficients of $p_1,q_1,p_2,q_2$ respectively and then check by gaussian elimination whether rank of $M$ is $3$ or not.  If yes, the same gaussian elimination can be used to obtain the one dimensional space of solutions to $Mv = 0$ for $v\in \F^4$. Fixing one such non-zero solution $u = (\alpha_1, \alpha_2, \alpha_3, \alpha_4)^T$ then gives us a scalar multiple of $\ell$ as $\alpha_1 p_1 + \alpha_2 q_1$. If no scalar multiple of $\alpha_1 p_1 + \alpha_2 q_1$ is already present in $\mathcal{L}$, then we add it to $\mathcal{L}$.

            \item For each $\ell\in \mathcal{L}$, check whether $NonLin(f)$ restricted to $\mathbb{V}(\ell)$ factorizes into a non-zero product of linear forms (See Definition \ref{defn:factorize-codim-1-subspace}). This can be done by defining an isomorphism $\Phi$ mapping $\ell\mapsto x_1$, simulating black-box computing $\Phi(NonLin(f))_{|_{x_1=0}}$. Using Lemma \ref{lemma:randomized-polynomial-identity-test}, check if this black-box computes the $0$ polynomial. If 'yes', remove $\ell$ from $\mathcal{L}$. Otherwise, using Algorithm \ref{algorithm:sps(1)-reconstruction}, with inputs as this restricted black-box and integers $n,t$, compute list of linear factors and check whether there are $t$ of them. If not, then remove $\ell$ from $\mathcal{L}$. Finally, {\bf Return $\mathcal{L}$}.

        \end{enumerate}
\end{algorithm} 

Now we prove the correctness of Algorithm \ref{algorithm:computing-candidate-set-of-linear-forms}. By correctness of Algorithm \ref{algorithm:sps(1)-reconstruction}, we know that Step $1$ correctly obtains black-box access to $NonLin(f)$, it's degree $t$ and linear factors (with multiplicity) of $f$ with probability $1-o(1)$. Similarly by correctness of Algorithm \ref{algorithm:compute-annihilating-subspaces}, we know that with probability $1-o(1)$, the set $\mathcal{S}$ representing elements of $\mathcal{S}(NonLin(f))$ is correctly computed. We prove correctness of the next two steps in the following lemma.

\begin{lemma}
Assuming Step $1$ works correctly, with probability $1-o(1)$, the output $\mathcal{L}$ of Algorithm \ref{algorithm:computing-candidate-set-of-linear-forms} is the same\footnote{the linear forms in this output are correct upto scalar multiplication} as $\mathcal{L}(NonLin(f))$.
\end{lemma}
\begin{proof}
Consider any $\ell \in \mathcal{L}(NonLin(f))$. By definition of the set $\mathcal{L}(NonLin(f))$, we know that there are linear forms $\ell_1, \ell_2$ with $\ell, \ell_1, \ell_2$ linearly independent, such that the co-dimension $2$ subspaces $\mathbb{V}(\ell, \ell_1), \mathbb{V}(\ell, \ell_2) \in \mathcal{S}(NonLin(f))$. So some tuples $(p_1, q_1)$ and $(p_2, q_2)$ corresponding to these two subspaces will be present in $\mathcal{S}$ and will be encountered in Step $2$. Note that $\mathbb{V}(p_1, q_1) = \mathbb{V}(\ell, \ell_1)$ and $\mathbb{V}(p_2, q_2) = \mathbb{V}(\ell, \ell_2)$ implies that $sp\{p_1, q_1\} = sp\{\ell, \ell_1\}$ and $sp\{p_2, q_2\} = sp\{\ell, \ell_2\}$ further implying that $sp\{p_1, q_1\}\cap sp\{p_2, q_2\} = sp\{\ell\}$. This implies that there are scalars $\alpha_1, \alpha_2, \alpha_3, \alpha_4$ such that $\alpha_1 p_1 + \alpha_2 q_1 + \alpha_3 p_2 + \alpha_4 q_2 = 0$, giving us the system of equations as described in the algorithm. In order for the intersection to be one dimensional, the matrix $M$ should have rank $3$. We check that using gaussian elimination which also gives the one dimensional set of solutions. Any non-zero solution $(\alpha_1,\alpha_2,\alpha_3,\alpha_4)$ will then give a linear form $\alpha_1 p_1 + \alpha_2 q_1$ in the intersection which will be a scalar multiple of $\ell$. Thus, Step $2$ identifies a  scalar multiple of $\ell$ and adds it to $\mathcal{L}$. Step $3$ just checks whether $NonLin(f)$ factorizes as a product of non-zero linear forms on $\mathbb{V}(\ell)$ (see Definition \ref{defn:factorize-codim-1-subspace}). Correctness of Step $3$ is implied by correctness of Lemma \ref{lemma:randomized-polynomial-identity-test} and Algorithm \ref{algorithm:sps(1)-reconstruction}. Since $\ell \in \mathcal{L}(NonLin(f))$, it will pass this test and remain in $\mathcal{L}$. Now consider any $\ell \in \mathcal{L}$ that is returned. In Steps $2$ and $3$ we have checked whether it satisfies the conditions required for it to be in $\mathcal{L}(NonLin(f))$ or not. Therefore we do not return any extra linear forms and correctly output 
$\mathcal{L}(NonLin(f))$ with high probability.
\end{proof}

Now we discuss the time complexity of the above algorithm.
\begin{lemma}
Algorithm \ref{algorithm:computing-candidate-set-of-linear-forms} takes $(nd\log|\F|)^{O(1)}$ time.
\end{lemma}

\begin{proof}
Time complexity of Algorithm \ref{algorithm:sps(1)-reconstruction} and Algorithm \ref{algorithm:compute-annihilating-subspaces} imply that Step $1$ takes $(nd\log|\F|)^{O(1)}$ time. By first part of Proposition \ref{propn:annihilating-planes}, we know that $|\mathcal{S}|\leq 3d^7$ and therefore going over pairs of elements of $\mathcal{S}$ takes $O(nd\log|\F|)^{O(1)}$ time. Gaussian elimination on matrix $M$ takes $(n\log|\F|)^{O(1)}$ time for each pair of tuples. After Step $2$ we will have at most $|\mathcal{S}|^2$ many elements in $\mathcal{L}$ leading to a size of $d^{O(1)}$. In Step $3$ for every $\ell \in \mathcal{L}$, the construction of $\Phi$, simulation of black-box for $\Phi(NonLin(f))_{|_{x_1=0}}$ are done in $(n\log|\F|)^{O(1)}$ time. Time complexity of algorithm provided in Lemma \ref{lemma:randomized-polynomial-identity-test} tests in time $(nd\log|\F|)^{O(1)}$, whether this new polynomial is identically zero or not. Finally, time complexity of Algorithm \ref{algorithm:sps(1)-reconstruction} implies that in time $(nd\log|\F|)^{O(1)}$ we can check whether it has $t$ linear factors or not. Therefore overall Algorithm \ref{algorithm:computing-candidate-set-of-linear-forms} takes$(nd\log|\F|)^{O(1)}$ time.
\end{proof}

\subsection{Reconstruction when $T_1$ (or $T_2$) $= \alpha y_1^t$}
\label{subsection:corner-case}
 This is a corner case of our problem and needs slightly different techniques. Here is a lemma summarizing the reconstruction algorithm in this case.

\begin{lemma}
\label{lemma:corner-case-reconstruction-lemma}
If for some $i\in [2]$, $T_i = \alpha y_1^t$ for some linear form $y_1$ and $\alpha\in \F$, then there exists a randomized algorithm that takes as input integers $n,d$ and black-box access to polynomial $f$, runs in time $(nd\log|\F|)^{O(1)}$, and with probability $1-o(1)$ outputs a $\Sigma\Pi\Sigma(2,n,d,\F)$ circuit computing $f$.
\end{lemma}
\begin{algorithm}
    \caption{Corner case}
    \label{algorithm:corner-case}
    \hspace*{\algorithmicindent} \textbf{Input - } Black-box access to polynomial $f$, integers $n, d$. \\
    \hspace*{\algorithmicindent} \textbf{Output - } A $\Sigma\Pi\Sigma(2,n,d,\F)$ circuit or $\#$.
        \begin{enumerate}
            \item Using Algorithm \ref{algorithm:sps(1)-reconstruction} with inputs as black-box access to $f$ and integers $n,d$ compute linear factors $\hat{\ell}_1,\ldots,\hat{\ell}_s$ and get access to black-box computing $NonLin(f)$. Compute degree of $NonLin(f)$ as $t=d-s$. Using Algorithm \ref{algorithm:computing-candidate-set-of-linear-forms}, compute set $\mathcal{L}(NonLin(f))$.
            
            \item  Iterate  over linear forms $\ell_1\in \mathcal{L}(NonLin(f))$. Construct an isomorphism $\Phi$ mapping $\ell_1\mapsto x_1$.
            \begin{enumerate}
                \item Simulate black-box for $\Phi(NonLin(f))_{|_{\{x_1=0\}}}$ and using Algorithm \ref{algorithm:sps(1)-reconstruction} identify two linearly independent factors say $\ell_2,\ell_3$. Construct another isomorphism $\Delta$ mapping $x_1\mapsto x_1, \ell_2\mapsto x_2, \ell_3\mapsto x_3$. Pick $\alpha_4, \ldots,\alpha_n$ uniformly randomly from $\F$. Simulate black-box for
                \[
                g(x_1,x_2,x_3) = \Delta(\Phi(NonLin(f)))_{|_{\{x_4=\alpha_4,\ldots,x_n=\alpha_n\}}}
                \]
                
                \item Using Lemma \ref{lemma:interp}, interpolate $g$ in monomial basis of $\F[x_1,x_2,x_3]$. Substitute $x_2=y x_1$ in all monomials and rearrange to get a representation in $\F[y][x_1,x_3]$. Equate coefficient polynomials of monomials containing $x_3$ to $0$ and solve the resulting system of equations using Lemma \ref{lemma:polynomial-equations}. If all $\ell_1$'s have been tried and no solution was obtained, {\bf Return $\#$}. Otherwise, for each solution, evaluate coefficient polynomial of $x_1^t$, creating a set of scalars.

                \item Iterate over all $\delta$'s in the set of scalars obtained above. Simulate black-box for $NonLin(f) - \delta \ell_1 ^t$ and
                using Algorithm \ref{algorithm:sps(1)-reconstruction} check if it has $t$ linear factors say $\ell_{s+1}, \ldots, \ell_{s+t}$.
                If not, then go to the next $\delta$. If all $\delta$ have been tried, go to next $\ell_1 \in \mathcal{L}(NonLin(f))$. If all $\ell_1$'s have been tried, {\bf Return \#}. Otherwise, simulate black-box for $f-f^\prime$, where 
                \[
                f^\prime = \hat{\ell}_1\times\ldots\times\hat{\ell}_s\times(\delta \ell_1^t + \ell_{s+1}\times\ldots\times\ell_{s+t})
                \]
                 and using Lemma \ref{lemma:sps(k)-pit} for $\Sigma\Pi\Sigma(4,n,d,\F)$ circuits, check if $f-f^\prime \equiv 0$. If output is 'yes', construct $\Sigma\Pi\Sigma(2,n,d,\F)$ circuit $C$ computing $f^\prime$. {\bf Return $C$}. If not, then go to next $\delta$. If all $\delta$ have been tried, go to next $\ell_1 \in \mathcal{L}(NonLin(f))$. If all $\ell_1$'s have been tried, {\bf Return $\#$}.
                
            \end{enumerate}
        \end{enumerate}
\end{algorithm} 
Now we prove the correctness of Algorithm \ref{algorithm:corner-case}. By correctness of Algorithm \ref{algorithm:sps(1)-reconstruction}, with probability $1-o(1)$, Step $1$ correctly obtains the black-box for $NonLin(f)$, it's degree $t$ and the multi-set of all linear factors of $f$. If we assume that these are correct, then by correctness of Algorithm \ref{algorithm:computing-candidate-set-of-linear-forms}, with probability $1-o(1)$, Step $1$ also correctly computes the set $\mathcal{L}(NonLin(f))$\footnote{all linear forms are correct up to scalar multiple.} of linear forms. In order to prove the correctness of Step $2$ we give two claims, both of which are proved in Appendix \ref{appendix:general-case-claims}. The first claim says that in this corner case, $NonLin(f)$ is actually the same as $T_1+T_2$ (up to scalar multiplication) and the second claim guarantees that some scalar multiple of $y_1$ actually belongs to the set $\mathcal{L}(NonLin(f))$. Here are the formal statements.

\begin{claim}
    \label{claim:corner-case-factors}
    Assume $T_i = \alpha y_1 ^t$, for some $i\in [2]$, $\alpha\in \F$ and linear form $y_1$. Then $Lin(f) = G$ (up to scalar factor). This also means that $NonLin(f)$ and $T_1 + T_2$ are equal up to a scalar factor.
\end{claim}

\begin{claim}
     \label{claim:corner-case-candidate}
     Assume $T_i = \alpha y_1 ^t$, for some $i\in [2]$, $\alpha\in \F$ and linear form $y_1$, then some scalar multiple of $y_1$ belongs to $\mathcal{L}(NonLin(f))$.
\end{claim}
We proceed in our correctness proof assuming that these claims are true. Assuming that Step $1$ was correct, we show that Step $2$ returns the correct circuit with high probability. Note that in Step $2(c)$, using Lemma \ref{lemma:sps(k)-pit}, we check whether the reconstructed circuit is correct or not. This ensures that we only return a correct circuit. Our algorithm in Steps $2(b),2(c)$ tries all linear forms in $\mathcal{L}(NonLin(f))$ and for each such linear form it constructs a set of scalars. So basically the algorithm iterates over possibilities of $\ell_1, \delta$ with the hope of finding one such that $T_i=\delta \ell_1^t$. If we can show that for some value of $\ell_1, \delta$ with high probability a correct $\Sigma\Pi\Sigma(2,n,d,\F)$ circuit is reconstructed, we will be done. We show this in the following lemma. We show this for $\ell_1$ being the scalar multiple of $y_1$ that belongs to $\mathcal{L}(NonLin(f))$ (guaranteed by Claim \ref{claim:corner-case-candidate}).  
\begin{lemma}
For $\ell_1$, the scalar multiple of $y_1$ in $\mathcal{L}(NonLin(f))$, the set of scalars constructed in Step $2(b)$ contains a scalar $\delta$ such that $T_i = \alpha y_1^t = \delta \ell_1^t$ and with probability $1-o(1)$ correctly reconstructs a $\Sigma\Pi\Sigma(2,n,d,\F)$ circuit computing $f$.
\end{lemma}

\begin{proof}
We know that $NonLin(f)$ restricted to the co-dimension $1$ subspace $\mathbb{V}(\ell_1)$ factors into a non-zero product of linear forms. By correctness of Algorithm \ref{algorithm:sps(1)-reconstruction}, we know that all linear factors of $\Phi(NonLin(f))_{|_{x_1=0}}$ can be computed. By Claim \ref{claim:corner-case-factors}, we know that this is the same as $\Phi(T_{3-i})_{|_{x_1=0}}$ up to scalar multiplication. Since $rank(f) = \Omega(\log^3 d)$ and linear factors of $T_i$ span a $1$ dimensional space, factors of this polynomial will span an $\Omega(\log^3 d)$ dimensional space and therefore we will be able to find at least two linearly independent factors $\ell_2, \ell_3$ in $\F[x_2,\ldots,x_n]$. This means that the polynomial $\Phi(NonLin(f))$ looks like
\[
\Phi(NonLin(f)) = \delta \ell_1^t + (\ell_2-\beta x_1)(\ell_3 - \gamma x_1) \prod\limits_{i=4}^{t+1}\ell_i,
\]
for some scalars $\beta, \gamma$ and linear forms $\ell_4,\ldots,\ell_{t+1}$ in $\F[x_1,\ldots,x_n]$. Recall the isomorphism $\Delta$ used in the algorithm, mapping $x_1\mapsto x_1, \ell_2\mapsto x_2, \ell_3\mapsto x_3$. Black-box computing the polynomial $\Delta(\Phi(NonLin(f)))$ can be constructed by taking every input of blackbox through the isomorphisms. The new polynomial now looks like
\[
\Delta(\Phi(NonLin(f))) = \delta x_1^t + (x_2-\beta x_1)(x_3 - \gamma x_1) \prod\limits_{i=4}^{t+1}\Delta(\ell_i),
\]
Finally, we plug in uniform random values for the variables $x_4,\ldots,x_n$. By Lemma \ref{lemma:szlemma} we know that with probability $1-o(1)$ the polynomial $\prod\limits_{i=4}^{t+1}\Delta(\ell_i)$ will not be identically zero and we will be left with a non-zero polynomial $g(x_1,x_2,x_3)$ computable by  a $\Sigma\Pi\Sigma(2,3,d,\F)$ circuit.
\[
g(x_1,x_2,x_3) = \delta x_1^t + (x_2-\beta x_1)(x_3-\gamma x_1)\prod\limits_{i=4}^{t+1}u_i,
\]
where $u_i$ are affine forms in $\F[x_1,x_2,x_3]$. Using the above black-box, we get access to black-box for $g$ and then using deterministic multivariate interpolation (Lemma \ref{lemma:interp}), interpolate it as a degree $t$ polynomial in the monomial basis of $\F[x_1,x_2,x_3]$.  $g$ depends on variable $x_3$. So substituting $x_2= y x_1$ for a fresh variable $y$, and solving for common zeros of all coefficient (of monomials involving $x_3$) univariate polynomials in $\F[y]$ would give us a set of scalars containing $\beta$. Note that, since our system has only univariate polynomials, all of degree $d^{O(1)}$, it can have at most $d^{O(1)}$ solutions. By correctness of algorithm in Lemma \ref{lemma:polynomial-equations}, with probability $1-o(1)$, this set would be correctly computed. Now substitution of $x_2 = \beta x_1$ would recover $\delta$ as coefficient of $x_1^t$. By correctness of Algorithm \ref{algorithm:sps(1)-reconstruction}, with probability $1-o(1)$, we will be able to completely factorize the black-box $NonLin(f) - \delta \ell_1^t$ into a product of $t$ linear factors giving us the correct $T_{3-i}$. 
By correctness of Step $1$, we know all linear factors of $f$, were correctly computed and therefore for scalar multiple $\ell_1$ of $y_1$ and the computed scalar $\delta$, with probability $1-o(1)$, we reconstruct a correct $\Sigma\Pi\Sigma(2,n,d,\F)$ circuit for $f$. Hence Proved. 
\end{proof}
Now we discuss the time complexity of the above algorithms.
\begin{lemma}
Algorithm \ref{algorithm:corner-case} takes $(nd\log|\F|)^{O(1)}$ time.
\end{lemma}
\begin{proof}
Time complexity of Algorithm \ref{algorithm:sps(1)-reconstruction} and Algorithm \ref{algorithm:computing-candidate-set-of-linear-forms} imply that Step $1$ takes $(nd\log|\F|)^{O(1)}$ time. In Step $2$, the outer iteration is over all linear forms in $\mathcal{L}(NonLin(f))$ which has size $d^{O(1)}$ (clear from Definition \ref{definition:candidate-linear-form} and explanation given in Algorithm \ref{algorithm:computing-candidate-set-of-linear-forms}).
Step $2(a)$ involves simulations of black-boxes post application of isomorphism and setting values for some variables. It also involves using Algorithm \ref{algorithm:sps(1)-reconstruction} to compute all linear factors. All these steps take $(nd\log|\F|)^{O(1)}$ time. Finding linearly independent pair of linear forms out of all linear factors is also done in  $(nd\log|\F|)^{O(1)}$ time. Step $3$ involves trivariate interpolation (Lemma \ref{lemma:interp}) which takes $(d\log|\F|)^{O(1)}$ time and by time complexity of Lemma \ref{lemma:polynomial-equations} solutions of the system of univariate polynomials (all have degree $d^{O(1)}$) are also found in $(nd\log|\F|)^{O(1)}$ time. The set of solutions is $d^{O(1)}$ sized since a univariate polynomial of degree $d$ has at most $d$ roots over a field. Therefore Step $2(b)$ takes $(nd\log|\F|)^{O(1)}$ time and creates a set of scalars of size $d^{O(1)}$. Step $2(c)$ iterates over this $d^{O(1)}$ sized set. Simulation of black-box and factorization using Algorithm \ref{algorithm:sps(1)-reconstruction} take $(nd\log|\F|)^{O(1)}$ time. Blackbox for $f-f^\prime$ is constructed in $(nd\log|\F|)^{O(1)}$ time and by time complexity of algorithm in Lemma \ref{lemma:sps(k)-pit}, it can be checked to be $0$ or not in $(nd\log|\F|)^{O(1)}$ time. Therefore overall Algorithm \ref{algorithm:corner-case} takes $(nd\log|\F|)^{O(1)}$ time. 
\end{proof}

\subsection{Reconstruction with linearly independent set dividing $T_i$ given}
\label{subsection:reconstruction-li-set-given}
 
Suppose we are given linearly independent linear forms $u_1,\ldots,u_t$, $t>60\log d+61$, such that for some $i\in [2]$, all the $u_j$'s divide $T_i$. Then there exists an efficient reconstruction algorithm as summarized in lemma below.

\begin{lemma}
\label{lemma:large-li-set-known}
There exists a randomized algorithm which takes as input integers $n,d$, black-box access to polynomial $f$ computable by a $\Sigma\Pi\Sigma(2,n,d,\F)$ circuit and linearly independent linear forms $u_1,\ldots,u_t$, $t>60\log d + 61$ (for some $i\in [2]$, all $u_j$'s divide $T_i$), runs in time $(nd\log|\F|)^{O(1)}$ and with probability $1-o(1)$ outputs a $\Sigma\Pi\Sigma(2,n,d,\F)$ circuit computing $f$. 
\end{lemma}

\begin{algorithm}
    \caption{Linearly independent linear factors of a multiplication gate are known}
    \label{algorithm:general-case}
    \hspace*{\algorithmicindent} \textbf{Input - } Black-box access to polynomial $f$, integers $n, d$, linear forms $u_1,\ldots,u_{t}$, $t>60\log d + 61$.\\
    \hspace*{\algorithmicindent} \textbf{Output - } A $\Sigma\Pi\Sigma(2,n,d,\F)$ circuit $C$ or $\#$.
        \begin{enumerate}
            \item Construct isomorphism $\Phi$ mapping $u_i\mapsto x_i, i\in [t]$ and simulate black-box computing $\Phi(f)$. Using Algorithm \ref{algorithm:sps(1)-reconstruction} with inputs as black-box computing $\Phi(f)$ and integers $n,d$, obtain all it's linear factors (with multiplicity) along with access to black-box computing $\Phi(NonLin(f))$. By traversing through the factors identify $e_i$, the largest power of $x_i$ that divides $\Phi(f)$. Using this set of factors and black-box computing $\Phi(f)$, simulate black-box computing $ g = \Phi(f)/\prod\limits x_i^{e_i} $. 
            
            \item For each $i\in [t]$, simulate black-box computing $g_{|_{\{x_i=0\}}}$ and using Algorithm \ref{algorithm:sps(1)-reconstruction} with inputs as this black-box, compute it's factors. If there are non linear factors, {\bf Return $\#$}. Otherwise, store factors in multi-set $\mathcal{U}_i$. Using Algorithm $5$ in \cite{Shpilka2007} merge the multi-sets $\mathcal{U}_i$ together to obtain a multiset $\mathcal{U}$ comprising of all linear factors of one of the product gates in the $\Sigma\Pi\Sigma(2, n, s, \F)$ circuit computing $g$ (here $s$ is some integer $\leq d$).

            \item Construct the multi-set $\mathcal{U}^\prime = \{\ell_{|_{\{x_1=0\}}} : \ell\in \mathcal{U}\}$. Check if this multi-set $\mathcal{U}^\prime$ and $\mathcal{U}_1$ contain same linear forms (upto multiplicity). If not, {\bf Return $\#$}. Otherwise compute scalar
            $
            \alpha =   \prod\limits_{\ell\in \mathcal{U}_1}\ell \Bigg/ \prod\limits_{\ell\in \mathcal{U}^\prime}\ell
            $
            by matching linear forms between $\mathcal{U}^\prime, \mathcal{U}_1$.
            
            \item Simulate black-box computing $g-\alpha \prod\limits_{\ell\in \mathcal{U}}\ell$ and factorize this polynomial using Algorithm \ref{algorithm:sps(1)-reconstruction}. If all factors are not linear, {\bf Return $\#$}. Otherwise, store factors in multi-set $\mathcal{V}$. Apply $\Phi^{-1}$ to all linear forms in $\mathcal{U,V}$.
            Simulate black-box for $f-f^\prime$, where 
                \[
                f^\prime = \prod\limits_{i=1}^{t} u_i ^{e_i}\times(\alpha \prod\limits_{\ell\in \mathcal{U}}\ell + \prod\limits_{\ell\in \mathcal{V}}\ell)
                \]
                 Using Lemma \ref{lemma:sps(k)-pit} for $\Sigma\Pi\Sigma(4,n,d,\F)$ circuits, check if $f-f^\prime \equiv 0$. If output is 'yes', construct $\Sigma\Pi\Sigma(2,n,d,\F)$ circuit $C$ computing $f^\prime$ and {\bf Return $C$}. If not, then {\bf Return $\#$}. 

            \end{enumerate}
\end{algorithm}

We present the algorithm for proving the above lemma in Algorithm \ref{algorithm:general-case}. We use Algorithm $5$ of \cite{Shpilka2007} in Step $2$. More details on this merge algorithm can be found in Algorithm $5$ and Theorem $29$ of \cite{Shpilka2007}. Now we prove correctness of the above algorithm. Black-box computing $\Phi(f)$ is simulated by passing every input through $\Phi$ first. Correctness of Algorithm \ref{algorithm:sps(1)-reconstruction} imply that with probability $1-o(1)$, all linear factors of $\Phi(f)$ and black-box access to $\Phi(NonLin(f))$ are correctly computed. From these linear forms, we remove any linear form $\ell$ that are divisible by some $x_i$. However we will keep the scalar $\ell / x_i$. The black-box obtained by multiplying the black-box of $\Phi(NonLin(f))$ returned by Algorithm \ref{algorithm:sps(1)-reconstruction} with these scalars and black-boxes computing the remaining linear factors simulates black-box access to $g = \Phi(f)/ \prod\limits_{i=1}^t x_i^{e_i}$. $g$ is a $\Sigma\Pi\Sigma(2, n, s, \F)$ circuit for some integer $s\leq d$. Assuming that Step $1$ is correct, simulation of black-boxes $g_{|_{x_i=0}}, i\in [t]$ can be done. Correctness of Algorithm \ref{algorithm:sps(1)-reconstruction} implies that with probability $1-o(1)$ all multi-sets $\mathcal{U}_i$ are correctly computed. By correctness of Algorithm $5$ in \cite{Shpilka2007}, we know that these multi-sets are glued together to obtain a multi-set $\mathcal{U}$ containing all linear factors of one of the product gates $S_2$ of $g$ (we are assuming that $g = S_1 + S_2$ where $S_1, S_2$ are product of linear forms and $x_i\mid S_1$ for $i\in [t]$.). Note that the algorithm only recovers all linear factors of $S_2$ and therefore it still needs to recover an appropriate scalar $\alpha$ (see algorithm) to completely recover $S_2$. Note that $g_{|_{x_1=0}} = {S_2}_{|_{x_1=0}}\neq 0$. Therefore we can compare the multi-set of linear forms in $\mathcal{U}_1$ with the multi-set of linear forms $\mathcal{U}^\prime = \{\ell_{|_{x_1=0}} : \ell \in \mathcal{U}\}$. All linear forms will match up to scalar multiplication giving us the scalar $\alpha$. By correctness of Algorithm \ref{algorithm:sps(1)-reconstruction}, we know that with probability $1-o(1)$, we will be able to correctly factor $g-\alpha \prod\limits_{\ell\in \mathcal{U}}\ell$ and collect them in multi-set $\mathcal{V}$.  Finally at the end, we can apply $\Phi^{-1}$ and multiply by $\prod\limits_{i=1}^t u_i^{t_i}$ and correctly recover the $\Sigma\Pi\Sigma(2,n,d,\F)$ circuit with probability $1-o(1)$. Note that in Step $4$, by correctness of Lemma \ref{lemma:sps(k)-pit}, we know that we can deterministically check whether the constructed circuit is correct or not and only return a correct circuit. Now we discuss the time complexity of the above algorithm.
\begin{lemma}
Algorithm \ref{algorithm:general-case} runs in time $(nd\log|\F|)^{O(1)}$ time.
\end{lemma}
\begin{proof}
Isomorphism $\Phi$ is constructed in $(n\log|\F|)^{O(1)}$ time. Time complexity of Algorithm \ref{algorithm:sps(1)-reconstruction} implies that $(nd\log|\F|)^{O(1)}$ time is spent on factorizing $\Phi(f)$. Removing powers of $x_i, i\in [t]$ again requires scanning through the linear factors and takes $(nd\log|\F|)^{O(1)}$. Black-box for $g = \Phi(f)/\prod\limits_{i=1}^t x_i^{e_i}$ is then created by multiplying outputs of all the black-boxes for any input and therefore is also simulated in $(nd\log|\F|)^{O(1)}$ time. Therefore Step $1$ takes $(nd\log|\F|)^{O(1)}$ time. Restrictions of black-box $g$ to $x_i=0, i\in [t]$ can be simulated by passing inputs through the restriction and therefore takes $(nd\log|\F|)^{O(1)}$ time. Time complexity of Algorithm \ref{algorithm:sps(1)-reconstruction} implies that factorization of $g_{|_{x_i=0}}$ can be done in $(nd\log|\F|)^{O(1)}$ time. Running time of Algorithm $5$ in \cite{Shpilka2007} is $(nd\log|\F|)^{O(1)}$ and therefore the multi-set $\mathcal{U}$ is created in $(nd\log|\F|)^{O(1)}$ time. Therefore Step $2$ takes $(nd\log|\F|)^{O(1)}$ time overall. Step $3$ involves iterating through the linear forms in $\mathcal{U}$, restricting them to $x_1=0$, giving multi-set $\mathcal{U}^\prime$, and then comparing the $d^{O(1)}$ sized multi-sets $\mathcal{U}^\prime$ and $\mathcal{U}$ to obtain the appropriate scalar $\alpha$. All these steps can be executed in polynomial time leading to a time complexity of $(nd\log|\F|)^{O(1)}$ for Step $3$. Black-box computing polynomial $g-\alpha\prod\limits_{\ell\in \mathcal{U}}\ell$ can be simulated in $(nd\log|\F|)^{O(1)}$ time by going through each of the involved (black-boxes) polynomials and then computing the output after algebraic operations. Time complexity of Algorithm \ref{algorithm:sps(1)-reconstruction} implies that the factorization of this black-box can be done in $(nd\log|\F|)^{O(1)}$ time. Finally computing the black-box for $f^\prime$ and simulating black-box for $f-f^\prime$ can similarly be done in $(nd\log|\F|)^{O(1)}$ time. By time complexity of algorithm in Lemma \ref{lemma:sps(k)-pit}, we know that in time $(nd\log|\F|)^{O(1)}$, we can deterministically test whether $f-f^\prime$ is the zero polynomial or not. Therefore Step $4$ also takes time $(nd\log|\F|)^{O(1)}$. So, overall Algorithm \ref{algorithm:general-case} runs in time $(nd\log|\F|)^{O(1)}$. 
\end{proof}

\subsection{Identify Linearly Independent Set Dividing $T_i$}
\label{subsection:identify-li-set}
In this subsection, our goal is to provide proof of Lemma \ref{lemma:general-case-search}. It plays a crucial role in Algorithm \ref{algorithm:high-rank-reconstruction} as explained in Section \ref{analysis-high-rank-reconstruction}, by optimizing the search for a large linearly independent set of linear forms dividing one of $T_1, T_2$. As we mentioned earlier, \cite{Shpilka2007} compute such an independent set by using a brute force search (Algorithm $4$, \cite{Shpilka2007}) on the space of linear forms over many variables, and therefore take quasi-polynomial time even before using this set in Algorithm $5$ (of \cite{Shpilka2007}). We significantly improve the search using candidate linear forms $\mathcal{L}(NonLin(f))$ and ordinary lines (see Definition \ref{defn:ordinary-line}) among them. First, in Section \ref{subsubsection:candidate-approximation} below we give intuition about why set $\mathcal{L}(NonLin(f))$ approximates the set of linear factors of $T_1\times T_2$ and then in in Lemma \ref{lemma:general-case-search}, Section \ref{subsubsection:proof-lemma-general-case-search} use this set to construct the required linearly independent set.

\subsubsection{Candidate set approximates set of linear forms dividing $T_1, T_2$}
\label{subsubsection:candidate-approximation}
 In order to quantify how close the candidate set $\mathcal{L}(NonLin(f))$ is to the set of linear forms in the input circuit, we define some new sets.
\[
\mathcal{L}_{good} = \{\ell \in \mathcal{L}(NonLin(f)) : \ell \mid T_1\times T_2\}, \hspace{2em} \mathcal{L}_{bad} = \mathcal{L}(NonLin(f))\setminus\mathcal{L}_{good},
\]
\[
\mathcal{L}_{others} = \{\ell \mid T_1\times T_2 : sp(\ell)\cap \mathcal{L}(NonLin(f)) = \phi\}\hspace{1em}\text{and}\hspace{1em} \mathcal{L}_{factors} = \{\ell: \ell \mid T_1+T_2\}
\]
For all sets, we only keep linear forms upto scalar multiplication and therefore treat them as proper sets (Definition \ref{defn:proper-set}). $\mathcal{L}_{good}$ contains all candidate linear forms which also divide one of the two gates $T_1,T_2$. $\mathcal{L}_{bad}$ are candidates which do not divide $T_1$ or $T_2$. $\mathcal{L}_{other}$ are linear forms dividing one of the gates but not captured (even up to scalar multiplication) in the candidate set and $\mathcal{L}_{factors}$ contain linear forms that divide $T_1+T_2$.  In the following claim, we show that
$\mathcal{L}_{good}$ is high dimensional and $\mathcal{L}_{bad}, \mathcal{L}_{other}$ are low dimensional quantifying the closeness of $\mathcal{L}(NonLin(f))$ to the set of linear forms dividing $T_1\times T_2$. We also show that $\mathcal{L}_{factors}$ is low dimensional. For better exposition, proof is provided in Appendix \ref{appendix:general-case-claims}.

\begin{claim}
\label{claim:candidate-approximation}
The following claim is true about these newly constructed sets.
 \begin{enumerate}
    \item $dim(sp(\mathcal{L}_{factors})) \leq \log d + 2$,
     \item $dim(sp(\mathcal{L}_{good}))\geq rank(f)-2$,
     \item $dim(sp(\mathcal{L}_{bad}))\leq \log d + 2$, and
     \item $dim(sp(\mathcal{L}_{others})) \leq 2$.
     
 \end{enumerate}
\end{claim}

\subsubsection{Proof of Lemma \ref{lemma:general-case-search}}
\label{subsubsection:proof-lemma-general-case-search}
In this subsection, we prove Lemma \ref{lemma:general-case-search} which was used by Algorithm \ref{algorithm:high-rank-reconstruction}.
 Recall that $rank(f)=\Omega(\log^3 d)$. We use definitions of $\mathcal{L}_{good}, \mathcal{L}_{bad}, \mathcal{L}_{other}, \mathcal{L}_{factors}$ given in Section \ref{subsubsection:candidate-approximation}. Recall the definition of the set of ordinary lines from Definition \ref{defn:ordinary-line}.
\begin{lemma}
\label{lemma:general-case-search}
The following are true.
\begin{enumerate}
    \item \label{item:general-case-search-1} $\exists$ $\ell \in \mathcal{L}_{good}$ such that the set of linear forms $\ell^\prime \in \mathcal{L}(NonLin(f))\setminus \{\ell\}$ for which $sp\{\ell, \ell^\prime\}$ intersects $\mathcal{L}(NonLin(f))$ only at $\{\ell, \ell^\prime\}$\footnote{basically $sp\{\ell, \ell^\prime\}$ is an ordinary line into $\mathcal{L}(NonLin(f))$.}, spans a space of dimension at least $\Omega(\log^2 d)$. Let $\mathcal{X}$ be some maximal independent subset $\Rightarrow |\mathcal{X}| = \Omega(\log^2 d)$.
    
    \item \label{item:general-case-search-2}   Every partition of $\mathcal{X}$ into $\Omega(\log d)$ equal parts of size $\Omega(\log d)$ each, contains a part $\mathcal{B}$ such that $\mathcal{B} \subset \mathcal{L}_{good}$ and for every $\ell ^\prime \in \mathcal{B}$, $sp\{\ell, \ell^\prime\}$ is an ordinary line into $\mathcal{L}_{good}, \mathcal{L}_{bad}, \mathcal{L}_{others}, \mathcal{L}_{factors}$.
    \item \label{item:general-case-search-3} Let $\ell ^\prime \in \mathcal{B}$ and assume $\ell \mid T_i$. Let $\Phi$ be an isomorphism mapping $\ell\mapsto x_1, \ell^\prime \mapsto x_2$, then,
\[
\Phi(NonLin(f))_{|_{x_1=0, x_2 =0}} = 0 \Leftrightarrow \ell ^\prime \text{ divides }T_{3-i}.
\]
\end{enumerate}
\end{lemma}

\begin{proof}
We prove all parts one by one.
\begin{enumerate}
    \item Let $\mathcal{T}\subset \mathcal{L}_{good}$ be a linearly independent set of size $126\log d +2$ (exists by Claim \ref{claim:candidate-approximation}). Applying Proposition \ref{propn:annihilating-planes} on $\mathcal{L}(NonLin(f))$ and $\mathcal{T}$ implies that there exists $\ell\in \mathcal{T}$ such that
    \[
    dim(\sum_{W\in \mathcal{O}(\ell, \mathcal{L}(NonLin(f)))} W) \geq \frac{dim(sp(\mathcal{L}(NonLin(f))))}{126\log d + 2} \geq
    \frac{dim(sp(\mathcal{L}_{good}))}{126\log d + 2} =\Omega(\log^2 d)
    \]
    Thus, the set of linear forms $\ell^\prime \in \mathcal{L}(NonLin(f))\setminus \{\ell\}$ for which $sp\{\ell, \ell^\prime\}$ intersects $\mathcal{L}(NonLin(f))$ only at $\{\ell, \ell^\prime\}$, spans a space of dimension at least $\Omega(\log^2 d)$. Let $\mathcal{X}$ be a maximal independent subset $\Rightarrow |\mathcal{X}| = \Omega(\log^2 d)$.

    \item Consider any partition of $\mathcal{X}$ into $\Omega(\log d)$ parts of size $\Omega(\log d)$ each.
    
    \begin{enumerate}
        \item We first claim that $\Omega(\log d)$ parts in this partition are inside $\mathcal{L}_{good}$. If not, then $\Omega(\log d)$ parts intersect $\mathcal{L}_{bad} \Rightarrow  dim(sp(\mathcal{L}_{bad}))=\Omega(\log d)$, contradicting Claim \ref{claim:candidate-approximation}. Now we will only deal with these $\Omega(\log d)$ parts inside $\mathcal{L}_{good}$. Since $\mathcal{L}_{good}, \mathcal{L}_{bad}\subset \mathcal{L}(NonLin(f))$, we see that for all $\ell^\prime$ in any of these parts $sp\{\ell, \ell^\prime\}$ is an ordinary line in $\mathcal{L}_{good}, \mathcal{L}_{bad}$ as required.
        
        \item Next we show that out of the $\Omega(\log d)$ parts inside $\mathcal{L}_{good}$, there is a part $\mathcal{B}$ such that for all $\ell^\prime \in \mathcal{B}$, $sp\{\ell, \ell^\prime\}$ is an ordinary line in $\mathcal{L}_{others}, \mathcal{L}_{factors}$, thereby completing the proof. If not then there are $\Omega(\log d)$ many $\ell ^\prime$'s, each belonging to a different part among the $\Omega(\log d)$ parts, such that $sp\{\ell, \ell^\prime\}$ intersects 
        $\mathcal{L}_{others}\cup \mathcal{L}_{factors}$
        at a linear form outside $sp\{\ell\}\cup sp\{\ell^\prime\}$ say $\ell ^{\prime\prime}$. Since all the $\Omega(\log d)$ $\ell^\prime$s are independent, the $\ell^{\prime\prime}$s span a space of dimension $\Omega(\log d) \Rightarrow$  
        $dim(sp(\mathcal{L}_{others}\cup \mathcal{L}_{factors})) = \Omega(\log d)$, contradicting Claim \ref{claim:candidate-approximation}.
    \end{enumerate}
    Therefore, we have shown the existence of a part $\mathcal{B}$ as desired. 
    
    \item Since $\ell \mid T_i$, we know that $x_1\mid \Phi(T_i)$. Therefore, the following equation holds in $\F[x_3,\ldots, x_n]$. 
    \[
    \Phi(L)_{|_{x_1=0, x_2 =0}}\Phi(NonLin(f))_{|_{x_1=0, x_2 =0}}  = \Phi(T_i)_{|_{x_1=0, x_2 =0}} + \Phi(T_{3-i})_{|_{x_1=0, x_2 =0}} = \Phi(T_{3-i})_{|_{x_1=0, x_2 =0}}.
    \]
    Here $L$ is the product of all linear factors of $T_1+T_2$ i.e. $L = Lin(T_1+T_2)$. First, we assume that $\Phi(NonLin(f))_{|_{x_1=0, x_2 =0}} = 0$. This implies using the above equation that $\Phi(T_{3-i})_{|_{x_1=0, x_2 =0}} = 0$. Therefore there is a linear form $\ell^{\prime\prime} \mid T_{3-i}$ such that $\ell^{\prime\prime} \in sp\{\ell, \ell^\prime\}$. If $\ell^{\prime\prime}$ is not a scalar multiple of $\ell$ or $\ell^\prime$, by construction of $\ell, \ell^\prime$ in parts $1$ and $2$ of this lemma, we know that no scalar multiple of $\ell^{\prime\prime}$ can belong to $\mathcal{L}_{good}$ or $\mathcal{L}_{others}$ and therefore it cannot divide $T_1\times T_2$ which is a contradiction since it divides $T_{3-i}$. Therefore, $\ell^{\prime\prime}$ has to be a scalar multiple of $\ell$ or $\ell^\prime$. It cannot be scalar multiple of $\ell$ since $\ell\mid T_i$ and $gcd(T_i, T_{3-i})=1$. Therefore $\ell^{\prime\prime}$ and $\ell^\prime$ are scalar multiples implying that $\ell^\prime$ divides $T_{3-i}$ as needed. Next, for the converse, we assume that $\ell^\prime \mid T_{3-i}$. Again, using the equation we gave at the beginning of this part, we get that,
    \[
    \Phi(L)_{|_{x_1=0, x_2 =0}}\Phi(NonLin(f))_{|_{x_1=0, x_2 =0}}  = \Phi(T_i)_{|_{x_1=0, x_2 =0}} + \Phi(T_{3-i})_{|_{x_1=0, x_2 =0}} = 0.
    \]
    Therefore, since $\F[x_3,\ldots,x_n]$ is an integral domain, either polynomial $\Phi(L)_{|_{x_1=0, x_2 =0}} = 0 $ or polynomial $\Phi(NonLin(f))_{|_{x_1=0, x_2 =0}} = 0$. Assume that $\Phi(L)_{|_{x_1=0, x_2 =0}} = 0$. This implies that there is some linear factor $\ell^{\prime\prime}$ of $T_1+T_2$ such that $\ell^{\prime\prime} \in sp\{\ell, \ell^\prime\}$. Since $gcd(T_1,T_2)=1$ and $\ell\mid T_i$,  $\ell^\prime \mid T_{3-i}$, the linear form $\ell^{\prime\prime}$ cannot be a scalar multiple of $\ell$ or $\ell^\prime$. So we found a linear form on $sp\{\ell, \ell^\prime\}$ different from scalar multiples of $\ell, \ell^\prime$, such that some scalar multiple of $\ell^{\prime\prime}$ belongs to $\mathcal{L}_{factors}$. By construction of $\ell, \ell^\prime$ in parts $1$ and $2$ of this lemma, we know that this cannot hold. Therefore our assumption is wrong and polynomial $\Phi(NonLin(f))_{|_{x_1=0, x_2 =0}} = 0$ completing the proof.
\end{enumerate}
\end{proof}

\section{Proof of Proposition \ref{propn:annihilating-planes}}
\label{section:codim-two-subspaces-f-vanishes}

In this section we prove Proposition \ref{propn:annihilating-planes}. Part \ref{item:annihilating-subspaces-size} is proved in Section \ref{subsection:annihilating-planes-size}.  Algorithm proving Part \ref{item:compute-annihilating-subspaces} is presented in Algorithm \ref{algorithm:compute-annihilating-subspaces} and it's correctness/complexity are analyzed in Section \ref{subsection:annihilating-planes-algorithm}.

\subsection{Proof of Part \ref{item:annihilating-subspaces-size}}
\label{subsection:annihilating-planes-size}

Let $W = \mathbb{V}(\ell, \ell^\prime) \subset \F^n$ be a co-dimension $2$ subspace on which $NonLin(f)$ vanishes i.e. $W\in \mathcal{S}(NonLin(f))$. Let $\Phi$ be an isomorphism mapping $\ell\mapsto x_1, \ell^\prime\mapsto x_2$. Since $NonLin(f)$ divides $T_1 + T_2$ we get that ${\Phi(T_1)}_{|_{x_1=0,x_2=0}} + {\Phi(T_2)}_{|_{x_1=0,x_2=0}} = 0$. This implies that either ${\Phi(T_1)}_{|_{x_1=0,x_2=0}} = {\Phi(T_2)}_{|_{x_1=0,x_2=0}} = 0$, or ${\Phi(T_1)}_{|_{x_1=0,x_2=0}} = -{\Phi(T_2)}_{|_{x_1=0,x_2=0}}\neq 0$. We prove the following lemma which implies the bound.
\begin{lemma}
\label{lemma:size-bound}
The following are true.
\begin{enumerate}
    \item \label{item:restriction-zero} $\#\{W\in \mathcal{S}(NonLin(f)) : {\Phi(T_1)}_{|_{x_1=0,x_2=0}} = {\Phi(T_2)}_{|_{x_1=0,x_2=0}} = 0\} \leq d^2$.
    \item \label{item:restriction-nonzero} $\#\{W\in \mathcal{S}(NonLin(f)): {\Phi(T_1)}_{|_{x_1=0,x_2=0}} = -{\Phi(T_2)}_{|_{x_1=0,x_2=0}}\neq 0 \} \leq d^5 + d^7$.
\end{enumerate}

\end{lemma}

\paragraph{Proof of \ref{item:restriction-zero}:} The statement implies that there are linear forms $\ell_1\mid T_1$ and $\ell_2\mid T_2$ such that ${\Phi(\ell_1)}_{|_{x_1=0,x_2=0}} =  {\Phi(\ell_2)}_{|_{x_1=0,x_2=0}} =0$. Also, $\ell_1, \ell_2$ are linearly independent since $gcd(T_1,T_2)=1$ implying that $sp\{\Phi(\ell_1), \Phi(\ell_2)\} = sp\{x_1, x_2\}$. On inverting via $\Phi$ this implies that $sp\{\ell_1, \ell_2\} = sp\{\ell, \ell^\prime\}$, which further implies that $\mathbb{V}(\ell_1, \ell_2) = \mathbb{V}(\ell, \ell^\prime) = W$. There can be at most $d^2$ such $W$'s completing the proof.

\paragraph{Proof of \ref{item:restriction-nonzero}:} We use the following lemma to prove this part. For clarity of presentation, we move it's proof to Appendix \ref{section:proof-annihilating-lemmas}.
\begin{lemma}
\label{lemma:one-dimension-found}
There exists a set $\mathcal{A}$ of co-dimension $1$ subspaces of $\F^n$ with $|\mathcal{A}|\leq d^4 + d^6$ such that for every $W\in \mathcal{S}(NonLin(f))$ satisfying ${\Phi(T_1)}_{|_{x_1=0,x_2=0}} = -{\Phi(T_2)}_{|_{x_1=0,x_2=0}}\neq 0$ , $\exists$ $V\in \mathcal{A}$ with $W\subset V$.
\end{lemma}
Assuming Lemma \ref{lemma:one-dimension-found}, we complete the proof as follows. For every $W\in \mathcal{S}(NonLin(f))$ satisfying ${\Phi(T_1)}_{|_{x_1=0,x_2=0}} = -{\Phi(T_2)}_{|_{x_1=0,x_2=0}}\neq 0$, we consider the co-dimension $1$ subspace $V$ given by Lemma \ref{lemma:one-dimension-found} such that $W\subset V$. Without loss of generality we assume $V = \mathbb{V}(x_1)$. We can now find a linear form $\ell_3$ such that $W = \mathbb{V}(x_1, \ell_3)$ and coeffcient of $x_1$ in $\ell_3$ is $0$ i.e. $\ell_3 = {\ell_3}_{|_{x_1=0}}$. Since $NonLin(f)$ vanishes on $W$ we know that 
$\Psi(NonLin(f))_{|_{x_1=0, x_2=0}}$ for isomorphism $\Psi$ mapping $x_1\mapsto x_1, \ell_3\mapsto x_2$. This also implies that $x_2$ divides $\Psi(NonLin(f))_{|_{x_1=0}}$. Since $\Psi$ keeps $x_1$ fixed this polynomial is same as $\Psi(NonLin(f)_{|_{x_1=0}})$. Inverting $\Psi$ we get that $\ell_3$ divides $NonLin(f)_{|_{x_1=0}}$. There are at most $d$ linear factors (upto scalar multiplication) of any degree $d$ polynomial, thus there are $\leq d$ such possible $\ell_3$. By going ever all choices of $V$ we get that there are at most $(d^4 + d^6)\times d$ many such $W$, completing our proof.

\subsection{Analysis of Algorithm \ref{algorithm:compute-annihilating-subspaces}}
\label{subsection:annihilating-planes-algorithm}

\begin{algorithm}
    \caption{Compute co-dimension $2$ subspaces on which $NonLin(f)$ vanishes}
    \label{algorithm:compute-annihilating-subspaces}
     \hspace*{\algorithmicindent} \textbf{Input - } Black-box access to polynomial $f$, integers $n, d$. \\
 \hspace*{\algorithmicindent} \textbf{Output - } A set $\mathcal{S}$  of tuples of independent linear forms in $\F[x_1,\ldots,x_n]$.
 
    \begin{enumerate}
        \item Create $n$ linear forms $\hat{\ell}_1, \ldots, \hat{\ell}_n$, such that the $n^2$ scalars used as coefficients in them are sampled uniformly randomly independently from $\F$. If these linear forms are linearly independent, define isomorphism $\Phi$ mapping $x_i\mapsto \hat{\ell}_i, i\in [n]$. Simulate black-box for $g = \Phi(f)$. 
        For $i\in [5,n]$, simulate black-box access for the following restricted polynomials in $\F[x_1,x_2,x_3,x_4,x_i]$.
        \[
        g_i = g_{|_{x_5=0, \ldots, x_{i-1}=0, x_{i+1}=0,\ldots,x_n=0}}
        \]
        Next, for each $i\in [5,n]$ using Algorithm \ref{algorithm:sps(1)-reconstruction} with inputs as black-box access to $g_i$ along with integers $5,d$ obtain black-box access to $NonLin(g_i)$ and integer $s$ denoting the number of linear factors of $g_i$. Define $t=d-s$. Using multivariate interpolation (Lemma \ref{lemma:interp}), interpolate $NonLin(g_i)$ as a degree $t$ polynomial in the monomial basis of $\F[x_1,x_2,x_3,x_4,x_i]$.
        
        \item Substitute $x_1 = y_3x_3 +y_4x_4 + y_ix_i$, and $x_2 = z_3x_3 + z_4x_4 + z_ix_i$ in $NonLin(g_i)$ to obtain a polynomial in $\F[y_3,y_4, y_i, z_3,z_4,z_i][x_3,x_4]$. Find common solutions to the system of polynomial equations defined by setting all coefficient polynomials ($\in \F[y_3,y_4, y_i, z_3,z_4,z_i]$) to zero. Initialize a set $\mathcal{S}_i \gets \phi$ and for each solution $(y_3,y_4,y_i,z_3,z_4,z_i)$ of the system above add tuple 
            $(x_1 - y_3x_3 - y_4x_4 - y_ix_i, x_2 - z_3x_3 - z_4x_4 - z_ix_i)$ to $\mathcal{S}_i$.

        \item Construct isomorphism $\Delta$ mapping $x_1\mapsto x_1, x_2\mapsto x_2, x_3\mapsto x_3, x_4\mapsto x_4$ and for $i\in [5.n]$, $x_i\mapsto x_i + \alpha_{i,3}x_3 + \alpha_{i,4}x_4$. The scalars $\alpha_{i,3}, \alpha_{i,4}, i\in [5,n]$ are sampled uniformly randomly independently from $\F$. Note that $\Delta$ can be viewed as an isomorphism on $\F[x_1,\ldots,x_n]$ as well as on each $\F[x_1,x_2,x_3,x_4, x_i]$ for $i\in [5,n]$.
        
        \item \label{item:iteration-level5}  Initialize a set $\mathcal{S}\gets \phi$. Iterate over all tuples $(x_1-\ell_1^5,x_2-\ell_2^5) \in \mathcal{S}_5$. Initialize $\ell_1\gets \ell_1^5, \ell_2\gets \ell_2^5$. Iterate over $i\in [6,n]$. Search for tuple $(x_1-\ell_1^i, x_2-\ell_2^i) \in \mathcal{S}_i$ such that tuples
        \[
        ({x_1-\Delta(\ell_1^5)}_{|_{x_5=0}}, {x_2-\Delta(\ell_2^5)}_{|_{x_5=0}}) = ({x_1-\Delta(\ell_1^i)}_{|_{x_i=0}}, {x_2 - \Delta(\ell_2^i)}_{|_{x_i=0}})
        \]
        If multiple or none such tuples are found in $\mathcal{S}_i$ then break out of this loop and go to the next tuple in the outer iteration. If only one such tuple is found then update $\ell_1 \gets \ell_1 - \alpha x_i$ and $\ell_2\gets \ell_2 - \beta x_i$ where $\alpha, \beta$ are coefficients of $x_i$ in $x_1-\ell_1^i, x_2-\ell_2^i$ respectively. At the end of iteration on $i$, update $\mathcal{S}\gets \mathcal{S}\cup \{(x_1-\ell_1, x_2-\ell_2)\}$.
        \item For each $(\ell_1, \ell_2) \in \mathcal{S}$, construct isomorphism $\Psi$ mapping $\ell_1\mapsto x_1, \ell_2\mapsto x_2$. Simulate black-box access to polynomial
        \[
        \Psi(NonLin(g))_{|_{x_1=0, x_2=0}}.
        \]
        Using randomized polynomial identity test given in Lemma \ref{lemma:randomized-polynomial-identity-test} with input as the above black-box and integer $n$,  check if it is identically the zero polynomial.
        If 'no', remove the tuple from $\mathcal{S}$, else replace it with $(\Phi^{-1}(\ell_1), \Phi^{-1}(\ell_2))$. {\bf Return $\mathcal{S}$}.
    \end{enumerate}
\end{algorithm} 
Before going to the correctness of Algorithm \ref{algorithm:compute-annihilating-subspaces}, we state a few useful lemmas. These are repeatedly used in our correctness and time complexity proofs.

\begin{lemma}
\label{lemma:annihilating-spaces-details-1}
With probability $1-o(1)$ over the random choices in Step $1$, the following hold.
\begin{enumerate}
\item The $n$ linear forms constructed in Step $1$ with the random coefficients are linearly independent.
\item $NonLin(f)$ vanishes on $\mathbb{V}(\ell_1,\ell_2)$ if and only if $NonLin(g)$ vanishes on $\mathbb{V}(\Phi(\ell_1), \Phi(\ell_2))$.
\item Polynomial $g_i$ has a $\Sigma\Pi\Sigma(2, 5, d, \F)$ circuit and $rank(g_i) = 5$. 
\item $NonLin(g_i) = NonLin(g)_{|_{x_5=0,\ldots, x_{i-1}=0, x_{i+1}=0,\ldots,x_n=0}}$.
\item For all $\mathbb{V}(\ell_1, \ell_2) \in \mathcal{S}(NonLin(g))$, there exist linear forms $\ell_1^\prime, \ell_2^\prime\in \F[x_3,\ldots,x_n]$ such that
\[
\mathbb{V}(\ell_1, \ell_2) = \mathbb{V}(x_1-\ell_1^\prime, x_2-\ell_2^\prime)
\]
\item Let $\mathbb{V}(x_1-\ell_1, x_2-\ell_2) \in \mathcal{S}(NonLin(g))$ with $\ell_1,\ell_2\in \F[x_3,\ldots,x_n]$. Then, $NonLin(g_i)$ vanishes on the co-dimension $2$ subspace $\mathbb{V}(x_1-\ell_1^i, x_2-\ell_2^i)$. Here $\ell_j^i = {\ell_j}_{|_{x_5=0,\ldots, x_{i-1}=0, x_{i+1}=0,\ldots,x_n=0}}$.
\end{enumerate}
\end{lemma}

\begin{lemma}
\label{lemma:annihilating-spaces-details-2}
With probability $1-o(1)$ over the random choices in Step $3$, the following holds. For all $i\in [5,n]$ and for all pairs of distinct tuples $(x_1-\ell_1, x_2-\ell_2)$, $(x_1-\ell_1^\prime, x_2-\ell_2^\prime)$ in $\mathcal{S}_i$,
\[
(x_1-\Delta(\ell_1)_{|_{x_i=0}}, x_2-\Delta(\ell_2)_{|_{x_i=0}}) \neq (x_1-\Delta(\ell_1^\prime)_{|_{x_i=0}}, x_2-\Delta(\ell_2^\prime)_{|_{x_i=0}})
\]
\end{lemma}

For better presentation we prove these lemmas in Appendix \ref{appendix:proofs-lemmas-algorithm-annihilating}. Now, we prove correctness of Algorithm \ref{algorithm:compute-annihilating-subspaces}.
By Part $1$ of Lemma \ref{lemma:annihilating-spaces-details-1}, the linear forms constructed in Step $1$ are linearly independent and therefore isomorphism $\Phi$ can be correctly constructed using them. Using this isomorphism, simulation of black-box for $g$ (by passing every input through the isomorphism) is straight forward. Further simulation of black-boxes computing the $g_i$s is also straight forward (by setting $x_5=0,\ldots,x_{i=1}=0, x_{i+1}=0,\ldots,x_n=0$ in the input to black-box). From Parts $4,5$ of Lemma \ref{lemma:annihilating-spaces-details-1}, we know that $g_i$ exhibits $\Sigma\Pi\Sigma(2,5,d,\F)$ circuit of rank $5$ and $NonLin(g_i) = NonLin(g)_{|_{x_5=0,\ldots,x_{i=1}=0, x_{i+1}=0,\ldots,x_n=0}}$, implying that all $g_i$ and $g$ have the same number of linear factors $s$ and degree of all polynomials $NonLin(g_i)$ are equal ($=t$) which is also the same as degree of $NonLin(g)$. By correctness of Algorithm $1$, with probability $1-o(1)$, Step $1$ correctly obtains black-box computing $NonLin(g_i)$ and it's degree $t$. Since all $g_i$ are $5-$ variate using deterministic multivariate interpolation (Lemma \ref{lemma:interp}), we can interpolate their black-boxes as degree $t$ polynomials in the monomial basis of $\F[x_1,x_2,x_3,x_4,x_i]$. Therefore, at the end of Step $1$, we would have correct monomial representations of all the $g_i$. Next, using Part $5$ of Lemma \ref{lemma:annihilating-spaces-details-1}, we know that any co-dimension $2$ subspace on which $NonLin(g)$ vanishes has the form $\mathbb{V}(x_1-\ell_1, x_2-\ell_2)$ with $\ell_1, \ell_2\in \F[x_3,\ldots,x_n]$. In Part $6$ of Lemma \ref{lemma:annihilating-spaces-details-1}, we show that $NonLin(g_i)$ vanishes on the co-dimension $2$ space $\mathbb{V}(x_1-\ell_1^i, x_2-\ell_2^i)$, where for $j\in [2]$ and $i\in [5,n]$, $\ell_j^i$ are restrictions of $\ell_j$ to $x_5 = 0, \ldots, x_{i-1}=0, x_{i+1}=0,\ldots,x_n=0$. Since these co-dimension $2$ subspaces have the particular form $\mathbb{V}(x_1-\ell_1^i, x_2-\ell_2^i)$, substituting $x_1=\ell_1^i, x_2=\ell_2^i$ in $NonLin(g_i)$ should give $0$. Step $2$ uses this observation and computes all possible $\ell_1^i, \ell_2^i$ by solving the system of polynomial equations we get on substitution. By correctness of Lemma \ref{lemma:polynomial-equations}, we can compute all such solutions. Therefore, the set $\mathcal{S}_i$ contain tuples corresponding to all co-dimension $2$ spaces of the form $\mathbb{V}(x_1-u_1, x_2-u_2)$ (with linear forms $u_1, u_2 \in \F[x_3,x_4,x_i]$) on which $NonLin(g_i)$ vanishes. In the next lemma, we show that these $\mathcal{S}_i$ are then glued in Steps $3$ and $4$ to create a set $\mathcal{S}$ which contains tuples corresponding to all elements of $\mathcal{S}(NonLin(f))$. 
\begin{lemma}
\label{lemma:gluing-correct}
Step $4$ outputs a set $\mathcal{S}$, such that with probability $1-o(1)$, it contains tuples of linear forms representing all co-dimension $2$ subspaces on which $NonLin(g)$ vanishes.
\end{lemma}

\begin{proof}
Let $\mathbb{V}(x_1-\ell_1, x_2-\ell_2)\in \mathcal{L}(NonLin(g))$. By Part $6$ of Lemma \ref{lemma:annihilating-spaces-details-1} we know that $NonLin(g_i)$ vanishes on the co-dimension $2$ subspace $\mathbb{V}(x_1-\ell_1^i, x_2-\ell_2^i)$ where for $j\in [2]$, $\ell_j^i = {\ell_j}_{|_{x_5=0,\ldots, x_{i-1}=0, x_{i+1}=0,\ldots,x_n=0}}$. Therefore the tuples $(x_1-\ell_1^i, x_2-\ell_2^i)$ belong to $\mathcal{S}_i$ computed at Step $2$. Observe that, for $i\in [6,n]$ we glue tuple $(x_1 - \ell_1^5, x_2-\ell_2^5)$ with tuple $(x_1 - \ell_1^i, x_2-\ell_2^i)$ only if the latter is the only tuple in $\mathcal{S}_i$ satsfying,
$ ({x_1-\Delta(\ell_1^5)}_{|_{x_5=0}}, {x_2-\Delta(\ell_2^5)}_{|_{x_5=0}}) = ({x_1-\Delta(\ell_1^i)}_{|_{x_i=0}}, {x_2 - \Delta(\ell_2^i)}_{|_{x_i=0}})$. Here $\Delta$ is the isomorphism constructed in Step $3$.
So all we need to show is that, there is no other tuple $(x_1-{\ell_1^i}^\prime, x_2 - {\ell_2^i}^\prime) \in \mathcal{S}_{i}$ with ${\ell_1^i}^\prime, {\ell_2^i}^\prime$ being linear forms in $\F[x_3,x_4,x_i]$ such that,
$({x_1-\Delta(\ell_1^5)}_{|_{x_5=0}}, {x_2-\Delta(\ell_2^5)}_{|_{x_5=0}}) = ({x_1-\Delta({\ell_1^i}^\prime)}_{|_{x_i=0}}, {x_2 - \Delta({\ell_1^2}^\prime)}_{|_{x_i=0}})$.
If there was such a tuple, comparing the two equations we got above gives $({x_1-\Delta(\ell_1^i)}_{|_{x_i=0}}, {x_2 - \Delta(\ell_2^i)}_{|_{x_i=0}}) =({x_1-\Delta({\ell_1^i}^\prime)}_{|_{x_i=0}}, {x_2 - \Delta({\ell_1^2}^\prime)}_{|_{x_i=0}})$, which contradicts Lemma \ref{lemma:annihilating-spaces-details-2}. Therefore tuple $(x_1 - \ell_1^5, x_2-\ell_2^5)$ gets correctly glued with each such tuple $(x_1 - \ell_1^i, x_2-\ell_2^i)$ for $i\in [6,n]$ leading to the tuple $(x_1-\ell_1, x_2-\ell_2)$ being constructed and added to $\mathcal{S}$. Hence Proved.  
\end{proof}
Assuming we have correctly glued the $\mathcal{S}_i$ into set $\mathcal{S}$, Step $5$, performs a final pruning by retaining tuples for which $NonLin(g)$ actually vanishes on the co-dimension $2$ subspace they represent. By correctness of Lemma \ref{lemma:randomized-polynomial-identity-test}, this is done correctly and only the right tuples are retained. By Part $1$ of Lemma \ref{lemma:annihilating-spaces-details-1}, in order to get set $\mathcal{S}(NonLin(f))$ from $\mathcal{S}(NonLin(g))$, we only need to invert all linear forms present in the elements (tuples) of $\mathcal{S}$. Therefore, with probability $1-o(1)$, the set of tuples representing co-dimension $2$ subspaces on which $NonLin(f)$ vanishes is correctly computed. Now we discuss the time complexity of the above algorithm.
\begin{lemma}
Algorithm \ref{algorithm:compute-annihilating-subspaces} runs in $(nd\log|\F|)^{O(1)}$ time.
\end{lemma}
\begin{proof}
Assuming that sampling of a uniformly random scalar from $\F$ takes $O(1)$ time, the $n$ linear forms are created in $(n\log|\F|)^{O(1)}$ time. Checking whether the linear are independent can be done in $(n\log|\F|)^{O(1)}$ time by gaussian elimination on the matrix defined by the $n^2$ coefficients of these linear forms. Black-boxes for $g$ and $g_i$ are simulated in $(n\log|\F|)^{O(1)}$ time by passing each input through $\Phi$ and then restricting to $x_5=0,\ldots,x_{i-1=0}, x_{i+1}=0, \ldots, x_n=0$. Time complexity of Algorithm \ref{algorithm:sps(1)-reconstruction} implies that black-box access to all $NonLin(g_i)$ along with their degrees $t=d-s$ can be obtained in $(nd\log|\F|)^{O(1)}$ time. Multivariate interpolation (Lemma \ref{lemma:interp}) on the $5$ variate polynomials of degree $t$ each is done in $(nd\log|\F|)^{O(1)}$ time. Therefore Step $1$ takes $(nd\log|\F|)^{O(1)}$ time. Each $g_i$ has $d^{O(1)}$ non-zero coefficients in the monomial representation. Substitutions lead to $d^{O(1)}$ many coefficient polynomials in $\F[y_3, y_4, y_i, z_3, z_4, z_i]$ with every polynomial having degree $d^{O(1)}$. By Part $2$ of Lemma \ref{lemma:annihilating-spaces-details-1}, every $g_i$ has a $\Sigma\Pi\Sigma(2,5,d,\F)$ circuit and has rank $5$, therefore, by Part $1$ of Proposition \ref{propn:annihilating-planes}, number of co-dimension $2$ subspaces on which they vanish are $d^{O(1)}$. Therefore our system of equations has at most $d^{O(1)}$ solutions since they characterize such co-dimension $2$ subspaces of a certain form. By time complexity of Lemma \ref{lemma:polynomial-equations}, for each $g_i$ all solutions to such a system can be computed in $(d\log|\F|)^{O(1)}$ time leading to $\mathcal{S}_i$. Therefore in time $(nd\log|\F|)^{O(1)}$ time all $\mathcal{S}_i$ are computed in Step $2$. Step $3$ involves sampling $O(n)$ many uniformly random scalars and construction of the isomorphism $\Delta$ can be done in $(n\log|\F|)^{O(1)}$ time. In Step $4$, we iterate over all tuples in $\mathcal{S}_5$ and then iterate over $i\in [6,n]$ trying to match our tuple with tuples in the $\mathcal{S}_i$. Since each tuple in $\mathcal{S}_5$ is matched to at most one tuple in each $\mathcal{S}_i$, for each tuple in $\mathcal{S}_5$, we go over all the set $\mathcal{S}_i, i\in [6,n]$ just once. Therefore, overall we take $(nd\log|\F|)^{O(1)}$ time in this step. Also, since each tuple in $\mathcal{S}_5$, creates at most one tuple $(x_1-\ell_1, x_2-\ell_2)$ to be added to $\mathcal{S}$, we create at most $d^{O(1)}$ such tuples leading to $|\mathcal{S}| = d^{O(1)}$. In Step $5$, for each tuple in $\mathcal{S}$, construction of isomorphism $\Psi$ and black-box access to $\Psi(NonLin(g))_{|_{x_1=0,x_2=0}}$ can be created in $(nd\log|\F|)^{O(1)}$ time. By time complexity of algorithm in Lemma \ref{lemma:randomized-polynomial-identity-test}, in time $(nd\log|\F|)^{O(1)}$ we can check whether this black-box computes the $0$ polynomial or not. Finally application of $\Phi^{-1}$ to tuples in $\mathcal{S}$ can be done in $(nd\log|\F|)^{O(1)}$ time. Our final set returned has size $d^{O(1)}$ as it is a subset of the set we created in Step $4$. Therefore, overall Algorithm \ref{algorithm:compute-annihilating-subspaces} takes $(nd\log|\F|)^{O(1)}$ time.
\end{proof}



\section{Proof of Proposition \ref{propn:ordinary-line-dim-main}}
\label{section:ordinary-lines}

In this section we present our proof of Proposition \ref{propn:ordinary-line-dim-main}. The proof is immediately implied by Lemma \ref{lemma:ordinary-line-dim} which is itself proved using Lemma \ref{lemma:two-sets-ordinary-lines}. Recall definition of set of ordinary lines(Definition \ref{defn:ordinary-line}).

\begin{lemma}
\label{lemma:ordinary-line-dim}
Let $\mathcal{S} \subset \F^n$ be a proper set (Definition \ref{defn:proper-set}) and
$\mathcal{T} \subset \F^n$ be any linearly independent set of size $\log|\mathcal{S}|+2$. Then, the following holds.
\[
sp(\mathcal{S}) \subseteq \sum\limits_{t\in \mathcal{T}} \sum\limits_{W\in \mathcal{O}(t, \mathcal{S})}W
\]
\end{lemma}

\paragraph{Proof of Proposition \ref{propn:ordinary-line-dim-main} using Lemma \ref{lemma:ordinary-line-dim}:}
By simply taking dimension of both sides in the containment, applying union bound on the right hand side and assuming $t\in\mathcal{T}$ maximizes $dim(\sum_{W\in \mathcal{O}(t, \mathcal{S})}W))$, we get
\[
dim(\sum_{W \in \mathcal{O}(t, \mathcal{S})} W) \geq \frac{dim(sp(\mathcal{S}))}{\log |\mathcal{S}| + 2}.
\]
which proves Proposition \ref{propn:ordinary-line-dim-main}. So we are left with proving Lemma \ref{lemma:ordinary-line-dim}.

\paragraph{Proof of Lemma \ref{lemma:ordinary-line-dim}}
Let $V$ be the vector space $\sum_{t\in \mathcal{T}} \sum_{W\in \mathcal{O}(t, \mathcal{S})}W$. We define set $\mathcal{S}^\prime = \mathcal{S}\setminus V$. $\mathcal{S}^\prime$ is a proper set. We will show that $\mathcal{S}^\prime = \phi \Rightarrow sp(\mathcal{S})\subset V$. If not, we show that there cannot be any ordinary line from $\mathcal{T}$ into $\mathcal{S}^\prime$. Suppose there is some such line $sp\{t,s\}$ where $t\in \mathcal{T}$ and $s\in \mathcal{S}^\prime$ are not scalar multiples. Since it is an ordinary line into $\mathcal{S}^\prime$, we get that $sp\{s,t\}\cap \mathcal{S}^\prime \subset sp\{s\}\cup sp\{t\}$. Then one of the following mutually exclusive statements will obviously be true.
\begin{enumerate}
    \item $sp\{s,t\}\cap V \subset sp\{s\}\cup sp\{t\}$
    \item $sp\{s,t\}\cap V \not\subset sp\{s\}\cup sp\{t\}$
\end{enumerate}
In the first case, since $\mathcal{S} = \mathcal{S}^\prime \cup (\mathcal{S}\cap V) \Rightarrow sp\{s,t\}\cap \mathcal{S} \subset sp\{s\}\cup sp\{t\}$. Therefore it is an ordinary line into $\mathcal{S}$. But all such lines are subsets of $V \Rightarrow s\in V$ which is a contradiction since $s\in \mathcal{S}^\prime$ which is disjoint from $V$. In the second case, there is some $v\in sp\{s,t\}\cap V$ such that $v\notin sp\{s\}\cup sp\{t\}$. Therefore $t,s,v$ are linearly dependent but $t,s$ and $s,v$ are not $\Rightarrow$ $s\in sp\{t,v\}$. Both $t,v$ are in $V$ by construction and thus $s\in V$ which is again a contradiction since $s\in \mathcal{S}^\prime$ which is disjoint from $V$. Therefore if $\mathcal{S}^\prime$ is non-empty, there are no ordinary lines from $\mathcal{T}$ into $\mathcal{S}$. Now we use Lemma \ref{lemma:two-sets-ordinary-lines} and complete the proof. We will prove Lemma \ref{lemma:two-sets-ordinary-lines} after the current proof.
\begin{lemma}
\label{lemma:two-sets-ordinary-lines}
Let $\mathcal{S} (\neq \phi) \subset \F^n$ be a proper set and $\mathcal{T} \subset \F^n$ be linearly independent such that for every $t\in \mathcal{T}$, there is no ordinary line (Definition \ref{defn:ordinary-line}) from $t$ into $\mathcal{S}$. Then 
$|\mathcal{T}| \leq \log|\mathcal{S}| + 1$.
\end{lemma}

Using Lemma \ref{lemma:two-sets-ordinary-lines} with $\mathcal{S}^\prime$ and $\mathcal{T}$, we get that $\log |\mathcal{S}|+2 = |\mathcal{T}| \leq \log |\mathcal{S}^\prime| + 1$ which is a contradiction since $\mathcal{S}^\prime \subset \mathcal{S}$. Therefore, the only conclusion left is $\mathcal{S}^\prime = \phi$, which completes the proof of our lemma as explained earlier.

\paragraph{Proof of Lemma \ref{lemma:two-sets-ordinary-lines}:} Let $|\mathcal{T}|=d$ and $|\mathcal{S}| = m$. We present a counting argument by building a one-to-one function mapping subsets of $[d-1]$ into $\mathcal{S}$. Such a function implies that $m\geq 2^{d-1}$ and we'll be done. The following describes this one-to-one function. Fix an element $s\in \mathcal{S}$ and let $\mathcal{T} = \{t_1,\ldots,t_d\}$. Without loss of generality we may assume that $s,t_1,\ldots,t_{d-1}$ are linearly independent.
\begin{claim}
\label{claim:one-one-map}
 For any subset $\mathcal{P}\subset[d-1]$, there exists  $s_\mathcal{P}\in \mathcal{S}$ in the interior\footnote{``interior'' means that when $s_\mathcal{P}$ is written as a linear combination of $\{\{t_i : i\in \mathcal{P}\}\cup \{s\}\}$, all coefficients are non-zero} of $sp\{\{t_i : i\in \mathcal{P}\}\cup \{s\}\}$.
\end{claim}

\begin{proof}
We prove by induction on $|\mathcal{P}|$. For $|\mathcal{P}| = 0$, define $s_{\mathcal{P}} = s$ and we are done. Let's assume the claim is true for $|\mathcal{P}| = k-1$. We prove it for $|\mathcal{P}| = k$. Consider any element $p\in \mathcal{P}$ and let $\mathcal{R} = \mathcal{P}\setminus \{p\}$. By induction, we know there exists $s_{\mathcal{R}}$ in the interior of $sp\{\{t_i : i\in \mathcal{R}\}\cup \{s\}\}$. Since there is no ordinary line from any $t\in \mathcal{T}$ into $\mathcal{S}$, the line $sp\{t_p, s_{\mathcal{R}}\}$ contains $s_{\mathcal{P}}\in \mathcal{S}$ such that $s_{\mathcal{P}} \notin sp\{t_p\}\cup sp\{s_{\mathcal{R}}\} \Rightarrow s_{\mathcal{P}} = \alpha t_p + \beta s_{\mathcal{R}}$ with $\alpha, \beta \in \F$ being non-zero scalars $\Rightarrow s_{\mathcal{P}}$ is in the interior of $sp\{\{t_i : i\in \mathcal{P}\}\cup \{s\}\}$ and the proof is complete.
\end{proof}

We can see that the function mapping $\mathcal{P}\subset [d-1]$ to $s_{\mathcal{P}}\in \mathcal{S}$, is one-to-one since for sets $\mathcal{P,Q}\subset [d-1]$, which differ at some $j\in [d-1]$, exactly one of $s_{\mathcal{P}}, s_{\mathcal{Q}}$ has a non-zero coefficient of $t_j$, implying they are different. This completes the proof.

\section{Acknowledgements}
We would like to thank Vineet Nair for helping with organization and presentation of the paper. He also provided multiple insights about the content which led to better presentation. We would also like to thank Neeraj Kayal and Chandan Saha for helpful comments on an early presentation of this work. Neeraj Kayal introduced the author to black-box reconstruction problems for depth three circuits. The simple idea behind proof of Lemma \ref{lemma:two-sets-ordinary-lines}, presented in this paper was shared with the author by Neeraj Kayal during a discussion. We would also like to thank Anuja Sharan for proofreading and helping in preparation of this paper.

\bibliographystyle{alpha}
\bibliography{ref}

\appendix

\section{Proof of Claims \ref{claim:corner-case-factors}, \ref{claim:corner-case-candidate} and \ref{claim:candidate-approximation}}
\label{appendix:general-case-claims}

\subsection{Proofs of Claims \ref{claim:corner-case-factors} and \ref{claim:corner-case-candidate}}
In these claims we are given that $T_i = \alpha y_1^t$ for some $i\in [2], \alpha\in \F$ and linear form $y_1$.
\begin{enumerate}
    \item To see the proof of Claim \ref{claim:corner-case-factors}, consider any linear factor $\ell$ of $T_1+T_2$. $\ell \nmid T_1,T_2$ since $gcd(T_1,T_2)=1$. Let $\Phi$ be an isomorphism mapping $\ell\mapsto x_1$. Setting $x_1=0$, we get that ${\Phi(T_1)}_{|_{x_1=0}} = -{\Phi(T_2)}_{|_{x_1=0}} \neq 0$. Both sides are non-zero products of linear forms in $\F[x_2,\ldots,x_n]$. Therefore, by unique factorization we can match factors (upto scalar multiplication). This implies that $dim(\{\text{linear form }\ell : \ell \mid T_1\})$ and $dim(\{\text{linear form }\ell : \ell \mid T_2\})$ cannot differ from each other by more than $1$. But since $rank(f) = \Omega(\log^3d)$, this cannot happen since one of the $T_i$'s spans a one dimensional space. Therefore $T_1+T_2$ has no linear factors and we are done.
    
    \item To see proof of Claim \ref{claim:corner-case-candidate}, without loss of generality assume $y_1\mid T_1$. Define isomorphism $\Phi$ mapping $y_1\mapsto x_1$. Using Claim \ref{claim:corner-case-factors} we know that
    \[
    0\neq {\Phi(T_2)}_{|_{x_1=0}} =(\Phi(T_1)+\Phi(T_2))_{|_{x_1=0}} = 
    \Phi(NonLin(f))_{|_{x_1=0}}
    \]
    
    So first condition of Definition \ref{definition:candidate-linear-form} is satisfied. As argued in Claim \ref{claim:corner-case-factors}, $rank(f)\geq \Omega(\log^3d) \Rightarrow$ linear forms dividing $T_2$, span a $\Omega(\log^3 d)$ dimensional space. Since ${\Phi(T_2)}_{|_{x_1=0}}$ is non-zero, it's factors also span $\Omega(\log^3 d)$ dimensional space and so there exist two linearly independent factors $y_2,y_3$ of $T_2$ such that $NonLin(f)$ vanishes on both $\mathbb{V}(y_1,y_2)$ and $\mathbb{V}(y_1, y_3)$. This implies that  second condition of Definition \ref{definition:candidate-linear-form} is also satisfied. Therefore, some scalar multiple of $y_1 \in \mathcal{L}(NonLin(f))$.
\end{enumerate}

\subsection{Proof of Claim \ref{claim:candidate-approximation}}
Recall definition of sets,
\[
\mathcal{L}_{good} = \{\ell \in \mathcal{L}(NonLin(f)) : \ell \mid T_1\times T_2\}, \hspace{2em} \mathcal{L}_{bad} = \mathcal{L}(NonLin(f))\setminus\mathcal{L}_{good}, 
\]
\[
\mathcal{L}_{others} = \{\ell \mid T_1\times T_2 : sp(\ell)\cap \mathcal{L}(NonLin(f)) = \phi\} \hspace{1em}\text{and}\hspace{1em} \mathcal{L}_{factors} = \{\ell: \ell \mid T_1+T_2\}
\]
For all sets, we keep linear forms upto scalar multiplication and therefore treat them as proper sets (Definition \ref{defn:proper-set}). Below we prove all parts of Claim \ref{claim:candidate-approximation}.
\begin{enumerate}

\item $dim(sp(\mathcal{L}_{factors})) \leq \log d + 2$: By definition $\mathcal{L}_{factors}$ is the set of all factors of $T_1+T_2$. Consider any linearly independent subset $\mathcal{Z}\subset\mathcal{L}_{factors}$ and let $\ell\in \mathcal{Z}$.
Define isomorphism $\Phi$ mapping $\ell\mapsto x_1$.
Setting $x_1=0$ in $\Phi(T_1)+\Phi(T_2)$ gives ${\Phi(T_1)}_{|_{x_1=0}} = -{\Phi(T_2)}_{|_{x_1=0}} \neq 0$. By unique factorization in ring $\F[x_2,\ldots,x_n]$, for every linear form $\ell_1 \mid T_1$ there exists $\ell_2\mid T_2$ such that
$\ell_2 \in sp\{\ell, \ell_1\}$. Since $\ell_2 \notin sp\{\ell\}\cup sp\{\ell_1\}$, this means that $sp\{\ell,\ell_1\}$ is not an ordinary line from $\ell$ into the proper set $\mathcal{L}$ containing linear factors of $T_1, T_2$. This set has size $\leq 2d$. Since $\ell$ was arbitrary in $\mathcal{Z}$, there are no ordinary lines from $\mathcal{Z}$ into $\mathcal{L}$. So using Lemma \ref{lemma:two-sets-ordinary-lines} we get that $|\mathcal{Z}|\leq \log |\mathcal{L}|+1 = \log d +2$, completing the proof. 

\item $dim(sp(\mathcal{L}_{good}))\geq rank(f)-2$ and $\mathcal{L}_{others}\leq 2$: Define $V_i = \{\text{linear form }\ell :\ell\mid T_i \}$. We break the proof into two cases. Note that linear forms dividing $T_1,T_2$ satisfy first condition of Definition \ref{definition:candidate-linear-form}. So whenever we are trying to show that they belong to $\mathcal{L}(NonLin(f))$, we only prove that they satisfy second condition of Definition \ref{definition:candidate-linear-form}.
\begin{enumerate}
    \item First we discuss the case $dim(V_i) \geq \log d+5$ for all $i\in [2]$. Let $H$ be such that $T_1+T_2 = H\times NonLin(f)$. Let $\ell_1\mid T_1$ and $\Phi$ be isomorphism mapping $\ell_1\mapsto x_1$, then, we see that ${\Phi(T_2)}_{|_{x_1=0}} = \Phi(H)_{|_{x_1=0}}\times \Phi(NonLin(f))_{|_{x_1=0}} \neq 0$. Dimension of span of linear factors of ${\Phi(T_2)}_{|_{x_1=0}}$ is at least $\log d+4$ by assumption in this case. By previous part, $dim(sp(\mathcal{L}_{factors}))\leq \log d +2 \Rightarrow \Phi(NonLin(f))_{|_{x_1=0}}$ has two independent linear factors. Using these we can satisfy second condition of Definition \ref{definition:candidate-linear-form} for $\ell_1 \Rightarrow$ some scalar multiple of $\ell_1 \in \mathcal{L}(NonLin(f))$. The same argument can be repeated for a linear factor $\ell_2\mid T_2$. Thus all linear factors of $T_1\times T_2$ are in $\mathcal{L}(NonLin(f))$ (upto scalar multiplication) $\Rightarrow$ $dim(\mathcal{L}_{good}) = rank(f)$. This also implies that $dim(\mathcal{L}_{others}) = 0$.
    \item In the case when $dim(V_i) \leq \log d + 4$ for some $i\in [2]$, we know that $dim(V_{3-i}) = \Omega(\log^3d)$ and therefore by an argument similar to the one given in proof of Claim \ref{claim:corner-case-factors}, $NonLin(f) = T_1+T_2$. Consider any basis $\{\ell_1, \ldots,\ell_r\}$ of $V_1+V_2$. If $dim(V_i)\geq 3$ for all $i\in [2]$, then using a similar argument as before, we can show that all $\ell_i$ satisfy second condition in Definition \ref{definition:candidate-linear-form}  $ \Rightarrow dim(\mathcal{L}_{good}) = rank(f) \Rightarrow dim(\mathcal{L}_{others})=0$. In case for some $i\in [2]$, $dim(V_i) =2$ (recall we have assumed $dim(V_i)\geq 2$ in the statement of Claim \ref{claim:candidate-approximation}), then all linear forms dividing $T_{3-i}$ are not contained in $V_i$ and hence satisfy second condition of Definition \ref{definition:candidate-linear-form}. Thus $dim(\mathcal{L}_{good})\geq rank(f)-2$ and $dim(\mathcal{L}_{others})\leq 2$.
\end{enumerate}

\item $dim(sp(\mathcal{L}_{bad}))\leq \log d + 2$: Assume $dim(\mathcal{L}_{bad}) \geq \log d+3$. Consider the proper set $\mathcal{L}$ containing all linear factors of $T_1,T_2\Rightarrow |\mathcal{L}|\leq 2d \Rightarrow |\mathcal{L}_{bad}| \geq \log |\mathcal{L}| +2$. Let $\mathcal{T}\subset\mathcal{L}_{bad}$ be a linearly independent set of size $\log |\mathcal{L}| +2$. Then by Proposition \ref{propn:annihilating-planes}, there exists $t\in \mathcal{T}$ such that ordinary lines from $t$ into $\mathcal{L}$ span a space of dimension $\geq \frac{dim(sp(\mathcal{L}))}{\log |\mathcal{L}| +2} \geq \frac{rank(f)}{\log d +3} =\Omega(\log^2 d)$. Since $t\in \mathcal{L}_{bad}$, restricting $T_1+T_2$ to $\mathbb{V}(t)$ (see Definition \ref{defn:factorize-codim-1-subspace}) gives some non-zero product of linear factors, say $H$. Let $\Phi$ be an isomorphism mapping $t\mapsto x_1$. Then,
\[
{\Phi(T_1)}_{|_{x_1=0}}+{\Phi(T_2)}{|_{x_1=0}} - H =0
\]
This gives an identically zero $\Sigma\Pi\Sigma(3,n,d,\F)$ circuit. Since $t\in \mathcal{L}_{bad}$, it does not divide $T_1,T_2 \Rightarrow$ the above circuit is minimal (Definition \ref{defn:minimal}). After cancelling common linear forms from the three gates ${\Phi(T_1)}_{|_{x_1=0}}, {\Phi(T_2)}_{|_{x_2=0}}, H$, we have a simple (Definition \ref{defn:simple}) and minimal, identically zero $\Sigma\Pi\Sigma(3,n,d,\F)$ circuit.  The $\Omega(\log^2d)$ ordinary lines from $t$ into $\mathcal{L}$ imply that after cancelling the common linear forms, the simple minimal circuit has rank $ \Omega(\log^2d)$ which is a contradiction to Lemma \ref{lemma:sps(k)-rank-bound}. Thus we conclude that $dim(sp(\mathcal{L}_{bad})) \leq \log d +2$.

\end{enumerate}

\section{Proof of Lemma \ref{lemma:one-dimension-found}}
\label{section:proof-annihilating-lemmas}

 Let $T_i = \prod\limits_{j=1}^m\ell_{i,j}$ where $\ell_{i,j}$ are linear forms. We know that,
    \[
    \prod\limits_{j=1}^m{\Phi(\ell_{1,j})}_{|_{x_1=0, x_2=0}} = -\prod\limits_{j=1}^m{\Phi(\ell_{2,j})}_{|_{x_1=0, x_2=0}} \neq 0.
    \]
    Note that ${\Phi(\ell_{i,j})}_{|_{x_1=0, x_2=0}}$ can be thought of as linear forms over $\F$ in $n-2$ variables, and by using unique factorization of polynomials over $\F$, without loss of generality we can assume
    ${\Phi(\ell_{1,j})}_{|_{x_1=0, x_2=0}} = \beta_j {\Phi(\ell_{2,j})}_{|_{x_1=0, x_2=0}}$ for some $0\neq \beta_j \in \F$. This implies\footnote{since $\Phi$ is an isomorphism.} $U_j = sp\{\ell_{1,j}, \ell_{2,j}\}$\footnote{$\ell_{1,j}, \ell_{2,j}$ are linearly independent since $gcd(T_1,T_2)=1$} intersects $U = sp\{\ell_1, \ell_2\}$ non-trivially. Since ${\Phi(\ell_{i,j})}_{|_{x_1=0, x_2=0}}\neq 0$, we know that $U\neq U_j \Rightarrow U\cap U_j$ is $1$ dimensional\footnote{since both $U, U_j$ are $2$ dimensional}. 
    We split the proof into two cases:
    \begin{itemize}
        \item {\bf There exist two distinct spaces, say $U_i, U_j$ such that $U\cap U_i = U\cap U_j$ :} This implies $U\cap U_i \subset U_i\cap U_j$. The space $U_i\cap U_j$ is $1$ dimensional  since $U_i,U_j$ are distinct, say $U_i\cap U_j = sp\{\ell\}$. Both sides of the containment $U\cap U_i \subset U_i\cap U_j$ are $1$ dimensional implying $U_i\cap U_j = U\cap U_i \subset U = sp\{\ell_1, \ell_2\}$. This further implies that $\ell \in U \Rightarrow W\subset \mathbb{V}(\ell) = V$. There are $\leq d^4$ choices for such $U_i,U_j$ and therefore $d^4$ possibilities for such $V$.
        
        \item {\bf For all distinct $U_i, U_j$, $U\cap U_i \neq U\cap U_j$ :} Vector space $U\cap U_i + U\cap U_j$ is $2$ dimensional, since it is a sum of disjoint $1$ dimensional spaces. $U$ is also $2$ dimensional $\Rightarrow U = U\cap U_i + U\cap U_j \subset U_i + U_j$. Using statement of Proposition \ref{propn:annihilating-planes}, we know that 
        \[
        5 \leq rank(f) = dim(sp\{\ell_{i,j}\}) = dim(\sum\limits_{j=1}^m U_j) \leq \sum\limits_{j=1}^m dim(U_j).
        \]
        $dim(U_i + U_j) \leq 4$, thus there exists $U_k$ such that $U_k \not\subset U_i + U_j$. Note that this would imply  that $U_k\cap (U_i + U_j)$ has dimension $\leq 1$. Since $U\subset U_i + U_j$, we get that $U_k\cap U \subset U_k\cap (U_i+U_j)$. Both sides are $1$ dimensional. Writing $U_k\cap (U_i + U_j) = sp\{\ell\} \Rightarrow \ell \in U \Rightarrow W \subset \mathbb{V}(\ell) = V$. There are $\leq d^6$ choices for $U_i, U_j, U_k$ and so $\leq d^6$ possibilities for such $V$.
    \end{itemize}
    $\mathcal{A}$ is collection of all $V$'s obtained above. $|\mathcal{A}|\leq d^4+d^6$ and $\mathcal{A}$ satisfies the required conditions.

\section{Proofs of Lemmas in Algorithm \ref{algorithm:compute-annihilating-subspaces}}
\label{appendix:proofs-lemmas-algorithm-annihilating}
In this appendix, we provide proofs to lemmas that were stated and used in Algorithm \ref{algorithm:compute-annihilating-subspaces} but proofs were not provided.

\subsection{Proof of Lemma \ref{lemma:annihilating-spaces-details-1}}
We prove each part one by one below. Let $\hat{\ell}_i = \sum\limits_{j=1}^n\alpha_{i,j}x_j, i\in [n]$ be the $n$ linear forms that were constructed using the uniformly randomly independently samples $\alpha_{i,j}, i\in [n], j\in [n]$. Recall that $\Phi$ maps $x_i\mapsto \hat{\ell}_i$. Let $\Gamma$ be a homomorphism from $\F[x_1,\ldots,x_n]\rightarrow \F[x_1,x_2,x_3,x_4,x_i]$ that sets $x_5=0,\ldots,x_{i-1}=0, x_{i+1}=0, \ldots,x_n=0$.

\begin{enumerate}
    \item 
    Showing that these linear forms are independent is equivalent to showing that with probability $1-o(1)$, the matrix $(\alpha_{i,j})_{(i,j)\in [n]\times [n]}$ is invertible. This is equivalent to saying that the determinant polynomial of this matrix is non-zero. Applying Lemma \ref{lemma:szlemma} on the determinant polynomial, we get this result.
    
    \item Consider any isomorphism $\Psi$ mapping $\ell_1\mapsto x_1, \ell_2\mapsto x_2$, then $\Psi\circ \Phi^{-1}$ is an isomorphism mapping $\Phi(\ell_1)\mapsto x_1, \Phi(\ell_2)\mapsto x_2$. Further, $\Psi(NonLin(f)) = \Psi \circ \Phi^{-1}(\Phi(NonLin(f)))$. Restricting both sides to $x_1=0, x_2=0$ gives,
\[
\Psi(NonLin(f))_{|_{x_1=0,x_2=0}} = \Psi \circ \Phi^{-1}(\Phi(NonLin(f)))_{|_{x_1=0,x_2=0}},
\]
implying that $NonLin(f)$ vanishes on $\mathbb{V}(\ell_1,\ell_2)$ if an only if $\Phi(NonLin(f))$ vanishes on subspace $\mathbb{V}(\Phi(\ell_1), \Phi(\ell_2))$. Since $\Phi$ is an isomorphism, irreducible factors of $f$ remain irreducible on applying $\Phi$, thereby implying that $\Phi(NonLin(f))  = NonLin(\Phi(f)) = NonLin(g)$. Hence claim is proved.

    \item Recall $f = G\times (T_1+T_2)$, with $G,T_1,T_2$ being product of linear forms and $gcd(T_1, T_2)=1$. Since $\Phi$ is an isomorphism, we get that $g = \Phi(G) \times (\Phi(T_1) + \Phi(T_2))$. Since $\Phi$ is an isomorphism, $gcd(\Phi(T_1), \Phi(T_2))=1$.  Therefore we get,
    \[
    g_i = \Gamma(g) = \Gamma(\Phi(G))(\Gamma(\Phi(T_1)) + \Gamma(\Phi(T_2))).
    \]
    Next, consider linear forms $\ell = \sum\limits_{j=1}^n a_j x_j$ and $\ell^\prime = \sum\limits_{j=1}^n a_j^\prime x_j$ such that $\ell \mid T_1$ and $\ell^\prime \mid T_2$. Applying $\Phi$ to these linear forms we get, $\Phi(\ell) = \sum\limits_{k=1}^n \sum\limits_{j=1}^n  a_j\alpha_{j,k}x_k$. Therefore coefficients of $x_1, x_2$ in $\Gamma(\Phi(\ell))$ are  $\sum\limits_{j=1}^n  a_j\alpha_{j,1}, \sum\limits_{j=1}^n  a_j\alpha_{j,2}$ respectively and those in $\Gamma(\Phi(\ell^\prime))$ are $\sum\limits_{j=1}^n  a_j^\prime\alpha_{j,1}, \sum\limits_{j=1}^n  a_j^\prime\alpha_{j,2}$. We argue that vectors $(\sum\limits_{j=1}^n  a_j\alpha_{j,1}, \sum\limits_{j=1}^n  a_j\alpha_{j,2})$ and $(\sum\limits_{j=1}^n  a_j^\prime\alpha_{j,1}, \sum\limits_{j=1}^n  a_j^\prime\alpha_{j,2})$ are not scalar multiples with probability $1-o(1)$. This is equivalent to showing that the following determinant is non-zero.
    \[
    \begin{vmatrix}
    \sum\limits_{j=1}^n  a_j\alpha_{j,1} &  \sum\limits_{j=1}^n  a_j\alpha_{j,2}\\
    \sum\limits_{j=1}^n  a_j^\prime\alpha_{j,1} & \sum\limits_{j=1}^n  a_j^\prime\alpha_{j,2}
    \end{vmatrix}
    \]
If $\ell, \ell^\prime$ are not scalar multiples, this determinant is not an identically zero polynomial in the $\alpha_{j,k}, j\in [n], k\in [2]$ and therefore probability (over the random choices of $\alpha_{j,k}$) that the determinant is non-zero $=1-o(1)$. Therefore with probability $1-o(1)$, $\Gamma(\Phi(\ell))$ and $\Gamma(\Phi(\ell^\prime))$ are not scalar multiples. Since $\ell, \ell^\prime$ are arbitrary linear factors of $T_1, T_2$ respectively, by union bound with probability $1-o(1)$, $gcd(\Gamma(\Phi(T_1)), \Gamma(\Phi(T_2))) = 1$ implying that all $g_i$ exhibit $\Sigma\Pi\Sigma(2,5,d,\F)$ circuit. Since $rank(f) = \Omega(\log^2 d)$, we know that $dim(sp\{\text{linear form }\ell : \ell \mid T_1\times T_2\}) \geq 5$, therefore, a similar argument (again using Lemma \ref{lemma:szlemma}), can be used to say that $\{\Gamma(\Phi(\ell)) : \text{linear form }\ell\mid T_1\times T_2\}$ spans a $5$ dimensional  space. This set is the same as $\{\text{linear form }\ell : \ell \mid \Gamma(\Phi(T_1))\times \Gamma(\Phi(T_2))\})\}$, proving that $rank(g_i)=5$ for all $i\in [5,n]$.

\item By effective Hilbert's irreducibility theorem (Lemma \ref{lemma:effectivehilbert}), we know that with probability $1-o(1)$ over the $\alpha_{i,j}, i\in [n],j\in [n]$, the irreudicible factors of $\Phi(f)(x_1,\ldots,x_n) = f(\Phi(x_1), \ldots,\Phi(x_n))$ remain irreducible on setting $x_5 =0, \ldots, x_{i-1}=0, x_{i+1}=0, \ldots,x_n=0$ i.e. on applying $\Gamma$. Example if $h$ is an irreducible factor of $f$, then $\Gamma(\Phi(h))$ is an irreducible factor of $\Gamma(\Phi(f))$. The same will apply to product (with multiplicity) of all non-linear irreducible factors implying that,
\[
NonLin(\Gamma(\Phi(f))) = \Gamma(NonLin(\Phi(f))).
\]
The left hand side is same as polynomial $NonLin(g_i)$ and right hand side is same as polynomial $NonLin(g)_{|_{x_5=0,\ldots, x_{i-1}=0, x_{i+1}=0,\ldots,x_n=0}}$. Hence Proved.

\item  Let $\mathbb{V}(\hat{\ell}_1, \hat{\ell_2})$ belong to $  \mathcal{S}(NonLin(f))$. Assume $\hat{\ell}_1 = \sum\limits_{j=1}^n a_jx_j$ and $\hat{\ell}_2 = \sum\limits_{j=1}^n b_jx_j$, Then we get that $\Phi(\hat{\ell}_1) = \sum\limits_{k=1}^n (\sum\limits_{j=1}^n a_j \alpha_{j,k})x_k$ and $\Phi(\hat{\ell}_2) = \sum\limits_{k=1}^n (\sum\limits_{j=1}^n b_j \alpha_{j,k})x_k$. We define $c_k = \sum\limits_{j=1}^n a_j \alpha_{j,k}$ an $d_k = \sum\limits_{j=1}^n b_j \alpha_{j,k}$ for $k\in [n]$. Therefore $\Phi(\hat{\ell}_1) = \sum\limits_{k=1}^n c_k x_k$ and $\Phi(\hat{\ell}_2) = \sum\limits_{k=1}^n d_k x_k$. Now we define new linear forms as follows:

\begin{gather}
 \begin{bmatrix} \ell_3 \\ \ell_4 \end{bmatrix}
 =
  \begin{bmatrix}
   d_2 & -c_2\\
  -d1 & c_1
   \end{bmatrix}
   \begin{bmatrix} \Phi(\hat{\ell}_1) \\ \Phi(\hat{\ell}_2) \end{bmatrix}
\end{gather}

Determinant of the matrix is $d_2c_1-c_2d_1$. This defines a polynomial in the $\alpha_{j,k}, j\in [n], k\in [2]$. Like in the previous part, unless $\hat{\ell}_1, \hat{\ell}_2$ are linearly dependent this polynomial is not identically $0$. Therefore with probability $1-o(1)$ over the uniformly randomly chosen linear forms in Step $1$, the determinant is non-zero implying that $d_2c_1-c_2d_1 \neq 0$. This also means that $\ell_3, \ell_4$ are linearly independent and $\mathbb{V}(\ell_3, \ell_4) = \mathbb{V}(\Phi(\hat{\ell}_1), \Phi(\hat{\ell}_2))$. Analyzing $\ell_3, \ell_4$ we see that,
\[
\ell_3 = (d_2c_1-c_2d_1)x_1 + \sum\limits_{k=3}^n (d_2c_k - c_2d_k)x_k, and
\]
\[
\ell_4 = (d_2c_1-c_2d_1)x_2 + \sum\limits_{k=3}^n (d_kc_1 - c_kd_1)x_k.
\]
Define $\ell_1^\prime = -\sum\limits_{k=3}^n \frac{d_2c_k - c_2d_k}{d_2c_1-c_2d_1}x_k$, and $\ell_2^\prime = -\sum\limits_{k=3}^n \frac{d_kc_1 - c_kd_1}{d_2c_1-c_2d_1}x_k$, further implying that $\mathbb{V}(\ell_1, \ell_2) = \mathbb{V}(x_1-\ell_1^\prime, x_2-\ell_2^\prime)$ with $\ell_1^\prime, \ell_2^\prime\in \F[x_3,\ldots,x_n]$. Now since $\mathcal{S}(NonLin(f))$ has size $d^{O(1)}$, by union bound, with probability $1-o(1)$, we can prove all of this for every $\mathbb{V}(\hat{\ell}_1, \hat{\ell_2}) \in \mathcal{S}(NonLin(f))$. 

Now, given any $\mathbb{V}(\ell_1, \ell_2) \in \mathcal{S}(NonLin(g))$,
by Part $2$ of this Lemma, we know that $\mathbb{V}(\ell_1, \ell_2) \in \mathcal{S}(NonLin(g))$ if and only if 
$\mathbb{V}(\Phi^{-1}(\ell_1), \Phi^{-1}(\ell_2)) \in \mathcal{S}(NonLin(g))$. So we can use our argument for $\hat{\ell}_1 = \Phi^{-1}(\ell_1)$, and $\hat{\ell}_2 = \Phi^{-1}(\ell_2)$, thereby completing the proof.

\item Let $\mathbb{V}(\ell_1, \ell_2) \in \mathcal{S}(NonLin(g))$ and $\ell_j^i = {\ell_j}_{|_{x_5=0,\ldots, x_{i-1}=0, x_{i+1}=0,\ldots,x_n=0}}$. By previous part we know that there exists $\ell_1^\prime, \ell_2^\prime\in \F[x_3,\ldots,x_n]$ such that $\mathbb{V}(\ell_1, \ell_2) = \mathbb{V}(x_1-\ell_1^\prime, x_2-\ell_2^\prime)$. Let $\Theta$ be an isomorphism mapping $x_1-\ell_1^\prime\mapsto x_1, x_2-\ell_2^\prime\mapsto x_2$ and for $j\in [3,n]$, $x_j\mapsto x_j$. Similarly let $\Theta^\prime$ be isomorphism on $\F[x_1,x_2, x_3,x_4,x_i]$ mapping $x_1-\ell_1^{\prime\prime}\mapsto x_1, x_2-\ell_2^{\prime\prime}\mapsto x_2$ and for $j\in \{3,4,i\}$, $x_j\mapsto x_j$. Finally let $\Gamma$ be the homomorphism from $\F[x_1,\ldots,x_n]$ to $\F[x_1,x_2,x_3,x_4,x_i]$ mapping $x_j\mapsto 0$ for all $j\in [5,i-1]\cup [i+1,n]$. The following diagram commutes.
\[ \begin{tikzcd}
\F[x_1,\ldots,x_n] \arrow{r}{\Theta} \arrow[swap]{d}{\Gamma} & \F[x_1,\ldots,x_n] \arrow{d}{\Gamma} \\%
\F[x_1,x_2, x_3,x_4,x_i] \arrow{r}{\Theta^\prime}& \F[x_1,x_2, x_3,x_4,x_i]
\end{tikzcd}
\]
We know that $NonLin(g)$ vanishes on $\mathbb{V}(x_1-\ell_1^\prime, x_2-\ell_2^\prime)$, therefore $\Theta(NonLin(g))_{|_{x_1=0,x_2=0}} = 0$, implying that $\Gamma(\Theta(NonLin(g))_{|_{x_1=0, x_2=0}})=0$. We know that $\Gamma$ fixes $x_1, x_2$ therefore we can set $x_1=0, x_2=0$ after applying $\Gamma$, thereby giving $\Gamma(\Theta(NonLin(g)))_{|_{x_1=0, x_2=0}} = 0$. Using the above commutative diagram we get, $\Theta^\prime(\Gamma(NonLin(g)))_{|_{x_1=0,x_2=0}}=0$. Now, Part $4$ of this lemma gives $NonLin(g_i) = \Gamma(NonLin(g))$. Using this we get, $\Theta^\prime(NonLin(g_i))_{|_{x_1=0,x_2=0}} = 0$. Therefore $NonLin(g_i)$ vanishes on the co-dimension $2$ subspace $\mathbb{V}(x_1-{\ell_1}^{\prime\prime}, x_2-{\ell_2}^{\prime\prime})$ of $\F^5$, thereby completing the proof. 

\end{enumerate}
\subsection{Proof of Lemma \ref{lemma:annihilating-spaces-details-2}}
 Fix $i\in [6,n]$. Consider a pair of distinct tuples $(x_1-\ell_1, x_2-\ell_2)$, $(x_1-\ell_1^\prime, x_2-\ell_2^\prime)$ in $\mathcal{S}_i$. By construction, $\ell_1, \ell_2, \ell_1^\prime, \ell_2^\prime \in \F[x_3,x_4,x_i]$. So we assume that, $\ell_1 = a_3 x_3 + a_4x_4 + a_ix_i$, $\ell_2 = b_3x_3+b_4x_4+b_ix_i$, $\ell_1^\prime = a_3^\prime x_3 + a_4^\prime x_4 + a_i^\prime x_i$ and $\ell_2^\prime = b_3^\prime x_3 + b_4^\prime x_4 + b_i^\prime x_i$. Therefore,
 \[
 \begin{split}
     & \Delta(\ell_1) = (a_3 + \alpha_{i,3}a_i)x_3 + (a_4 + \alpha_{i,4}a_i)x_4 + a_ix_i,\\ 
     & \Delta(\ell_2) = (b_3 + \alpha_{i,3}b_i)x_3 + (b_4 + \alpha_{i,4}b_i)x_i + b_ix_i,\\
     & \Delta(\ell_1^\prime) = (a_3^\prime + \alpha_{i,3}a_i^\prime)x_3 + (a_4^\prime + \alpha_{i,4}a_i^\prime)x_4 + a_i^\prime x_i,\\
     & \Delta(\ell_2^\prime) = (b_3^\prime + \alpha_{i,3}b_i^\prime)x_3 + (b_4^\prime + \alpha_{i,4}b_i^\prime)x_4 + b_i^\prime x_i
 \end{split}
 \]
If $(\Delta(\ell_1)_{|_{x_i=0}}, \Delta(\ell_2)_{|_{x_i=0}}) = (\Delta(\ell_1^\prime)_{|_{x_i=0}}, \Delta(\ell_2^\prime)_{|_{x_i=0}})$, then we get a system of linear equations in $ \alpha_{i,3}, \alpha_{i,4}$ which can be simplified to get

\[
  \begin{bmatrix}
  \alpha_{i,3}(a_i-a_i^\prime)\\
  \alpha_{i,4}(a_i-a_i^\prime)\\
  \alpha_{i,3}(b_i-b_i^\prime)\\
  \alpha_{i,4}(b_i-b_i^\prime)
  \end{bmatrix}
  = 
  \begin{bmatrix} 
  a_3^\prime - a_3 \\ 
  a_4^\prime - a_4 \\ 
  b_3^\prime - b_3 \\ 
  b_4^\prime - b_4 
  \end{bmatrix}
\]
Since tuples $(\ell_1, \ell_2)$ and $(\ell_1^\prime, \ell_2^\prime)$ are distinct, at least one of $(a_3^\prime-a_3), (a_i-a_i^\prime), (a_4^\prime - a_4), (b_i-b_i^\prime), (b_3^\prime - b_3), (b_4^\prime - b_4)$ is non-zero implying that at least one linear equation is not identically zero. By Lemma \ref{lemma:szlemma}, we then know that with probability $1-o(1)$ over the uniformly random choices of $\alpha_{i,3}, \alpha_{i,4}$ the equation cannot be zero. Therefore with probability $1-o(1)$, $(\Delta(\ell_1)_{|_{x_i=0}}, \Delta(\ell_2)_{|_{x_i=0}}) \neq (\Delta(\ell_1^\prime)_{|_{x_i=0}}, \Delta(\ell_2^\prime)_{|_{x_i=0}})$. Using Part $3$ of Lemma \ref{lemma:annihilating-spaces-details-1}, we know that $rank(NonLin(g_i))=5$ implying that $|\mathcal{S}_i| = d^{O(1)}$. So we can take a union bound over all pairs of tuples in $\mathcal{S}_i$. Finally, we take a union bound over all $i$ and guarantee that with probability $1-o(1)$, the statement in this lemma holds.

\end{document}